\documentclass[12pt]{article}
\usepackage{amsmath,amsfonts,amsthm,amssymb,amscd}
\usepackage[running,displaymath,mathlines]{lineno}
\usepackage[toc,page]{appendix}
\numberwithin{equation}{section}

\newcommand{\be}{\begin{equation}}
\newcommand{\ee}{\end{equation}}
\newcommand{\bs}{\begin{split}}
\newcommand{\es}{\end{split}}
\newcommand{\ba}{\begin{align}}
\newcommand{\ea}{\end{align}}
\newcommand{\basl}[1]{\begin{align}\begin{split}\label{#1}}
\newcommand{\bas}{\begin{align}\begin{split}}

\newtheorem{theo}{Theorem}[section]
\newtheorem{prop}[theo]{Proposition}
\newtheorem{lemm}[theo]{Lemma}

\newtheorem{defi}[theo]{Definition}

\newcommand\fpr{\hfill$\Box$\null}

\newcommand\N{\mathbb{N}}

\newcommand\R{\mathbb{R}}
\newcommand\C{\mathbb{C}}

\title{ Time evolution for the Pauli-Fierz operator \\
(Markov approximation  and Rabi cycle)  }

\author{ L. Amour, J. Nourrigat}

\date{Laboratoire de Math\'ematiques de Reims - UMR CNRS 9008  \\ \vskip 0.15cm Universit\'e de Reims Champagne-Ardenne, France}

\begin{document}


\maketitle

\begin{abstract}

\

\noindent

This article is concerned with a system of particles interacting with the quantized electromagnetic field (photons) in the non relativistic Quantum Electrodynamics (QED) framework and governed by the Pauli-Fierz Hamiltonian.
We are interested not only in  deriving approximations of  several quantities when the coupling constant is small but also in obtaining different controls of the error terms.
First, we investigate the time dynamics approximation in two situations,  the Markovian (Theorem 1.4 completed by Theorem 1.16) and non Markovian (Theorem 1.6) cases. These two contexts differ in particular regarding the approximation leading terms, the error control and the initial states.
Second, we examine two applications. The first application is the study of marginal transition probabilities related to those analyzed by Bethe and Salpeter in \cite{B-S}, such as proving the exponential decay in the Markovian case assuming the Fermi Golden Rule (FGR) hypothesis (Theorem 1.17 or Theorem 1.15) and obtaining a FGR type approximation in the non Markovian case (Theorem 1.5). The second application, in the non Markovian case, includes the derivation  of Rabi cycles  from QED (Theorem 1.7). All the results  are established under the following assumptions at some steps of the proofs: an ultraviolet and an infrared regularization are imposed, the quadratic terms of the Pauli-Fierz Hamiltonian are dropped, and the dipole approximation is assumed but only to obtain  optimal error controls.
\end{abstract}
\parindent=0pt

\

\


\

\parindent=0pt
\parindent = 0 cm

\tableofcontents

\parskip 3pt
\baselineskip 17pt

 \section{Introduction and results.}
We consider in this article the Pauli-Fierz Hamiltonian in the non relativistic Quantum Electrodynamics (QED) \cite{BFS98a,B-F-S,B-F-S-99}  (see also \cite{F73}).
The time evolution of a system of one or several non  relativistic moving quantum particles in interaction with the quantized electromagnetic field (photons) can be described  by the Pauli-Fierz  Hamiltonian operator. This operator is depending on a positive real parameter denoted here $g$, the coupling constant, related to the electric charge and supposed small \cite{B-F-S,B-F-S-99,IMS}. Our objective  throughout this paper consists in deriving different time evolution approximations  when the coupling parameter $g$ is small together with a control of the error terms.

 Bethe and  Salpeter \cite{B-S} study a marginal transition  probability notion,   from particle energy level states  towards lower level states, at each time $t$, when the particle and photon states are initially in  the photon  vacuum.
We shall prove that these marginal transition  probabilities are well approximated by Markov processes (as the coupling parameter $g$ tends to zero). This is the first application of the time evolution approximation proved in this work  with different error controls.

The second   application derived in this paper  is the time dynamics approximation with error control when initial states  are superpositions of the preceding states (particle energy level states) with different energy levels, with {at least} two states and still in the photon vacuum. In that case, we observe approximate periodic time evolutions, very likely related to Rabi cycles. It is a non Markov approximation.

The proofs of these approximations and  the error control results will exploit technical tools sometimes comparable to the methods developed for studying open quantum systems (such as Lindblad operators).

In this work,   ultraviolet and infrared regularizations are imposed and the square order terms in the Pauli-Fierz Hamiltonian are dropped  in order to get the results.

The dipole  approximation is also required  when a very precise control of the error is intended
in the Markov approximation. It is not used for the Rabi cycle.

\subsection{Pauli-Fierz Hamiltonian (simplified).}

  The Hilbert space of the model is the completed tensor product of the two Hilbert spaces of the elements constituting the system, that is, the quantized electromagnetic field (photons) and matter (particles):
 $$ {\cal H} _{\rm tot} =  {\cal H} _{\rm ph}  \otimes   {\cal H} _{\rm mat}. $$

The photon state Hilbert space denoted here by ${\cal H} _{\rm ph}$ is the the symmetrized  Fock space ${\cal F} _s ({\cal H }_{\rm ph}^{(1)})$ over the single photon state Hilbert space  ${\cal H }_{\rm ph}^{(1)}$ (See \cite{RS}, Volume II). The space ${\cal H }_{\rm ph}^{(1)}$ is the divergence free in momentum variables of vector fields $u$ in $L^2 (\R^3, \R^3 ) $, namely,
    $k\cdot u(k) = 0$ almost everywhere for $k\in\R^3$. This takes account of the photon polarization (see \cite{L-L}). The subspace ${\cal H} _{\rm ph}^{\rm reg}$
of  ${\cal H} _{\rm ph}$ stands for the set of all finite linear combinations of tensor products of elements of  ${\cal H }_{\rm ph}^{(1)}$  belonging to ${\cal S} (\R^3, \R^3)$.
The photon vacuum state in the Fock  space is denoted by $\Psi_0$.

We denote  by  $k \mapsto a(k)$ (annihilation operator) the map from $\R^3$   into
  the set of linear mappings  ${\cal L}({\cal H }_{\rm ph}^{\rm reg},({\cal H }_{\rm ph}^{\rm reg})^3 )$ defined almost everywhere by,
    \be\label{def-a-k}  a(k) ( u_1 \otimes \dots\otimes  u_m )= \sqrt {m} \  u_1 (k) ( u_2 \otimes \cdots \otimes u_m ),\ee
for non zero integers $m$ and $0$ when $m=0$. See \cite{RS} for more details.

We recall that the free photon Hamiltonian is a nonnegative self-adjoint operator  $( H_{\rm ph} , D(H_{\rm ph}))$
   in the Hilbert space  ${\cal H} _{\rm ph}$. Also,   ${\cal H }_{\rm ph}^{\rm reg}\subset D(H_{\rm ph})$ and we have,

    \be\label{def-Hph}  < H_{\rm ph} f , g > = \int _{\R^3 } |k| < a(k) f , a(k) g > dk, \ee
 for all $f,g$ in ${\cal H }_{\rm ph}^{\rm reg}$. See also \cite{CTD01,CTD02}.

 We set  ${\cal H} _{\rm mat}= L^2(\R^3)$  for the space of matter (quantum) states.
 The standard Pauli-Fierz operator is written as,
   $$ H_1(g) =  \sum _{j= 1} ^3 ( D_j - g A_{j} (x) )^2 +  I \otimes V(x) + H_{\rm ph} \otimes I, $$
  where $H_{\rm ph}$ is the above photon Hamiltonian, $V$ stands for the electric potential and the  $ A_{j}(x)$ are operators  defined below, acting in ${\cal H}_{\rm ph}$ and depending on the the position variable  $x\in \R^3$. Also, $D_j$ is the derivative with respect to $x_j$, $j=1,2,3$, multiplied by the factor $-i$ and $I$ denotes identity operators in the photon or matter space.

The coefficient $g$ will be considered as a parameter going to zero and since we are interested in obtaining an asymptotic expansion in power of the small parameter $g$, we  choose here to omit the terms of order $g^2$ in the above Pauli-Fierz Hamiltonian $H_1(g)$, that is, to drop the square of the operators $ A_{j} (x) $, $j=1,2,3$.
Note that, this simplification is also effectuated in different works  concerning the Pauli-Fierz operator, see {\it e.g.}, \cite{D-J}. Thus,
    \be\label{def-Hg}  H(g) =   H_{\rm ph} \otimes I + I \otimes H_{\rm mat} + g H_{\rm int},\ee
 with  $ H_{\rm ph}$ being the the photon Hamiltonian, $H_{\rm mat}$ the matter Hamiltonian, that is, the Schr\"odinger operator,
  \be\label{2C-ham}  H_{\rm mat} = - \Delta + V(x) \ee 
 and  $H_{\rm int}$ is the simplified interaction Hamiltonian,
 \be\label{def-H-int}  H_{\rm int} =   -2 \sum _{j= 1}^3 A_j (x)  D_j. \ee 
 See below for details on  $H_{\rm mat}$, $A_j(x)$  and  $H_{\rm int}$.

   \vskip 2mm

    {\bf Matter Hamiltonian.}

       \vskip 2mm

   We denote by  $H_{\rm mat}$
the self-adjoint extension in $L^2(\R^3)$ of the differential operator
defined in (\ref{2C-ham}), where  $V$ is a  $C^{\infty}$ function defined 
on  $\R^3$ and taking real values, bounded together with all of its derivatives  
and tending to zero at infinity.
Choose a real number   $E_0 < 0 $ not lying in the spectrum of
 $H_{\rm mat}$. Using the smoothness of the potential $V$ and its vanishing at infinity, we observe from
 Cwikel-Lieb-Rosenbljum Theorem  (\cite{RS}, Volume IV) that the spectrum of  $H_{\rm mat }$ restricted to $(-\infty , E_0)$ is discrete. It is a finite set of eigenvalues of finite multiplicity and the lowest eigenvalue is simple
  (\cite{RS},  Volume IV, Theorem XIII.46). The set $ S _{\rm inf}$ refers  as the spectrum of $H_{\rm mat }$ restricted to the interval $(-\infty , E_0)$. We define $ {\cal H}_{\rm inf}$ as the spectral subspace of  $H_{\rm mat }$ associated with $(-\infty , E_0)$. Similarly, $ S _{\rm sup}$ is standing for the the spectrum of $H_{\rm mat }$ restricted to $ [ E_0, + \infty )$ and  ${\cal H} _{\rm sup}$ for  the spectral subspace of  $H_{\rm mat }$ associated with $ [ E_0, + \infty )$. According to Agmon inequalities \cite{A82} and since  $E_0$ is (strictly) smaller than  the limit at infinity of the potential $V$,  we notice that ${\cal H}_{\rm inf}$ is included in  the Schwartz space
  ${\cal S}(\R^3)$.
There exists $C>0$ such that $H_{\rm mat } + C I \geq 0$ and  the domain of the operator $( H_{\rm mat } + C I ) ^{m/2}$ is the standard
   Sobolev space for all integers   $m\geq 0$.  These Sobolev spaces are denoted by $W_m^{\rm mat}$ throughout the rest of the article for consistency with the other domain notations, namely,  
  $$ W_m^{\rm mat} = \{ u \in L^2(\R^3),\   D^{\alpha }_ju \in L^2 (\R^3),\  j=1,2,3,\ |\alpha | \leq m \},$$
  for $m\in\N$.

   \vskip 2mm

     {\bf Interaction operator.}

        \vskip 2mm

The   $A_j (x)$ are unbounded operators in
    ${\cal H} _{\rm ph}$, for each $x\in \R^3$ and $j=1,2,3$ (electromagnetic vector potential). For all $x\in \R^3$ and $j=1,2,3$, the  $A_j (x)$ can be formally written as,    
    $$ A_j (x) = \int _{\R^3}  ( a(k) \otimes A_j (x , k) ^{\star }
    +  a^{\star }(k) \otimes A_j (x , k)  ) dk $$
    where 
    $$ A_j (x , k) =  \frac {\phi (|k|)} {|k|^{1/2} } 
    e^{-i k\cdot x} \pi _{k^{\perp} } (e_{j }). $$
    See \cite{RS} (Volume II)  for   creation and annihilation operators $a^{\star} (k)$ and $a(k)$ (see also \cite{G-R}).
    Here,  $(e_1 , e_2 , e_3)$ is the canonical basis of $\R^3$ and $ \pi _{k^{\perp} }$ stands for the orthogonal projection on the  set orthogonal to $k$, for any non zero  $k\in\R^3$.
The real-valued function $\phi $ is the ultraviolet smooth cut-off and is taken in the Schwartz space ${\cal S}(\R)$.

    The   photon-matter interaction is the unbounded operator acting in $ {\cal H} _{\rm tot}$
 given formally by (\ref{def-H-int}).

    In order to define more easily $H_{\rm int}$, it is convenient to introduce the following function $k \mapsto E(k)$ on $\R^3$, taking values in the set of (unbouded) operators in $H_{\rm mat}$ and defined by,
    $$ E(k) = \sum _{j = 1}^3 A_j (\cdot , k) D_j,$$
    for all  $k\in \R^3$.
 This function   $E(\cdot)$ is often called {\sl form factor} \cite{D-J,SP}.
 Then, two expressions for $E(\cdot)$ are considered here. The first one is close to the standard
 Pauli Fierz operator,

 \be\label{stand}  (E(k)f) (x) =  \frac {\phi (|k|)} {|k|^{1/2} }
 \sum_ {\alpha = 1}^3 e^{-i k\cdot x}
 \pi _{k^{\perp} } (e_{\alpha })
 D_{{\alpha } }f(x), \ee
  and the second one called  {\sl dipole approximation} \cite{SP} will be required for the proof of some propositions in the sequel,
 \be\label{dipol} (E(k)f) (x) =  \frac {\phi (|k|)} {|k|^{1/2} }
 \sum_ {\alpha = 1}^3
 \pi _{k^{\perp} } (e_{\alpha })
 D_{{\alpha } }f(x).\ee
Note that the factor $e^{-i k\cdot x}$ is replaced by the factor $1$ in the dipole  approximation.

We can equally use  (\ref{stand}) or  (\ref{dipol}) in order to obtain the approximation expressions  themselves for the
 time dynamics that we consider (Markovian, non Markovian and Rabi). That is, the dipole  approximation is not involved to get the time dynamics approximations themselves but it is needed in order to 
 have a better control of  the error terms.

Then, in the two cases, the photon-matter interaction is defined by the following quadratic form $ Q_{\rm int}$ on the algebraic tensor product ${\cal H} _{\rm ph} ^{\rm reg} \otimes {\cal S}(\R^3)$,
   \be\label{Q-int} Q_{\rm int} (f , g) = \int _{\R^3 } ( < ( a(k) \otimes I )  f ,(I \otimes E(k))  g >
     + < ( I  \otimes E (k)) f , ( a(k) \otimes I) g > ) dk, \ee
        for all $f$ and $g$ in ${\cal H} _{\rm ph} ^{\rm reg} \otimes {\cal S}(\R^3)$. The above scalar product is the  $({\cal H} _{\rm tot})^3$ scalar product.

  \begin{theo}\label{Kato-Rell}  We have the following properties under the above hypotheses.

    i) The operator $H(0)$ defined in (\ref{def-Hg}) for $g= 0$
     has a unique  self-adjoint extension. Its  domain is denoted
      $W_2^{\rm tot}$ throughout the paper.  There exists a real number   $C>0$  satisfying $H(0) + C I \geq 0$. We denote by $W_m ^{\rm tot}$ the domain of the operator $ (H(0) + CI ) ^{m/2}$
    for any  $m\geq 0$.
     The operator $e^{it H(0)}$ is bounded in
       $W_m ^{\rm tot}$ independently of  $t\in\R$.

    ii) Let $ Q_{\rm int}$ be  the above quadratic form on ${\cal H} _{\rm ph} ^{\rm reg} \otimes {\cal S}(\R^3)$ with the form factor $E(k)$ defined
in either   (\ref{stand}) or  (\ref{dipol}).
Then there exists an   operator
   $H_{\rm int}$ bounded from $W_2^{\rm tot} $
   into  ${\cal H} _{\rm tot}$ satisfying,   
   $$ Q_{\rm int} (f , g) = < H_{\rm int}f , g >, $$
   for all $f$ and $g$ in ${\cal H} _{\rm ph} ^{\rm reg} \otimes {\cal S}(\R^3)$.

   Moreover, this operator is bounded
    from $W_{p+2}^{\rm tot}$ into $W_{p}^{\rm tot}$ for all $p\geq 0$.

iii)  The operator $H(g)$ with domain  $W_2^{\rm tot}$
    is self-adjoint for sufficiently small $g$. There exists a real number  $C_1$ such that $H(g) + C_1 I \geq 0$.
  The domain of the self-adjoint operator
    $( H(g) + C_1 I)^{m/2}$ is denoted by  $W_m^{\rm tot}$  for each $m\geq 0$.
    The operator $e^{it H(g)}$ is bounded in
       $W_m ^{\rm tot}$ independently of  $t\in\R$ for all small enough coupling parameter $g$.

 \end{theo}

 Theorem \ref{Kato-Rell} is  proved in  the Appendices B and C.  In particular, we observe that  the operators $e^{it H(0)}$ and $e^{it H(g)}$ are uniformly
 bounded in the spaces $W_m^{\rm tot}$, $m\geq 0$, for any coupling parameter $g$ sufficiently small. Also, we formally write,
 \be\label{H-int-def}  H_{\rm int} = \int _{\R^3}   ( a(k) \otimes E^{\star} (k)
 + a^{\star} (k)  \otimes E (k))dk. \ee
 See also \cite{H02} for self-adjointness results for the Pauli-Fierz operator.

\subsection{Statement of  results.}

We are concerned in this work with values  of quadratic forms   on the evolution states $e^{ - i t H(g)} F$, that is with $ < Z e^{ - i t H(g)} F , e^{ - i t H(g)} F >$, rather than $e^{ - i t H(g)} F$ itself, where $F$ is the initial state and $Z$ is a self-adjoint operator that can be chosen bounded.
 Moreover, we concentrate on operators $Z$ for the matter (particle) dynamics only, and thus for $Z$ written as  $Z = I \otimes X$ with $X$ operator in ${\cal H} _{\rm mat}$. Also, we focus on initial states $F$ only being in  the photon vacuum, namely,
$F = \Psi_0 \otimes u$ with $\Psi_0$ being the vacuum in  ${\cal H} _{\rm ph}$
 and $u\in {\cal H} _{\rm mat}$.

 As a consequence of the foregoing points, we are led to the following definition.  For all $t>0$ and $g>0$, for each $X\in {\cal H}_{\rm mat}$, we denote by $S^{\rm mat} (t , g )  X$ the operator in
${\cal H}_{\rm mat}$  defined by,
\be\label{S-mat}  <  ( S^{\rm mat} (t , g )  X  ) u, v > =  < (I \otimes X) e^{-it H(g)} (\Psi_0 \otimes u),
  e^{-it H(g)} (\Psi_0 \otimes v) > ,\ee
  for all $u$ and $v$ in ${\cal H}_{\rm mat}$.

We are then interested in deriving several approximations of  $S^{\rm mat} (t , g )  X$
as $g$ tends to zero. Let us mention at this stage another reduced time dynamics definition in   \cite{SP} (Chapter 17, Section 2).

Concerning spectral issues instead of time evolution problems, we refer to the works of  \cite{B-F-S,B-F-S-99} for the relation between Schr\"odinger eigenvalues and Pauli-Fierz resonances, which is obtained without any infrared regularization.

The work of Breit \cite{BR} for relativistic corrections is similar to some of  our results here. In \cite{BR}, starting from relativistic considerations, the observable time evolution is derived and is a correction in $1/c$ (where $c$ is the speed of light) of the two-body Heisenberg-Pauli equation.
This correction is also summarized in \cite{B-S} (page 181).  See also \cite{CFO,A-J-N-3} for related issues concerning Pauli-Fierz Hamiltonians. For spectral problems in the non relativistic limit and in particular for the relation between Schr\"odinger eigenvalues and Dirac  resonances, see, {\it e.g}, \cite{Hun,Bal-Hel-1,Bal-Hel-2,Par-1,Par-2,A-B-N}.

We also mention the works \cite{Haa73,Dav,Dav76,H-S,SP,RH12,VH55,VH57,Aga}
 for  models similar to the Pauli-Fierz operator and simpler model such as the spin-boson model.

\subsubsection{Markovian approximation.}

 A first result in the direction of Markovian approximations for this system of particles framework is established in \cite{B-S}
 (also \cite{W}) partly dedicated to marginal transition probabilities. Let us give our definition of these probabilities studied in the sequel. Recall that for any two unit vectors
 $F$  and $G$ in ${\cal H} _{\rm tot}$, the transition probability from $F$ to $G$  is commonly given by the scalar product $  | < e^{-it H(g)} F , G > |^2$.
  Then, if   $u$ and $v$ are two unit elements of  ${\cal H}_{\rm inf}$, we can call  {\it marginal transition probabilities from $u$ to $v$ with initial photon vacuum} (at time $t>0$), the expression,
   $$ P(t, g , u, v) =  \sum _{\alpha } |   < e^{-it H(g)} (\Psi_0 \otimes u )
   , (e_{\alpha } \otimes v)  > |^2, $$
with  $(e_{\alpha })$ being an Hilbertian basis of  ${\cal H} _{\rm ph}$.

Then,  we see that,
  $$ P(t, g, u, v) = < e^{it H(g)} ( I \otimes \pi_v) e^{-it H(g)}
   (\Psi_0 \otimes u ) , (\Psi_0 \otimes u ) >, $$
 with  $\pi_v$ standing for the orthogonal  projection  in ${\cal H} _{\rm mat}$ on
the line  spanned by  the vector  $v$.  That is, we have,

\be\label{P-trans}  P(t, g, u, v) = <  ( S^{\rm mat} (t , g) \pi_v  ) u , u >. \ee
Choose temporarily as in \cite{B-S}, an Hilbertian basis $(u_j)$ of ${\cal H} _{\rm inf}$, with each $u_j$ being an  eigenfunction of  $H_{\rm mat}$,
  $$ H_{\rm mat} u_j = \mu_j u_j, $$
  where $\mu_j \in S_{\rm inf}$. Bethe and Salpeter \cite{B-S} seem to consider that the matrices
  $ P(t, g , u_j, u_m)$ are defining a Markov  process and compute
its infinitesimal  generator. More precisely, a matrix  $L = ( L_{jm})$  is defined in \cite{B-S} and the transition marginal probabilities seem  to satisfy in \cite{B-S},
  $$ P(t, g, u_j, u_m)=  ( e^{- t L}  ) _{jm}.$$
 Then, \cite{B-S} provides a numerical computation of the matrix $L$ in the case of the hydrogen atom for the $N$ lowest eigenvalues of
 $H_{\rm mat}$.

 Therefore, our first objective here is precisely  to prove that this idea of Bethe and Salpeter is approximatively accurate, and in addition, to control the error.

 For that purpose, we shall prefer to state the result in a form that is not depending on a particular choice of a basis of    ${\cal H}_{\rm inf}$. To this end, we use the definitions and notations given in  Definition \ref{K-P} below.

     \begin{defi}\label{K-P} We denote by
${\cal L} ({\cal H}_{\rm inf})$ the set of  operators in ${\cal L} ( {\cal H} _{\rm mat})$
vanishing in  ${\cal H}_{\rm sup}$ and mapping ${\cal H}_{\rm inf}$ into itself.
Similarly, we also use the notation
${\cal L} ({\cal H}_{\rm sup})$ for the set of operators in ${\cal L} ( {\cal H} _{\rm mat})$
which vanish on  ${\cal H}_{\rm inf}$ and map ${\cal H}_{\rm sup}$ into itself.
The  set ${\cal K}$  stands for the algebra of  operators $X\in {\cal L} ( {\cal H} _{\rm inf})$
commuting with  the restriction  of $H_{\rm mat}$ and endowed with
 the restriction of the ${\cal L} ( {\cal H} _{\rm mat})$ norm.
 We denote by $\Pi (\mu)$  the orthogonal projection  on  $E(\mu )  = {\rm Ker} (H_{\rm mat} - \mu I)$,
for every $\mu \in S_{\rm inf}$.
We define ${\cal P}_{\cal K} : {\cal L}( {\cal H} _{\rm mat}) \rightarrow
{\cal K}$ as the  projection given by,
\be\label{proj-P} {\cal P}_{\cal K}  X = \sum _{\mu \in { S} _{\rm inf}} \Pi (\mu) X \Pi (\mu),\  X  \in {\cal L} ( {\cal H} _{\rm mat}).\ee

 \end{defi}
 We use the following Definition for a semigroup to be Markovian.

      \begin{defi}\label{def-Markov}   Let   ${\cal K}$ be any
unital $C^{\star}$-algebra with the unit denoted by $I$. Let ${\cal K}_{\R}$
be the space of self-adjoint elements of ${\cal K}$.
We say that a semigroup $G(t)$ ($t\geq 0$) acting
 in ${\cal K}_{\R}$ is a  Markov semigroup if the following properties are satisfied:

1.  $G(t)$ ($t\geq 0$) is a contraction semigroup in ${\cal K}_{\R}$.

2. We have $G(t) I = I$,  for all $t>0$.

3. If  $X\in {\cal K} $ is  self-adjoint and nonnegative then $G(t)X$
is also self-adjoint and nonnegative,  for all $t>0$.

\end{defi}
Note that a related class of semigroups called dual dynamic semigroups is studied by 
 Kossakowski \cite{Kos72}.

Also, we shall use the  hypothesis below on the photon-matter interaction in the statement of the results.

{\bf Hypothesis (FGR). } There is $\gamma  >0$ such that,
$$ \sum _{\mu \in S_{\rm inf} \atop \mu < \lambda } \int _{|k| = \lambda - \mu }
\Vert \Pi (\mu ) E (k) f \Vert ^2 d \sigma (k) \geq \gamma \Vert f \Vert ^2 ,$$
 $d \sigma  $ being the surface measure on spheres, for
all $\lambda \in S_{\rm inf}$ excepted for the infimum of the spectrum of  $H_{\rm mat}$ and for any $f\in {\rm Ker } ( H_{\rm mat} - \lambda I)$.

Note that a similar hypothesis appears in  \cite{B-F-S,FS14,DGK15} for various reasons.

Our first result is the Markovian approximation of the time dynamics.

\begin{theo}\label{approx-SG-prov} Let  the form factor $E(k)$ be given
either by (\ref{stand}) or by (\ref{dipol}). Suppose  that the function
 $\phi$ in (\ref{stand}) or in (\ref{dipol})  is vanishing at the origin and assume  that
the hypothesis (FGR) is satisfied. Then, there exists a Markov semigroup $G(\cdot )$ in ${\cal K}$ such that,

i) If the form factor $E(k)$ is defined by (\ref{stand}) then,
$$ \Vert {\cal P} _{\cal K}   ( S^{\rm mat} (t , g) X  ) -    G(t g^2) X \Vert
 \leq C g (1+ t^2) \Vert X \Vert,  $$
for some $C>0$, for all $X\in {\cal K}$,
for every $t>0$ and for any $g>0$ sufficiently small.

ii) If the form factor $E(k)$ is defined by (\ref{dipol}) then
we have, with the same conditions,
$$ \Vert {\cal P} _{\cal K}   ( S^{\rm mat} (t , g) X  ) -    G(t g^2) X \Vert
 \leq C g  \Vert X \Vert.  $$

\end{theo}

The definition and various properties of the semigroup $G(\cdot)$ in Theorem \ref{approx-SG-prov}  are postponed to Section 1.5 after stating additional notations. Theorem \ref{approx-SG-prov} will be restated as Theorem \ref{approx-SG} once these informations become available.

The starting point of the proof of Theorem  \ref{approx-SG-prov}
    is the  differential system (\ref{Sys-Dif-Init})-(\ref{defi-R2}) satisfied by $ S^{\rm mat} (t , g) X$.
    In this system, the two error terms are estimated in Sections 4.1 and 4.2.
    The main term has an approximation defined in Section 1.4. After these approximation issues,
    we get an approximate system directly related to the infinitesimal generator
    and obtain the  exponential behavior with the hypothesis (FGR).
    Then, the   Duhamel principle is involved in order to estimate the error between the evolution and its approximation by $G(tg^2)$ (Section 4.4).  If one uses the dipole approximation
    (form factor $E(k)$ defined by (\ref{dipol})) then one has a better estimate of the error term
    (a bound independent of $t$).

Next, we turn to     marginal transition probabilities. For each eigenvector $u_m$
associated with an eigenvalue in $S_{\rm inf}$, the orthogonal
projection $\pi _{u_m}$ is in ${\cal K}$.
As a consequence of  Theorem  \ref{approx-SG-prov}, we have,
   $$ \left |  P(t, g, u_j, u_m) - < ( G(tg^2 ) \pi _{u_m}u_j,u_j> \right |
   \leq  Cg, $$
  and since   the semigroup $G(\cdot )$ is defined as $G(t) = e^{- t {\cal L} }$,
 where  ${\cal L }$ is an operator in ${\cal K}$, we then see that,
 $$ \left |  P(t, g, u_j, u_m) -  (  e^{ - tg^2 {\cal L} } )  _{jm} \right |
   \leq  Cg.  $$
  That is to say,  the picture of  Bethe and  Salpeter described above and  concerning marginal transition probabilities is accurate up to an
  ${\cal O}(g)$ term. It is highly likely that the matrix
  computed in \cite{B-S} (page 266, table 15) for the hydrogen atom
  actually  is the matrix multiplied by the factor $g^2$ in some basis  of the operator ${\cal L}$ that we define (see Definition \ref{def-L} below).

Let us mention another consequence of
 Theorem \ref{approx-SG-prov}.
   If $\gamma >0$ is the constant in the hypothesis (FGR)
  and if  $ 0 < \delta  < \gamma $ then we see in Proposition \ref{relax}
 that the two  following properties hold true:

   If the state $u_m$ is orthogonal the ground state $u_0$ then we have,
    $$  \left | P(t , u_j, u_m , g )   \right | \leq C e^{-\delta g^2 t} + C g, $$
    If $u_m$ is the ground state $u_0$ then,
    $$   \left | P(t , u_j, u_0 , g ) -1   \right | \leq C e^{-\delta g^2 t} + C g. $$
  This can be viewed as the  relaxation to the ground state property.     Return to equilibrium for Pauli-Fierz is also studied in, {\it  e.g.},  \cite{A-J-N-2,BFSRE,DH1,DH2,DH3,D-J} and in \cite{Mer19,Mer22a,Mer22b,JP96} for related results including Markov approximations and oscillations for other coupled systems.

Let us also underline the following interesting remark. If $\lambda \in S_{\rm inf}$ is
  a non degenerate eigenvalue then
    \cite{B-F-S} and \cite{B-F-S-99} prove the existence of a resonance     $E_{\lambda } (g)$ converging to $\lambda$ when $g\to 0$. Let
    $u_{\lambda }$ be a unitary eigenvector and  $\Pi (\lambda)$ be the orthogonal projection
   on the span of  $u_{\lambda}$. Then,
    $ < (G(t g^2)\Pi (\lambda)) u_{\lambda }, u_{\lambda} > $ is exactly  the
 expression  in \cite{B-F-S} and \cite{B-F-S-99} up to some power of $g$ of the lifetime of the resonance $E_{\lambda } (g)$.

\subsubsection{Non Markovian approximation.}

The error for the Markovian approximation  of the marginal transition
probabilities  is estimated by ${\cal O} (g)$ with the dipole
approximation or ${\cal O} (g(1+ t^2))$ without it.
We shall give another   approximation of these  probabilities
 with an error bounded by  ${\cal O} ( g^3 ( t + t^3))$.
Thus, the non Markovian approximation is more precise than the Markovian approximation  if  $ t g^2  < 1$. This is the content of  the following result in which the dipole approximation is not used.

\begin{theo}\label{ANM} Let the form factor be given either by (\ref{stand}) or by
   (\ref{dipol}) and assume that the function $\phi$ in (\ref{stand}) or in (\ref{dipol}) is vanishing at the origin.  Let  $\lambda _j$ and $\lambda _m$
be two distinct eigenvalues belonging to   $S_{\rm inf}$. Set  $u_j$
 and $u_m$ two unitary eigenvectors associated to  $\lambda _j$ and $\lambda _m$. Then, the marginal transition probabilities $ P(t , u_j, u_m , g )$
satisfy,
$$ P(t , u_j, u_m , g ) = 2 ( ig )^2  \int _{\R^3} \frac { 1  - \cos ( t ( |k| + \lambda _j - \lambda _m )) }
  { ( |k| + \lambda _j - \lambda _m )^2 } \ | < E(k) u_j , u_m > | ^2 dk
  + {\cal O} ( g^3  (t^2 + t^3) ). $$

\end{theo}

Theorem \ref{ANM} will be proved in  Section 6 using results of  Sections
1.3, 1.4, 4.1 and 4.2, (see also a link with the Fermi Golden Rule in \cite{RS}, Volume IV, page 68).

\subsubsection{Rabi cycle.}

 We now turn  to  the investigation of $<  ( S^{\rm mat} (t , g)  X  ) u , u >$
 without assuming that   $u\in {\cal H}_{\rm inf}$ is an eigenfunction of  $H_{\rm mat}$. The vector  $u$ is still supposed to be in ${\cal H}_{\rm inf}$
    and we write its spectral decomposition with respect to  $H_{\rm mat}$ as,
   $$ u = \sum _{\lambda \in S_{\rm inf}}  u_{\lambda }, $$
 with $u_{\lambda }  = \Pi (\lambda ) u$. It is a common fact in quantum mechanics that, if a quantum particle system  is  interacting  with  photons and if it is initially a superposition of two eigenfunctions of the Hamiltonian with distinct eigenvalues, then  its time evolution exhibits a periodic behavior. This is known as Rabi cycle. We shall now give more precisions on that picture.

For any  such $u=\sum _{\lambda \in S_{\rm inf}}  u_{\lambda }$   and all $X\in {\cal K}$, one writes,
    $$ <  ( S^{\rm mat} (t , g)  X  ) u , u >
    = \sum _{\lambda , \mu} <  ( S^{\rm mat} (t , g)  X  )
     u_{\lambda} , u_{\mu} >. $$
  Then,   the terms with $\lambda = \mu$ are handled as before and the other terms for  $\lambda \not= \mu$ are discussed in the two following Theorems.

    \begin{theo}\label{approx-Rab} Suppose that the form factor is given either by  by
     (\ref{stand}) or by (\ref{dipol}) and that the function
 $\phi$ in (\ref{stand}) or in (\ref{dipol}) is vanishing at origin.  Fix $\lambda $
and $\mu $  in $S_{\rm inf}$ with $\lambda \not= \mu$ and let $\omega = \mu - \lambda$.
 Set $u\in {\rm Ker} (H_{\rm mat }- \lambda I)$
and $v\in {\rm Ker} (H_{\rm mat }- \mu I)$. Then,  we have for all $X\in {\cal K}$,
\be\label{approx-Rab-2} <  ( S^{\rm mat} (t , g)  X  )
     u , v >  = \frac { (ig)^2} {i\omega}
      ( e^{i\omega t} < ( L_{\infty }^0 X ) u, v >
     -  < ( L_{\infty }^{\omega}  X ) u, v > )  +
     < R(t, g, X ) u, v > , \ee
where $L_{\infty }^{\omega}  X$  is defined in   (\ref{L-omega})
and  (\ref{Lt-LI}) with,
$$ | < R(t, g, X ) u, v >  | \leq C \left (  g^3 (t + t^3)  +  \frac { g^2 } {1+t} \right ) \Vert X \Vert
\Vert u \Vert   \Vert v \Vert.  $$
\end{theo}

The operator $ L_{\infty }^{\omega}  X$  is
defined by (\ref{Lt-LI}) as a limit in some sense of an operator
$ L^{\omega} (t) X$ defined in  (\ref{L-omega}). This includes  the case $\omega = 0$. The proof of  this Theorem
 has a common part  with the proof of Theorem \ref{approx-SG-prov} (given in Sections 1.3, 4.1 and 4.2)
  and the more specific part  of the proof is given in Section 5.1.

Also note that since the main term of this asymptotic is in
$g^2$ then the other terms are assumed to be small if
$t$ is large and $tg$ is small. The Rabi oscillation is therefore a direct result of the Pauli-Fierz Hamiltonian, approximatively, and this approximation is relevant
if $t$ is large and $g$ small with respect ot $t^{-1}$.

\begin{theo}\label{approx-Rab-3} Under the hypotheses of Theorem \ref{approx-Rab}, assuming that  $\lambda \not = \mu$ and setting  $h_{\lambda \mu} = \frac{2 \pi} {|\lambda- \mu|}$, there exists
 $C>0$ satisfying,
\be\label{app-period} \left |    < ( S^{\rm mat} (t + h_{\lambda \mu}  , g)  X )
     u , v >    - <  ( S^{\rm mat} (t  , g)  X  )
     u , v > \right |   \leq C \left ( g^3 (1+ t^2)  +  \frac { g^2 } {1+t} \right )
\Vert X \Vert \Vert u  \Vert \Vert v \Vert.
 \ee
 \end{theo}

The  Rabi cycle was discovered in 1938 in \cite{Rabi} in the context of nuclear magnetic resonance.
The matter particules are supposed to be at rest with a spin interacting with a constant magnetic field. This model is not described by the Pauli-Fierz model but rather  by the spin-boson model.

Since then, various generalizations have been discussed  corresponding to different physical frameworks and studied with ad hoc models. Let us mention in particular
 \cite{JCM} (introducing a model largely used) and   \cite{SK93,SB09,LM21,Hir-Hir,T-L-DV}.

It is interesting to note that some of these works are concerned with bounded sets of $\R^3$ and not
the whole domain $\R^3$.
 Also, some of these models substitutes the discrete spectrum by a finite spectrum leading to simplifications.
Moreover, in our work, the   essentiel spectrum (of the
 Schr\"odinger operator) also creates serious technical  difficulties (that are partially overcome since we need to use the dipole  approximation to complete the proofs), even if this essential spectrum is not explicitly involved in the statement of the final  result.
 Besides, note that  O. Matte \cite{M17} and H. Spohn \cite{SP} (Chapter 13) studied analogues
  of  Pauli-Fierz operators  in bounded  domains (cavities).

We focus in this work on  the case of one electron but the case of $N$ particles could be probably also be handled  by adapting the methods here.
 Indeed,  the operator $L_{\infty }^{\omega}  X $ involved in
the  Rabi approximation  (\ref{approx-Rab-2}) can be computed in the case of a system of $N$ particles.  We give this calculus in Section 5.2. The main interest of the resulting formula is that we observe an interaction between  particles even if we cut this interaction in the Schr\"odinger Hamiltonian. See  Theorem \ref{Rabi-N} and the
  remark below for more precisions. Thus, there is an interaction between particules,
 emerging only in  the QED framework and being manifest in  the Rabi cycles for several particles.

\subsection{Local in time Dyson approximation.}

The purpose of this section is to prove Proposition \ref{deriv-prelim} below.
Proposition \ref{deriv-prelim}  will be important as the starting point of the proofs in Section 4 of the three main results of this article:   Theorem \ref{approx-SG-prov}, completed by Theorem \ref{approx-SG}  (Section 4),
Theorems  \ref{approx-Rab} and  \ref{approx-Rab-3}.
In addition, the  operator $L^{\omega}(t) X$  defined in  (\ref{L-omega}) will play an essential role in what follow.

The first step  refers as Dyson approximation with $H(0)$ as the free energy operator and $g H_{\rm int}$ as the perturbation
($H(0)$ and $ H_{\rm int}$ are the operators of Theorem \ref{Kato-Rell}).

 To do this, let us introduce some standard notations for Dyson expansions.

 First, set
$$ H_{\rm int} ^{\rm free} (t) = e^{it H(0)} H_{\rm int} e^{-it H(0)},$$
where $H_{\rm int}$ and $H(0)$ are  the operators given in Theorem \ref{Kato-Rell}.

Then, according to Theorem \ref{Kato-Rell}, the operator
$ e^{it H(0)}$ is  bounded in $W_{m}^{\rm tot}$,  uniformly in time  $t$,
 for any $m\geq 0$, and the operator $H_{\rm int}$  is bounded from
$W_{m+2}^{\rm tot}$ to $W_{m}^{\rm tot}$. Thus, the operator
$H_{\rm int} ^{\rm free} (t) $ is bounded from
$W_{m+2}^{\rm tot}$ to $W_{m}^{\rm tot}$ uniformly in time $t$.

Next, for each $Z\in {\cal L} ( {\cal H}^{\rm tot})$ and for every  $t\in \R$,
we define a quadratic form
 $A(t) Z$ on $W_{2}^{\rm tot}$ by,
$$ < ( A(t)  Z) u, v >  =
 <   Z u, H_{\rm int} ^{\rm free} (t) v >
 - < Z H_{\rm int} ^{\rm free} (t) u, v > ,  $$
for all $u$ and $v$  in  $W_{2}^{\rm tot}$. For each $t_1$ and $t_2$ in
$\R$, we define  a quadratic form $A(t_1) A(t_2) Z$ in $W_{4}^{\rm tot}$,
for all $u$ and $v$  in  $W_{4}^{\rm tot}$ by,
 \be\label{def-At}  
 < ( A(t_1) A (t_2) Z ) u, v > = < Zu, H_{\rm int} ^{\rm free} (t_2) H_{\rm int} ^{\rm free} (t_1) v >
 - < Z H_{\rm int} ^{\rm free} (t_1) u, H_{\rm int} ^{\rm free} (t_2) v > \ee
 $$- < Z H_{\rm int} ^{\rm free} (t_2 ) u, H_{\rm int} ^{\rm free} (t_1) v >
 + < Z H_{\rm int} ^{\rm free} ( t_2) H_{\rm int} ^{\rm free} (t_1) u, v > .$$
  Therefore $\sigma _0 ( A(t_1) A(t_2) Z)$ is well defined as a quadratic form on
   $W_{4}^{\rm mat}$ (Proposition \ref{norme-sigma}).

    We define an operator $\sigma _0 Z$ in ${\cal H} _{\rm mat}$ for each $Z$ in ${\cal H} _{\rm tot}$ by,
 \be\label{sigma-0}   < (\sigma _0 Z) u, v > = < Z ( \Psi_0 \otimes u) , ( \Psi_0 \otimes u) > ,\ee
for all $u$ and $v$ in ${\cal H} _{\rm mat}$.

Also let,
 \be\label{X-free}  X^{\rm free}(t) = e^{i t H_{\rm mat}} X e^{-i t H_{\rm mat}}, \ee
 for every operators $X\in {\cal H} _{\rm mat}$.

Set,
 \be\label{S-tot}  S^{\rm tot} (t , g) X = e^{it H(g)} ( I \otimes X ) e^{- it H(g)}.\ee

In particular, $S^{\rm mat}(t,g)X$ defined in (\ref{S-mat}) equals to $\sigma_0 S^{\rm tot}(t,g)X$ defined in (\ref{sigma-0}) and (\ref{S-tot}).

Now, we can state the main result of this section.

\begin{prop}\label{deriv-prelim} Set $X \in {\cal K}$, let  $u$ and $v$ be eigenfunctions of $H_{\rm mat}$ satisfying,
\be\label{lamb-mu}  H_{\rm mat} u = \lambda u,\   H_{\rm mat}v = \mu v, \ee
with $\lambda $ and $\mu $ belonging to $S_{\rm inf}$. Set $\omega = \mu - \lambda $. Then, the following identity holds true,
\begin{align}\label{Sys-Dif-Init}  \left (  \frac {d} {dt} - i \omega \right ) <  ( S^{\rm mat} (t , g) X  ) u, v > &=
  (ig)^2 < L^{\omega} (t)(S^{\rm mat} (t , g) X ) u , v>
 \\[4mm]&
  + < R_1 (t,  g, \omega , X ) u, v >  + < R_2 (t,  g, \omega , X ) u, v >  . \nonumber\end{align}
where $L^{\omega} (t) Z$
is the quadratic form on $W_4 ^{mat}$ defined with the notation  (\ref{def-At}) by,
 \be\label{L-omega}  L^{\omega} (t) Z =   \int _0^t e^{i \omega s}
 \sigma _0 \Big ( A(-s) A(0)  ( I \otimes Z)\Big )  ds,  \ee
for each $Z$ in ${\cal L} ({\cal H}_{\rm mat})$.
Recall that, if $Z \in {\cal L} ({\cal H}_{\rm mat})$ then
$(I\otimes Z) \in {\cal L} ({\cal H}_{\rm tot})$.

Moreover,
\be
\label{defi-R1} 
R_1 (t,  g, \omega , X ) = (ig)^2 \int _0^t  e^{i  \omega ( t-s)  }   \sigma _0  ( A( s - t)
A(0)   (  S^{\rm tot} (s, g) X - I \otimes S^{\rm mat} (s, g) X   )
    ds\ee
 and
\be\label{defi-R2}  R_2 (t,  g, \omega , X ) =
 (ig)^2 \int _0^t  e^{i  \omega ( t-s)  }   \sigma _0  ( A( s - t)
A(0)  \big( I \otimes  ( S^{\rm mat} (s, g) X - S^{\rm mat} (t, g) X  )   ds.\ee

\end{prop}

{\it Proof of Proposition \ref{deriv-prelim}. First step.} The first step is actually only an order two Dyson expansion.
We begin to check that,
\be\label{appro-bis}     S^{\rm mat} (t , g ) X   =
   X^{\rm free} (t)     + (ig)^2 E_2(t, g) X,  \ee
where
\be\label{reste-bis}   E_2(t, g) X = \sigma _0  \int _{0 < s_1 < s_2 <t}
 e^{i( t-s_1) H(0)}    ( A( s_1 - s_2)
A(0)( S^{\rm tot} ( s_1, g) X  )   )  e^{i( s_1 - t ) H(0)}  ds_1 ds_2, \ee
with $\sigma _0$  defined in (\ref{sigma-0}). Indeed, set,
$$ G_{\rm dys} (t) Z  = e^{-i t H(0)}  e^{i t H(g)}  Z     e^{-i t H(g)} e^{i t H(0)}, $$
for all operators $Z$ in ${\cal H} _{\rm tot}$.
We have,
$$ \frac {d} {dt} G_{\rm dys} (t) Z = ig  A( -t)  G_{\rm dys} (t) Z. $$
Thus, we get,
$$ G_{\rm dys} (t) Z = Z + ig \int _0^t  A( -s)  G_{\rm dys} (s) Z  ds. $$
Iterating this identity, we see that,
$$  G_{\rm dys} (t) Z = Z + ig \int _0^t  A( -s)   Z  ds
+ (ig)^2 \int _{0 < s_1 < s_2 <t}  A( -s_2) A(-s_1 ) G_{\rm dys} (s_1 ) Z
ds_1 ds_2.$$
Now, we use the above equality with $Z = I \otimes X$ where $X$
is an operator in ${\cal H} _{\rm mat}$. Then we apply
$e^{it H(0)}$ and  $e^{-it H(0)}$ respectively on the left and  on the right hand sides. Finally, we  complete the proof of (\ref{appro-bis}) by applying the operator $\sigma_0$ on the two sides while using,
$$ \sigma _0  ( e^{it H(0)} A( -s)   (I\otimes X) e^{-it H(0)}   ) = 0,$$
which comes from (\ref{H-int-def})(\ref{sigma-0}) and from $a(k) \Psi_0 = 0$.

{\it Second step.}  One gets differentiating (\ref{appro-bis}),
\be\label{SD-init-2} \left (  \frac {d} {dt} - i \omega \right ) <  ( S^{\rm mat} (t , g) X  ) u, v > =
\Phi ( t, g, \omega , X , u, v ), \ee
where
\be\label{SD-init-22} \Phi ( t, g, \omega , X , u, v ) =
(ig)^2 \int _0^t  e^{i  \omega ( t-s)  }  < \sigma _0  ( A( s - t)
A(0)  ( S^{\rm tot} (s, g) X  )
    ) u, v > ds.  \ee
Then one observes that,
$$ \Phi ( t, g, \omega , X , u, v ) =\hskip 13 cm$$
$$
\hskip 2 cm(ig)^2 < L^{\omega} (t)(S^{\rm mat} (t , g) X ) u , v>
  + < R_1 (t,  g, \omega , X ) u, v >   + < R_2 (t,  g, \omega , X ) u,  v >. $$

  \fpr

The first term in the right hand side of  (\ref{Sys-Dif-Init}) is the main one and the two others are error  terms  that will be  estimated in Section 4.1 and Section  4.2.

If  $Z$ is not in $ {\cal L} ({\cal H} _{\rm mat})$ but is belonging to  $ {\cal L} ( W_p ^{\rm mat}, {\cal H} _{\rm mat} )$ then the same reasoning 
shows that $ L^{\omega} (t) Z$ is well defined as a quadratic form on  $ W_{p+4}  ^{\rm mat}$.

We will be concerned with the issue of the existence of the limit as $t$ goes to infinity in the aim of getting a much simpler differential system. This is precisely the content of the next section that provides in addition an estimate that is independent of $t$.

 \subsection{Large time limits.}

For each $Z$ in ${\cal L}({\cal H}_{\rm mat})$,  recall that $L^{\omega} (t) Z$
given by (\ref{L-omega}) is  a quadratic form on $W_4^{\rm mat}$.
Since the elements of ${\cal H}_{\rm inf}$ belong to   $W_m^{\rm mat}$ for any $m$ and since
   ${\cal H}_{\rm inf}$ is finite dimensional then it follows that
 $L^{\omega} (t) Z$ is also a quadratic form on  ${\cal H}_{\rm inf}$,
 or also a bounded operator in ${\cal H}_{\rm inf}$, still denoted by  $L^{\omega} (t) Z$.

  We prove in the next proposition (point ii)  the existence of the limit as $t$
  goes to infinity of this operator $L^{\omega} (t) Z$ for $Z$  belonging to either
   ${\cal L}({\cal H}_{\rm inf} )$ or   ${\cal L}({\cal H}_{\rm sup} )$
   (see Definition \ref{K-P}).

   If  $Z$ is an arbitrary  operator belonging to ${\cal L}({\cal H} _{\rm mat} )$ then 
   we prove in point iii) of the next proposition that $<   (L^{\omega}(t)  Z) u, v >$
   remains bounded for suitable $u$, $v$ and $\omega$.  Even, $Z$ can be
   a bounded operator from $W_2 ^{mat}$ to ${\cal H} _{\rm mat}$. This is needed
   for applications in Section 4 and Section 5. Finally, in point iv), we prove that the above product has a  limit when $t$ goes to infinity under additional hypotheses on  $H_{\rm mat}$.
   
  In the sequel, one says that the function $\phi$ vanishes at the origin at the order $p\geq 1$, if $\phi^{j}(0)=0$ for $j=0,\dots, p-1$, and in particular, if $p=1$, the condition is only $\phi(0)=0$. Note that only the order $p=1$ will be  used to get the main Theorems stated in Section 1.

\begin{prop}\label{limite-L}  Suppose that the
form factor is defined by either (\ref{stand}) or (\ref{dipol}) and assume that  the function
$\phi$  of (\ref{stand}) or  (\ref{dipol}) is vanishing at the origin at the order $p\geq 1$.
Then:

i)  For all
$Z$ in  ${\cal L} ( {\cal H} _{\rm mat})$,  we have
\be\label{LtZuv}   < (L ^{\omega } (t) Z) u, v >
 =
 \int _{\R^3 \times (0, t)  } e^{i\omega s}   (e^{i s |k|}
<  E^{\rm free} (k, - s)^{\star}\ [ E(k) , Z] u, v >    \ee
$$  - e^{-is  |k|} <  [ E^{\star} (k) , Z]  E^{\rm free} (k, - s) u, v > )  dk ds  $$
for all $u$ and $v$ in ${\cal H} _{\rm inf}$ and for all $\omega \in \R$.

ii)   For each $Z$ which is either in ${\cal L} (  {\cal H} _{\rm inf})$ or in
${\cal L} (  {\cal H} _{\rm sup})$, there exists an operator $L_{\infty}^{\omega}  Z$ in $ {\cal L} ({\cal H}_{\rm inf})$ such that,
  \be\label{Lt-LI}  \left | <  ( L_{\infty}^{\omega}  Z -  L^{\omega} (t) Z) u , v > \right |
    \leq \frac {K} {1+t^{2p} }
  \Vert Z  \Vert  \   \Vert u  \Vert    \Vert v  \Vert ,   \ee
  for all time $t>0$
  and
  \be\label{majo-LI} \left | <   (L_{\infty}^{\omega} Z) u, v > \right |    \leq K
  \Vert Z  \Vert  \  \Vert u  \Vert    \Vert v  \Vert ,   \ee
   for some $K>0$ and all $u$ and $v$ in $ {\cal H}_{\rm inf}$.

   iii) Let  $u$ and $v$ be eigenfunctions of $H_{\rm mat}$ satisfying (\ref{lamb-mu})
with $\lambda $ and $\mu $ belonging to $S_{\rm inf}$. Set $\omega = \mu - \lambda $.
  Then, for all $Z\in {\cal L} ( W_2 ^{mat}, W_0 ^{mat} )$,
we have,
   \be\label{majo-LII}  \left | <   (L^{\omega}(t)  Z) u, v > \right |    \leq K
    \Vert Z  \Vert_{ {\cal L} ( W_2 ^{mat}, W_0 ^{mat} ) }     \  \Vert u  \Vert    \Vert v  \Vert ,   \ee
  where $K$ is a real number independent of $Z$, $t$, $u$ and $v$.

   iv) In addition to the hypotheses on the matter Hamiltonian of Section 1.1, we suppose that   $[0, +\infty )$
   is the continuous spectrum of $H_{\rm mat}$.
  Then, $ <   (L^{\omega}(t)  Z) u, v > $ has a  limit as $t$ goes to $+\infty$, for all
   $Z\in {\cal L} ( {\cal H}_{\rm mat})$, for any $u$ and $v$
 satisfying (\ref{lamb-mu}) with $\omega = \mu - \lambda $.

\end{prop}

Concerning examples with the hypothesis in   Point iv)  that  are satisfied, see \cite{RS} (Volume IV, Section XIII.3 and Section XIII.13) and \cite{CFKS} (Chapter 4). 

This proposition will be proved in  Section 2.3. Point ii)
will be used to define the operator ${\cal L}$
which is the infinitesimal generator of the  
semigroup $G(t)$. It is also implied in the proof of Proposition \ref{P-K-K} and in Section 5. Point iii)  is involved in the proof of Proposition \ref{P-K-K} and also in Section 5.  Point iv) is not used in the sequel but has its own interest.

In the following, when $\omega=0$,  $L (t)$ and
$L_{\infty} $ stand respectively for $L^{0} (t)$ and
$L_{\infty}^{0} $.

Let us underline the analogy between
 $L_{\infty}$ and Lindblad operators frequently   used for open quantum systems. The terminology may vary in the literature.
 In the survey \cite{C-P}, a  GKLS operator (belonging to ${\cal L} ( {\cal L} (H))$ is a linear combination of  $X \mapsto A_j X B_j + B_j^{\star} X
A_j^{\star} $ with  the $A_j$ and $B_j$  in $ {\cal L} (H)$ where $H$ is a  Hilbert space. Some works call them Lindblad or  master equations.  In  \cite{A-L,G-K-S,Lind}, some properties for these operators are examined, in particular, the possible semigroup  property. These operators are often used in the following settings:

  $\quad\bullet$ Open quantum systems \cite{Dav,Dav76},

     $\quad\bullet$  Spin-boson models \cite{A-N},  \cite{DRK} (formula (3.34)),   \cite{DH1,DH3},

     $\quad\bullet$ Simplified Pauli-Fierz operators (generalized spin-boson model) \cite{DH2},

      $\quad\bullet$ Decoherence issues \cite{SP2,D-S}.

      In some sense, $ L^0 (t) $ given by   (\ref{L-omega}) can be called GKLS or Lindblad operator in view of its  specific form.

\subsection{Resonance operators. }

For all $X$ in ${\cal  L}({\cal H }_{\rm inf})$ and any  $\lambda $ in
 $S_{\rm inf}$, we shall see in  Theorem \ref{Bet-Sap}  that there exist operators  $T_{\lambda }$ and $ R_{\lambda }X$ in  ${\rm Ker } ( H_{\rm mat} - \lambda I)$ such that 
the operator  $L_{\infty} ^0 X$   of Proposition \ref{limite-L}  is satisfying,
 \be\label{LXuv}  < (L_{\infty} ^0 X) u, v > =  < ( T_{\lambda} X + X T_{\lambda} ^{\star}) u, v >
  - < (R_{\lambda } X)u , v >, \ee
for all $u$ and $v$ in the subspace ${\rm Ker } ( H_{\rm mat} - \lambda I)$
and where
 $R_{\lambda } X$ has the following property concerning positivity preserving:
 if $X$ is a nonnegative self-adjoint  operator in ${\cal H }_{\rm inf} $ then
 $R_{\lambda } X$ is a nonnegative self-adjoint  operator in ${\rm Ker } ( H_{\rm mat}  - \lambda I)$.
 Regarding the operator $ T_{\lambda }$, it is define in the next Proposition, whereas the operator 
$ R_{\lambda }X $ will be defined in Theorem 3.1. The operator $T_{\lambda}$
will play an important role  in Section 3.

  In this section, we only study  $ < (L_{\infty} ^0 X) u, v >$ and not 
  $ < (L_{\infty} ^{\omega} X) u, v >$.

For all $t>0$, the equality below defines a quadratic form on  $W_4^{\rm mat}$.
\be\label{T(t)}  < T(t)u, v >  =   \int _{0 }^t
    <  \sigma _0 \Big (  e^{ - i  s H(0)  } H_{\rm int}  e^{  i  s H(0)  }  H_{\rm int}\Big ) u, v >    ds.  \ee
As already mentioned, it  can be associated with an element of
${\cal L}({\cal H}_{\rm inf})$.

\begin{prop}\label{limite-T}  i) We have,
\be\label{Tt-integ}  < T(t) u, v > = \int _{\R^3 \times (0, t)} e^{is|k|} <  E^{\rm free}(k, -s)^{\star} \ E(k) u , v > dk ds,\ee
for all $u$ and $v$ in ${\cal H}_{\rm inf}$, where
the operator $T(t)$ is defined in (\ref{T(t)}).

ii) If the
form factor is defined either by (\ref{stand}) or (\ref{dipol}) and if the function
$\phi$ in (\ref{stand}) or (\ref{dipol}) vanishes  the origin at order $p\geq 1$, then,
  there exist a bounded operator $T$   of  ${\cal L} ({\cal H}_{\rm inf})$
   and a constant $K>0$ such that,
\be\label{conv-T}  \left | < ( T(t) - T) u, v > \right  | \leq \frac {K} {1+t^{2p}}
  \   \Vert u  \Vert    \Vert v  \Vert, \ee
    for all $u$ and $v$ in $ {\cal H}_{\rm inf}$.

iii) Under the same  hypotheses, the above operator $T$ satisfies,
for any  $u$ any $v$ in  ${\cal H}_{\rm inf}$, with $v$ in
${\rm Ker} ( H_{\rm mat} - \mu I)$ (where $\mu \in S_{\rm inf}$),
$$  < T u , v > =   i  \lim _{\varepsilon \rightarrow 0^+}
 \int _{\R^3 }  < ( H_{\rm mat} + |k|  - \mu +  i \varepsilon ) ^{-1}
  E(k) u , E(k) v >   dk. $$
 For all $\lambda \in S _{\rm inf}$, we define an operator  $T_{\lambda }$  in
 ${\rm Ker } ( H_{\rm mat} - \lambda I)$
by,
$$ T_{\lambda } u = \Pi (\lambda ) Tu,\quad u \in {\rm Ker } ( H_{\rm mat} - \lambda I) $$
where $ \Pi (\lambda )$ is the orthogonal  projection on ${\rm Ker } ( H_{\rm mat} - \lambda I)$.

iv) The following identity holds true for $T_{\lambda } + T_{\lambda } ^{\star}$,
for any $u\in {\rm Ker } ( H_{\rm mat} - \lambda  I)$ with  $\lambda  \in S_{\rm inf}$:
$$ < ( T _{\lambda }+ T_{\lambda } ^{\star}) u , u > =  2 \lim _{\varepsilon \rightarrow 0^+} \int _{\R^3 \times \R_+  }
e^{ -\varepsilon s }
 < \cos ( s( H_{\rm mat} + |k| - \mu)) E(k) u, E(k) u > dk ds,$$
 that is,
\be\label{gamma} < ( T_{\lambda } + T_{\lambda }^{\star }) u , u >  = 2 \pi \sum _{ \rho \in S_{\rm inf} \atop  \rho < \lambda}
\int _{ |k| =  \lambda - \rho  }  \Vert \Pi (\rho ) E(k) u \Vert ^2 d\sigma (k). \ee
Recall that  $d\sigma (k)$  is the sphere surface measure
and that  $\Pi ( \rho)$ denotes the orthogonal projection  on ${\rm Ker} (H_{\rm mat} -  \rho I)$.

v) The expression below  for $T_{\lambda } - T_{\lambda } ^{\star }$ is valid,
for any $u\in {\rm Ker } ( H_{\rm mat} - \lambda  I)$ with  $\lambda \in S_{\rm inf}$:
\begin{align}
\label{Bethe}  (2i) ^{-1}  < ( T_{\lambda } - T_{\lambda } ^{\star }) u , u > &= \int _{\R^3}
< ( H_{\rm mat} + |k| - \lambda  )^{-1}  \Pi ^{\rm sup} (\lambda) E(k) u, \Pi ^{\rm sup} (\lambda) E(k) u > dk
\nonumber \\[4mm]
&+ \sum _{\rho \in S_{\rm inf} \atop \rho < \lambda}   {\rm PV}   \int _{\R^3 }
\frac { < \Pi (\rho ) E(k) u , E(k) u > } { |k|+ \rho -\lambda } dk,
\end{align}
where  $\Pi ^{\rm sup} (\lambda)$ is the spectral projection on the interval
$(\lambda , + \infty )$ and $  {\rm PV}  $ stands for the principal value of the singular integral  on $\R^3$.

 \end{prop}

 This proposition will be proved in Section 2.4.

It is precisely the operator  $T_{\lambda }$ that play a role in  (\ref{LXuv}). This operator also appears in  \cite{B-F-S-99}.

In  \cite{B-F-S-99}, the square of the  potential vectors are not dropped from the Hamiltonian but play no role here.
 The  operator $E(k)$ is denoted  $w_{1, 0} (k)$ in  \cite{B-F-S-99}. The parameter  $\lambda$ taking values $1$ or $2$ in formulae (3.8)(3.9) in  \cite{B-F-S-99} does not appear explicitly in $E(k)$ and is taken into account since $E(k)$  is a vector field on  $\R^3$ that is the Fourier transform of divergence free vector fields. 
 Also, the operator $E^{\star} (k)$ is  $w_{0, 1} (k)$    in \cite{B-F-S-99}. In \cite{B-F-S-99},  formula  (3.8) defines two operators $Z_j^{\rm od}$ and  $Z_j^{\rm d}$ in  ${\rm Ker}  ( H_{\rm mat} - \lambda _j )$ setting $\lambda = \lambda _j$,
 $$ < Z_j^{\rm od} u , v > = \int _{\R^3} < E^{\star} (k) \Pi(\lambda _j)^{\perp}
 ( H_{\rm mat} - \lambda _j + |k| - i 0)^{-1} \Pi (\lambda _j)^{\perp} E(k) u, v > dk $$
 and
 $$ < Z_j^{\rm d} u , v > = \int _{\R^3} < E^{\star} (k) \Pi (\lambda _j) E(k) u , v >
 \frac {dk}  {|k|}, $$
for all  $u$ and  $v$ in  ${\rm Ker}  ( H_{\rm mat} - \lambda _j )$ where 
$\Pi (\lambda _j)$ denotes the orthogonal projection on ${\rm Ker}  ( H_{\rm mat} - \lambda _j )$.
In point iii) of the above  proposition, we do not get the same limit resolvent as in  \cite{B-F-S-99} but the same argument shows that,
$$  < (T_{\lambda _j}) ^{\star}   u , v > =  - i  \lim _{\varepsilon \rightarrow 0^+}
 \int _{\R^3 }  < ( H_{\rm mat} + |k|  - \mu -  i \varepsilon ) ^{-1}
  E(k) u , E(k) v >   dk,$$
  for any $u$ and $v$ belonging to
${\rm Ker}  ( H_{\rm mat} - \lambda _j )$.

Thus, we have according to  point iii) the following identity,
$$ T(\lambda _j)^{\star}  = -i (   Z_j^{\rm od} +  Z_j^{\rm d}). $$
It is proved in \cite{B-F-S-99}, without infrared regularization,  that for every eigenvalue $ E_j$
 of multiplicity $m$ in  $S_{\rm inf}$ and for any $g$ sufficiently small, there exist resonances $E_{jp} (g)$
 which are, up to a  $o(g^2)$ term, the eigenvalues of the operator 
 $E_j + g^2 ( Z_j ^{\rm d} - Z_j^{\rm od} )$ acting in the space  ${\rm Ker } ( H_{\rm mat}  - E_j )$.   More precisely, the operator $T- T^{\star}$ is used for the real parts of the resonances (Bethe) and  $T + T^{\star}$
 for the imaginary parts of the resonances (Fermi).

 Point iii) of the above  proposition allows to rewrite the hypothesis (FGR)
 stated in Section 1.2.1. It amounts to the existence of  $\gamma >0$  satisfying,
 $$   < ( T_{\lambda } + T_{\lambda }^{\star }) u , u > \geq \gamma \Vert u \Vert ^2, $$
for every  $u\in  {\rm Ker } ( H_{\rm mat} - \lambda I) $ and each $\lambda\in S_{\rm  inf}$ excepted for $\lambda = {\rm inf} \sigma ( H_{\rm mat })$.
In view of  Point iv) in the above  proposition, one still has
 $< ( T_{\lambda } + T_{\lambda }^{\star }) u , u > \geq 0$ for all
 $u\in  {\rm Ker} ( H_{\rm mat} - \lambda I) $ and any  $\lambda \in S_{\rm  inf}$ even if the hypothesis (FGR) is not verified.
This hypothesis (FGR)  will play  a central role to investigate the exponential behavior of the semigroup  $G(\cdot)$  defined in Section 1.5.

Point v) of the above proposition is not used in the sequel. Its interest lies in the fact that in the historical article of  Bethe \cite{BET}  concerning the  Lamb shift, many logarithms  throughout the paper are probably hiding the Cauchy principal values of  Point v). Simply put, the operator $T_{\lambda }$ was actually  already involved in this article of 1947.

\subsection{Semigroup:  statement of properties.}

The objective of the Markovian approximation here is to approximate the operator ${\cal P}_{\cal K}( S^{\rm mat} (t , g) X) $ for each $X\in {\cal K}$ by a semigroup $G(t)$ acting ${\cal K}$.
More precisely, we shall see that,
 ${\cal P}_{\cal K}( S^{\rm mat} (t , g) X) $ is approximated 
by $G(t g^2) X$.

Recall that the operator algebra
${\cal K}$ and the projector ${\cal P}_{\cal K}$ are given by Definition \ref{K-P}
 and $ S^{\rm mat} (t , g) X$  is defined in (\ref{S-mat}).

  \begin{defi}\label{def-L}
 We denote by ${\cal L}$ the operator  defined for all $X\in {\cal K}$  by,
 \be\label{def-cal-L}  {\cal L} X = {\cal P}_{\cal K} ( L_{\infty} X )
 = \sum _{\lambda \in S_{\rm inf} } \Pi(\lambda ) ( L_{\infty} X ) \Pi(\lambda ),\ X\in {\cal K}, \ee
where   $L_{\infty} X = L_{\infty}^0 X $ is the operator given by Proposition \ref{limite-L}
and, for each $\lambda \in S_{\rm inf}$, $ \Pi(\lambda ) $ is the orthogonal
projection on ${\rm Ker} (H_{\rm mat}-\lambda I)$.
Then, we set for each $t>0$,
\be\label{def-Gt}  G(t) X = e^{ - t {\cal L} } X, \ee   
for all $X\in {\cal K}$.
\end{defi}

Markov semigroups considered here are introduced in
Definition \ref{def-Markov}. Note that there is also a complete positivity property  for Markov semigroup, see, {\it e.g.},  \cite{FFFS}. This property will not be used here.

 \begin{theo} \label{se-gr}The semigroup $G(\cdot)$ given by (\ref{def-cal-L})(\ref{def-Gt})  is a  Markov  semigroup in    ${\cal K}_{\R}$ (the set of self-adjoint elements of the $C^{\star}$-algebra ${\cal K}$). 
 \end{theo}

This Theorem will be proved in Section 3.3.

We then study the behavior of  $G(t) X$ for large time $t$ under the hypothesis  (FGR) (see Section 1.2).

\begin{theo}\label{expo} We suppose that the
form factor $E(k)$ is defined either by   (\ref{stand}) or  (\ref{dipol})  and we assume that the smooth ultraviolet cut-off
$\phi$ in (\ref{stand}) or (\ref{dipol}) is  also vanishing at $0$. We also make the hypothesis
 (FGR). Let  ${\cal K} _{\rm dec}$ be the set of all  $X\in {\cal K}$
satisfying  $< Xu_0 , u_0> = 0$ where $u_0$ is a unit ground state of  $H_{\rm mat}$.
Define ${\cal K} _{\rm inv}$ as the line of ${\cal K}$ spanned by $I_{\rm inf}$ the identity map on ${\cal H}_{\rm inf}$. Fix  any $\delta$ satisfying $0< \delta < \gamma$ where   $\gamma >0$ is given in hypothesis (FGR).
Then, the following properties hold true:

i) We have  ${\cal L} X\in {\cal K}_{\rm dec}$ for all  $X\in{\cal K}$ and  $G(t) X = X$    for any $X\in{\cal K} _{\rm inv}$ and all $t>0$.

ii) There exists $C>0$  such that,
\be\label{dex-Gt-2}  \Vert G(t) X \Vert \leq C e^{- \delta t} \Vert X \Vert,\  X \in {\cal K}_{\rm dec}, \ t>0. \ee
iii) We have the direct sum decomposition,
\be\label{som-dir}  {\cal K}=  {\cal K} _{\rm inv} \oplus  {\cal K} _{\rm dec},\ee
and we have,
\be\label{dex-Gt}  \Vert G(t) X - \pi _{\rm inv} X \Vert \leq C e^{- \delta t} \Vert X \Vert,\  X \in {\cal K}, \ee
where   $\pi_{\rm dec}$ and $\pi_{\rm inv}$ denote the  projections associated with the  decomposition (\ref{som-dir}). 

\end{theo}

Theorem \ref{expo} will proved in Section 3.3. 

Since the semigroup $G(\cdot)$ is defined and some of its properties are established, we can rewrite
 Theorem \ref{approx-SG-prov} with these informations.

\begin{theo}\label{approx-SG}   We suppose that the
form factor $E(k)$ is defined either by   (\ref{stand}) or  (\ref{dipol})
  and we assume also that the smooth ultraviolet cut-off
$\phi$ in (\ref{stand}) or (\ref{dipol}) is   vanishing at $0$.
Suppose that  the hypothesis  (FGR)  holds true and
let $G(\cdot)$ be the Markov semigroup defined in (\ref{def-cal-L})(\ref{def-Gt}) and satisfying  (\ref{dex-Gt-2})(\ref{dex-Gt}).
Then,

i) If $E(k)$ is defined by   (\ref{stand}) then  there exists $C>0$ such that we have for all
$X\in {\cal K}$, for each  $t>0$ and any   $g>0$ sufficiently small,
$$  \Vert {\cal P}_{\cal K} S^{\rm mat}  (t, g) X  - G (tg^2 )X \Vert \leq C g (1+ t^2)  \Vert X \Vert. $$

ii) If $E(k)$ is defined by   (\ref{dipol}) then there exists $C>0$ such that we have for all
$X\in {\cal K}$, for each  $t>0$ and any   $g>0$ sufficiently small,
$$  \Vert {\cal P}_{\cal K}  S^{\rm mat}  (t, g) X  - G (tg^2 )X \Vert \leq C g \Vert X \Vert. $$

\end{theo}

The starting point of the proof of Theorem  \ref{approx-SG-prov}
    is the  differential system (\ref{Sys-Dif-Init})-(\ref{defi-R2}) satisfied by $ S^{\rm mat} (t , g) X$.
  The two error terms    in  this system are estimated in Section 4.1 and Section 4.2.
    Then, the   Duhamel principle is involved in order to estimate the error between the evolution and its approximation by $G(tg^2)X$ (Section 4.4).

\vskip 2mm

 {\bf   Application to transition marginal probabilities.}

\vskip 2mm

 We denote by $\pi_v$
 the orthogonal projection  on the line of ${\cal H} _{\rm mat}$ spanned by  $v$,  for any $v\in {\cal H} _{\rm mat}$. Also recall that,
 for any vecteurs $u$ and $v$ of ${\cal H}_{\rm inf}$,  we agreed that the transition marginal probability $ P(t , u, v , g )$ from the state $u$ to the state $v$,  initially in the photon vacuum, is given by the expression (\ref{P-trans}).

In view of   (\ref{P-trans}) and according to Theorem \ref{approx-SG} and Theorem \ref{expo}, we have the following result.

   \begin{theo}\label{relax}  Assume that  the form factor $E(k)$ is defined, either by (\ref{stand})
     or by (\ref{dipol}),  and suppose that
the smooth  cutoff function $\phi$ in (\ref{stand}) or in (\ref{dipol}) is vanishing at the origin. Suppose  that the hypothesis (FGR) is satisfied and let   $\delta $ verifies  $0< \delta < \gamma$ with
    $\gamma$  given by the hypothesis  (FGR). Also suppose that  $g$ is sufficiently small.
Choose any Hilbertian basis  $(u_j)$ of ${\cal H}_{\rm inf}$ ($j\geq 0$) with
 $u_0$ being the ground state of $H_{\rm mat}$.

 i) If $E(k)$ is defined by (\ref{dipol}) then  there is $C>0$ such that the following estimates are valid for any $t>0$,
and for any $u_j$ and $u_m$,
   \be\label{A} \left | P(t , u_j, u_m , g )  - <  (  G(t g^2) \pi _{u_m}  ) u_j , u_j >
    \right | \leq C g. \ee
    If $u_m \not= u_0$, we have,
    \be\label{B} \left | P(t , u_j, u_m , g )   \right | \leq C e^{-\delta g^2 t} + C g. \ee
If $u_m = u_0$ then,
    \be\label{C} \left | P(t , u_j, u_0 , g ) -1   \right | \leq C e^{-\delta g^2 t} + C g. \ee
    ii) If $E(k)$ is defined by (\ref{stand}), we have,
   $$ \left | P(t , u_j, u_m , g )  - <  (  G(t g^2) \pi _{u_m}  ) u_j , u_j >
    \right | \leq C g (1+ t^2). $$

   \end{theo}

Indeed, if $u_m \not= u_0$ then $\pi _{\rm inv} \pi_{u_m} = 0$
   and if $u_m = u_0$ then $\pi _{\rm inv} \pi_{u_0} = I_{\rm inf}$.

Point (\ref{A}) expresses that  the approximation of the transition probabilities by a Markov process is accurate.
Points
 (\ref{B}) and (\ref{C}) reflect the  relaxation to the ground state.

 \section{Limits at infinity and consequences. }

\subsection{Operator integral representations. }

We shall use the equalities collected in  Proposition \ref{equ}  below in Section 2.33 and in Section 2.4
for the proofs of Proposition \ref{limite-L} and Proposition  \ref{limite-T},
and in Sections 4 and 5, for the proofs of the three main Theorems (Theorems 1.4, 1.6 and 1.7).
We use the notations    (\ref{def-At}),  (\ref{sigma-0}) and   (\ref{X-free}).

\begin{prop}\label{equ} We have,
\begin{align}\label{sig0-As-At}  \sigma _0 \left ( A(s) A(t) (I \otimes X) \right ) =
 \int _{\R^3}  &e^{i (t-s) |k|}  E^{\rm free} (k, s)^{\star} \
 [ E^{\rm free} (k, t),  X ] \nonumber \\[4mm]
 - &e^{i (s-t) |k|}  [ E^{\rm free} (k, t)^{\star} ,  X ] \
  E^{\rm free} (k, s) \,dk,\end{align}
 for any operator $X$ in ${\cal H} _{\rm mat}$ and for
all $(s,t)\in \R^2$, where the two sides of (\ref{sig0-As-At}) are   seen as  quadratic forms  on
 $W_4^{\rm mat}$.

 We also  have, for the  quadratic form  $L^{\omega} (t) X $  defined in   (\ref{L-omega}),
\begin{align} \label{integ-L-t} L^{\omega} (t) X =
 \int _{\R^3 \times (0, t) } e^{i \omega s }\big ( &e^{i s |k|}
 E^{\rm free} (k, - s)^{\star}\ [ E(k) , X]  \nonumber \\[4mm]
 - &e^{-is  |k|} [ E^{\star} (k) , X]  E^{\rm free} (k, - s)  \big) dk ds,
  \end{align}
Moreover,
\be\label{Tt-integ}  < T(t) u, v > = \int _{\R^3 \times (0, t)} e^{is|k|} <  E^{\rm free}(k, -s)^{\star} \ E(k) u , v > dk ds,\ee
for all $u$ and $v$ in ${\cal H}_{\rm inf}$, where
the quadratic form $T(t)$ is defined in (\ref{T(t)}).

 \end{prop}

 {\it Proof.}  From (\ref{H-int-def}) and (\ref{evol-a-k}), one has that,
  \begin{align}
  \label{H-int-fr} H_{\rm int}^{\rm free} (s) &= e^{ i s H(0)} H_{\rm int } e^{- i s H(0)} \nonumber \\[4mm]
&=  \int _{\R^3}
   e^{is|k|}  a^{\star} (k) \otimes E^{\rm free} (k, s) + e^{- is|k|}  a(k) \otimes E^{\rm free} (k, s) ^{\star} \,dk.
    \end{align}
Thus,
$$ [ H_{\rm int}^{\rm free} (t) , I \otimes X]=  \int _{\R^3}
   e^{it|p|}  a^{\star} (p) \otimes [ E^{\rm free} (p, t) , X] +  e^{- it|p|}  a(p) \otimes [ E^{\rm free} (p, t) ^{\star}, X] \, dp.  $$
One  writes the image under  $\sigma _0$ defined in (\ref{sigma-0}) of the  composition of these two operators. Then,  one uses the following properties,
$$ \sigma _0 \int _{ R^6}  a(k) a(p) \otimes F_1(k, p) + a^{\star } (k) a(p) \otimes F_2(k, p) +
a^{\star} (k) a^{\star } (p) \otimes F_3 (k, p) \,dk dp = 0 $$
and (see (\ref{COMM})),
$$  \sigma _0 \int _{ R^6}  a(k) a^{\star }(p) \otimes G(k, p) \, dk dp =
\int _{\R^3} G(k, k) dk. $$
Consequently, one deduces  (\ref{sig0-As-At}).
To derive the point (\ref{integ-L-t}), one replaces  $s$ and $t$ by
$-s$ and $0$, and integrates.
From (\ref{H-int-fr}) and since $a(k) \Psi_0 = 0$, we have,
$$  < T'(t) u , v > = \int _{\R^6} e^{it|k|} <  ( a(k) \otimes E^{\rm free}(k, -t)^{\star}  )
( a^{\star} (p) \otimes E(p)  ) ( \Psi_0 \otimes u) ,  ( \Psi_0 \otimes v) >  dk dp. $$
One obtains according to (\ref{COMM}),
$$  < T'(t)u , v > = \int _{\R^3} e^{it|k|} <  E^{\rm free}(k, -t)^{\star} \ E(k) u , v > dk,  $$
 and  therefore (\ref{Tt-integ}) is proved.

\fpr

To get useful expressions for  the  operator $e^{i s H(g)  } (a(k) \otimes I)  e^{-i s H(g)  }$,
we also use a   Dyson's type  integral. This expression will be used in
the proof of Proposition \ref{P-majo-R1} (estimation of the first error term
in Proposition \ref{deriv-prelim}).

   \begin{prop}\label{step-3-syst}  We have,
   \be\label{ak-prop}  e^{i s H(g)  } (a(k) \otimes I)  e^{-i s H(g)  } =
    e^{- i s |k| }   (a(k) \otimes I) - ig  \int _0  ^s e^{i (\sigma - s)  |k| }
     S(\sigma , g) ( I \otimes E(k) ) d\sigma \ee
     and
   \be\label{a-star-k-prop}  e^{i s H(g)  } (a^{\star} (k) \otimes I)  e^{-i s H(g)  } =
    e^{i s |k| }  (a^{\star} (k) \otimes I) + ig  \int _0  ^s e^{ i ( s- \sigma )  |k| }
     S(\sigma , g) ( I \otimes E^{\star}(k) ) d\sigma, \ee
    for all $k\in \R^3$ and $s>0$.
 \end{prop}

   {\it Proof.} According to (\ref{evol-a-k}), one computes,
\begin{align*}
 \frac {d } {dt} e^{it|k|} e^{it H(g)} (a(k) \otimes I) e^{-it H(g)}
   &= \frac {d } {dt}  e^{it H(g)}  e^{- it H(0)} (a(k) \otimes I) e^{ it H(0)}  e^{-it H(g)} \\[4mm]
   & = ig e^{it H(g)} \big [ H_{\rm int} ,  e^{- it H(0)} (a(k) \otimes I) e^{ it H(0)} \big ]
    e^{-it H(g)} \\[4mm]
    & = ig e^{it|k|} e^{it H(g)} [ H_{\rm int} ,   (a(k) \otimes I) ] e^{-it H(g)} \\[4mm]
    & =  -ig e^{it|k|} e^{it H(g)} ( I \otimes E(k) ) e^{-it H(g)}.
    \end{align*}
   Note that   (\ref{COMM}) is used. Thus, equality (\ref{ak-prop}) follows  and the proof of (\ref{a-star-k-prop}) is  similar.

    \fpr

 \subsection{Role of  Agmon inequalities. }

 If $E(k)$ is defined, either by (\ref{stand}) or by  (\ref{dipol}),
and if ${\phi }$ is in ${\cal S}(\R)$, we obtain
\be\label{majo-Ek}  \Vert E(k) f  \Vert _{W_m ^{\rm mat}}
 \leq C |k|^{-1/2} (1+ |k|)^{-N} \Vert f \Vert  _{W_{m+1}  ^{\rm mat}} \ee
\be\label{majo-Ek-star}  \Vert E^{\star} (k) f  \Vert
 \leq C |k|^{-1/2} (1+ |k|)^{-N} \Vert f \Vert  _{W_{1}  ^{\rm mat}}. \ee
One deduces,
\be\label{H-1-Ek}  \Vert ( H_{\rm mat} + i) ^{-1} E(k) f \Vert
 \leq C |k|^{-1/2} (1+ |k|)^{-N} \Vert f \Vert. \ee
 Indeed, inequality  (\ref{majo-Ek-star}) amounts to,
 $$  \Vert E^{\star} (k) ( H_{\rm mat} + i) ^{-1}    \Vert _{{\cal L} ( {\cal H} _{\rm mat} ) }
\leq C  |k|^{-1/2} (1+ |k|)^{-N} .$$
Taking the adjoint, ones obtains (\ref{H-1-Ek}).

Differentiating  $E(k)$ requires  some care. In that purpose, spherical coordinates  are needed setting   $k= \rho \omega$ ($\rho >0$ and 
$\omega \in S^2$). One observes that,
   \be\label{deriv-FF}   \Vert \partial _{\rho}^{\alpha  } \rho^{1/2}  E(\rho \omega ) u \Vert _{W_m^{\rm mat} }  ^2
   \leq C (1+ \rho)^{-N} \sum _{\gamma \leq m+1}  \int _{\R^3 }  (1+ |x|) ^{2\alpha }
    | D^{\gamma} u(x) |^2 dx, \ee
 for all $u\in {\cal S}(\R^3)$.

 Agmon inequalities \cite{A82} show that the above right hand side is finite for every $u\in {\cal H}_{\rm inf}$ since the supremum  $E_0$ of $S_{\rm inf}$ is smaller than the limit at infinity of the potential $V$. Then, one can write using equivalence of norms on ${\cal H}_{\rm inf}$,  
 \be\label{deriv-Ek}  \Vert(  \partial_{\rho}  ^{\alpha }\rho^{1/2}    E (\rho \omega )  ) u
 \Vert
 \leq   C (1+ \rho)^{-N} \Vert u \Vert  ,\ee
 for any $\alpha$ and $N$, and every  $u\in {\cal H}_{\rm inf}$.

Similarly,
$$  \Vert(  \partial_{\rho}  ^{\alpha }\rho^{1/2}    E (\rho \omega )  ) u \Vert_{W_m^{\rm mat} }
\leq C (1+ \rho )^{-N} \Vert u \Vert   $$
and
$$  \Vert(  \partial_{\rho}  ^{\alpha }\rho^{1/2}    E^{\star}  (\rho \omega )  ) u \Vert_{W_m^{\rm mat} }
\leq C (1+ \rho )^{-N} \Vert u \Vert ,  $$
for all  $u\in {\cal H}_{\rm inf}$.

One has, 
  \be\label{der-Ek-Pi-inf}  \Vert \partial _{\rho}^{\alpha } \big  ( \rho^{1/2}
  \Pi_{\rm inf}    E (\rho \omega ) f  \big ) \Vert
  \leq C_{\alpha N}  (1+ \rho)^{-N} \Vert f \Vert,  \ee
for any $f$ in ${\cal H} _{\rm mat}$.
To see this, let  $(v_j)$ be a Hilbertian basis of ${\cal H} _{\rm inf}$.
   Since  ${\cal H} _{\rm inf}$ is of finite dimension, one has,   
  $$ \Vert \partial _{\rho}^{\alpha }   ( \rho^{1/2}
  \Pi_{\rm inf}    E (\rho \omega ) f   ) \Vert
  \leq   C  \sum _j | <  \partial _{\rho}^{\alpha }
    ( \rho^{1/2}    E (\rho \omega ) f) , v_j  > | $$
   $$  \leq   C \Vert f \Vert   \sum _j
    \Vert  \partial _{\rho}^{\alpha }
     ( \rho^{1/2}    E^{\star}  (\rho \omega )  v_j )  \Vert .$$

 \subsection{Limits: proof of Proposition \ref{limite-L}. }

 {\it Point i)} is (\ref{integ-L-t}).

{\it Point ii)} Let  $Z$ be either in  ${\cal L} ( {\cal H}_ {\rm inf})$ or in 
${\cal L} ( {\cal H}_ {\rm sup})$. Fix  $u$ and  $v$ in  ${\cal H}_ {\rm inf}$. Set $\omega \in \R$.  Starting from (\ref{integ-L-t}), that is,
\begin{align*}  < (L ^{\omega } (t) Z) u, v >
 =
 \int _{\R^3 \times (0, t)  } e^{i\omega s}   (&e^{i s |k|}
<  E^{\rm free} (k, - s)^{\star}\ [ E(k) , Z] u, v >   \\[4mm]
 - &e^{-is  |k|} <  [ E^{\star} (k) , Z]  E^{\rm free} (k, - s) u, v > )  dk ds, \end{align*}
one observes that, $ < (L ^{\omega } (t) Z) u, v > $ is a linear combination of the following four functions:
$$  F_1 (t) = \int _{\R^3 \times (0, t) } < e^{is ( H_{\rm mat} + |k| + \omega) } E(k) Z u ,
 E(k) e^{is H_{mat }} v > dk ds $$
 $$  F_2 (t) = \int _{\R^3 \times (0, t) } < e^{is ( H_{\rm mat} + |k| + \omega) } Z  E(k)  u ,
 E(k) e^{is H_{mat }} v > dk ds  $$
 $$  F_3 (t) = \int _{\R^3 \times (0, t) } < e^{-is ( H_{\rm mat} + |k| - \omega) }  E(k) e^{is H_{mat }} u ,
 Z^{\star} E(k)  v > dk ds $$
 $$  F_4 (t) = \int _{\R^3\times (0, t) } < e^{-is ( H_{\rm mat} + |k| - \omega) }   E(k) e^{is H_{mat }} u ,
 E(k) Z^{\star}   v > dk ds.  $$
 One notes that if  $u$ and $v$ belong to ${\cal H}_{\rm inf}$, and if  $X$ is in either
 ${\cal L} ({\cal H}_{\rm inf})$ or  ${\cal L} ({\cal H}_{\rm sup})$ then  $Zu$ and  $Z^{\star } v$ lie in ${\cal H}_{\rm inf}$. If  $X\in {\cal L} ({\cal H}_{\rm sup})$ then  $Zu = Z^{\star } v = 0$.
 In the aim to prove the existence of limits, we shall bound the derivatives of these four functions.
We then use spherical coordinates setting  $k = \rho \omega$ with  $\rho >0$ and  $\omega \in S^2$. For example, 
$$ \frac {d} { dt}  F_1 (t) = \int _{\R_+ \times S^2} < e^{it ( H_{\rm mat} + \rho + \omega) } E(\rho \omega) Z u \ ,\
E(\rho \omega ) e^{it H_{\rm mat }} v > \rho ^2 d\rho d\sigma (\omega). $$
Next, we integrate by parts in the  variable $\rho$.
If the function $\phi$ in (\ref{stand}) or (\ref{dipol}) is vanishing at the origin at the order $ p$,  then  $2p+1$ integrations by parts lead to,
 $$ t ^{2p+1} \frac {d} { dt}  F_1(t)= \hskip 10 cm $$
 $$  \sum _{\alpha + \beta = 2p+1}
 \int _{\R_+ \times S^2}  a_{\alpha \beta }  < e^{it ( H_{\rm mat} + \rho + \omega) }
   \partial _{\rho}^{\alpha  } \Big ( \rho^{1/2}  E(\rho \omega )Z u \Big ) \ ,\
 \partial _{\rho}^{\beta  } \Big ( \rho^{1/2}  E(\rho \omega )e^{it H_{mat }} v  \Big ) >
 \rho  d\rho d\sigma (\omega)   $$
  $$ +  \sum _{\alpha + \beta = 2p}
   \int _{\R_+ \times S^2} b_{\alpha \beta }  < e^{it ( H_{\rm mat} + \rho + \omega) }
    \partial _{\rho}^{\alpha  } \Big ( \rho^{1/2}  E(\rho \omega )Z u \Big ) \ ,\
 \partial _{\rho}^{\beta  } \Big ( \rho^{1/2}  E(\rho \omega )e^{it H_{mat }} v  \Big ) >
  d\rho d\sigma (\omega),   $$
where the  $a_{\alpha \beta }$ and  $b_{\alpha \beta }$ are  real constants.
 One can apply  (\ref{deriv-Ek}) since   $Z u $ and  $e^{it H_{mat }} v $
are in  ${\cal H} _{\rm inf}$. One obtains,
 $$ \left  |  \frac {d} { dt} F_1(t) \right | \leq \frac {K} {1+t^{2p+1} }
  \Vert Z  \Vert  \   \Vert u  \Vert    \Vert v  \Vert.$$
 The others terms $F_j (t)$ ($2 \leq j \leq 4$) are similarly bounded.
As a consequence,
  $$ \left | \frac {d} {dt}  < (L ^{\omega } (t) Z) u, v > \right |
  \leq \frac {K} {1+t^{2p+1} }
  \Vert Z  \Vert  \   \Vert u  \Vert    \Vert v  \Vert, $$
proving the existence of the limit if  $p\geq 1$.

  {\it Point  iii)}  Set $Z\in {\cal L} ( W_2 ^{mat}, W_0 ^{mat} )$,
  $u \in {\rm Ker } ( H_{\rm mat} - \lambda I)$ and 
  $v \in {\rm Ker } ( H_{\rm mat} - \mu I)$, $\omega = \mu - \lambda $ with  $\lambda $ and $\mu $ in $S_{\rm inf}$.
  The four  terms $F_j(t)$ are the same as those above but  have more precise expressions in view of the hypotheses on $u$, $v$ and
   $\omega$. Namely, one has,
$$  F_1 (t) = \int _{\R^3 \times (0, t) } < e^{is ( H_{\rm mat} + |k| - \lambda) } E(k) Z u ,
 E(k)  v > dk ds $$
 $$  F_2 (t) = \int _{\R^3 \times (0, t) } < e^{is ( H_{\rm mat} + |k| - \lambda) } Z  E(k)  u ,
 E(k)  v > dk ds  $$
 $$  F_3 (t) = \int _{\R^3 \times (0, t) } < e^{-is ( H_{\rm mat} + |k| - \mu) }   E(k)  u ,
 Z^{\star} E(k)  v > dk ds $$
 $$  F_4 (t) = \int _{\R^3\times (0, t) } < e^{-is ( H_{\rm mat} + |k| - \mu) }   E(k) u ,
 E(k) Z^{\star}   v > dk ds. $$
First consider the term $F_1(t)$.
Let $\Pi_{\rm sup}$ (resp. $\Pi_{\rm inf}$)  be the spectral projection of 
 $H_{\rm mat}$ on the interval $[E_0 , + \infty )$ (resp.
   $(-\infty , E_0)$). One has $F_1 (t) = F_1^{\rm sup}  (t) + F_1^{\rm inf}  (t)$,
with 
  $$ F_1^{\rm sup}  (t)  = \int _{\R^3 \times (0, t) }
  < e^{is ( H_{\rm mat} + |k| -\lambda ) } \Pi_{\rm sup}  E(k) Z u ,
 E(k)  v > dk ds  $$
  and
  $$ F_1^{\rm inf}  (t)  = \int _{\R^3 \times (0, t) }
  < e^{is ( H_{\rm mat} + |k| -\lambda ) } \Pi_{\rm inf}  E(k) Z u ,
 E(k)  v > dk ds.  $$
Also, 
$$  F_1^{\rm sup}  (t)  = \int _{\R^3  }
< A(t, k , \lambda ) \Pi_{\rm sup}  E(k) Z u ,  E(k)  v > dk $$
with
\be\label{Atkl}  A(t, k , \lambda ) = \int _0 ^t e^{is ( H_{\rm mat} + |k| -\lambda ) } \Pi_{\rm sup} ds. \ee
There exists  $C>0$ such that for all  $t>0$,  $\lambda  \in S_{\rm inf}$
and $k\in \R^3$, 
$$ \Vert A(t, k , \lambda ) f \Vert  \leq C \Vert ( H_{\rm mat} + i) ^{-1} f \Vert. $$
Thus,
$$ |  F_1^{\rm sup}  (t) | \leq C \int _{\R^3 }  \Vert ( H_{\rm mat} + i) ^{-1} E(k) Z u \Vert \ \Vert E(k) v \Vert dk.$$
One then deduces according to  (\ref{majo-Ek}) and (\ref{H-1-Ek}) that,
$$ |  F_1^{\rm sup}  (t) | \leq C  \int _{\R^3 } |k|^{-1} (1+ |k|)^{-2N}
\Vert Zu \Vert \   \Vert v \Vert _{W_1^{\rm mat}} dk.  $$
One knows that  ${\cal H} _{\rm inf}$ is included in  $W_1^{\rm mat}$ and that the norm are equivalent on   the finite dimensional space  ${\cal H} _{\rm inf}$. Therefore, $Zu$
is well defined and the following equality holds,
  $$  |  F_1^{\rm sup}  (t) | \leq C  \Vert Zu \Vert \ \Vert v \Vert
  \leq C \Vert Z \Vert _{{\cal L} ( W_2 ^{mat}, W_0 ^{mat} )} \
  \Vert u \Vert \ \Vert v \Vert. $$
Now, in order to get a bound on  $ |  F_1^{\rm inf}  (t) |$, one checks that, 
  \be\label{deriv-Finf}  \left | \frac {d} {dt }  F_1^{\rm inf}  (t) \right | \leq
  \frac {K} {1+t^{2p+1} } \Vert Z u \Vert \     \Vert v  \Vert.\ee
  To this end, one integrates by parts in the variable  $\rho$ as in  point ii). The only difference is that one now  uses  inequality (\ref{der-Ek-Pi-inf})
applied with  $f= Zu$.  One then obtains,
   $$ | F_1(t) | \leq C   \Vert Z \Vert_{ {\cal L} ( W_2^{\rm mat},
  W_0^{\rm mat}) }   \  \Vert u \Vert \ \Vert v \Vert. $$
  Second, the bounds on $F_2(t)$ and  $F_3(t)$ are simpler. They are  effectuated as in Point ii), with a bound of the derivative, using integrations by parts, without splitting the expression into two terms. One then sees,
  $$  | F_2(t) | +   | F_3(t) | \leq C \Vert Z \Vert_{ {\cal L} ( W_2^{\rm mat},
  W_0^{\rm mat}) }   \  \Vert u \Vert \ \Vert v \Vert. $$
  Finally,  the term $F_4(t)$ is estimated as  $F_1(t)$. One gets,
   $$ | F_4(t)| \leq C \Vert u \Vert \ \Vert Z^{\star } v \Vert. $$
 Point iii) then follows.

  {\it Point iv) } Let  $Z$ in ${\cal L}({\cal H}_{\rm mat})$.
Set  $\lambda $ and  $\mu $ in  $S_{\rm inf}$, $\omega = \mu - \lambda$,
  $u\in {\rm Ker} ( H_{\rm mat} - \lambda I)$ and $v\in {\rm Ker} ( H_{\rm mat} - \mu I)$.
  We shall prove that under our hypothesis, the four  terms $F_j(t)$  have limits when  $t$ tends to  $+\infty$.
First consider $F_1(t)$. The term $F_1^{\rm inf}  (t)$ has a limit from inequality (\ref{deriv-Finf}) that is still satisfied here.
  Its proof is not using the fact that  $Z$ belongs to ${\cal L} ({\cal H}_{\rm sup})$,
  neither that it belongs to ${\cal L} ({\cal H}_{\rm inf})$.
   For the term $F_1^{\rm sup}  (t)$, one notices that the operator  $A(t, k , \lambda )$
   defined in  (\ref{Atkl}) satisfies,   
   $$   A(t, k , \lambda )= \big (  e^{it ( H_{\rm mat} + |k| -\lambda ) } - I
   \big )  \Pi_{\rm sup} ( H_{\rm mat} + |k| -\lambda )^{-1}. $$
   One uses   \cite{RS} (Volume III, page 24, Lemma 2). According to this Lemma,
   $< e^{it ( H_{\rm mat} + |k| -\lambda ) } \varphi,
   \psi >$ tends to 0 as $t$ goes to $+\infty$
 for every  $\varphi$ and $\psi $ belonging  to the absolutely continuous spectral subspace of  $H_{\rm mat}$.
Thus, 
 if $[0, \infty)$ belongs to the absolutely continuous spectrum of  $H_{\rm mat}$ then one gets,     
     $$ \lim _{t\rightarrow \infty  } < A(t, k , \lambda ) \Pi_{\rm sup}  E(k) Z u ,  E(k)  v > =
    -  <  ( H_{\rm mat} + |k| -\lambda )^{-1} \Pi_{\rm sup}   E(k) Z u ,  E(k)  v >,  $$
  for all  $k\in \R^3$.
Besides, one has,
    $$ \big | < A(t, k , \lambda ) \Pi_{\rm sup}  E(k) Z u ,  E(k)  v > \big | \leq
     C \Vert (H_{\rm mat } + i )^{-1} E(k) Zu \Vert \ \Vert E(k) v \Vert. $$
  The above right hand side is a function in   $L^1(\R^3)$ in view of 
     (\ref{majo-Ek}) and   (\ref{H-1-Ek}). This function is independent of  $t$. 
Point iv) then comes from the dominated convergence Theorem.

 \fpr

 The proof of Proposition \ref{limite-L} is then completed.

  \fpr

 \subsection{Limits: proof of Proposition  \ref{limite-T} }

 {\it Point i)} is (\ref{Tt-integ}).

{\it Point ii) }   One assumes that the function $\phi$ in (\ref{stand}) or (\ref{dipol}) is vanishing at the origin at the order  $p\geq 1$.
One has from (\ref{Tt-integ}),
$$  < T'(t)u , v > = \int _{\R^3} e^{it|k|}   <  e^{itH_{\rm mat}}  E(k) u ,
E(k)   e^{itH_{\rm mat}} v > dk, $$
$$ = \int _{\R_+ \times S^2}  <  e^{it(H_{\rm mat} + \rho ) }  E(\rho \omega ) u,
E(\rho \omega )  e^{itH_{\rm mat}} v > \rho ^2 d\rho d \sigma (\omega) $$
for all $u$ and $v$ in ${\cal H}_{\rm inf}$.  One estimates this integral as for the  term $F_1 (t)$ in the latter Section. One obtains,
$$  | < T'(t)u , v > | \leq  \frac {C}  { 1+  t^{2p+1}  \  }  \
\Vert u \Vert \ \Vert v \Vert, $$
for  all $u$ and $v$   in ${\cal H}_{\rm inf}$.
One thus deduces the existence of the limit if  $ p\geq 1$.  Point ii)  then follows.

  {\it  Point iii) } From Point i), if  $u\in {\cal H}_{inf}$ and $v\in {\rm Ker} (H_{\rm mat} - \mu I)$ ($\mu \in S_{inf}$) then the function
  $$  s  \mapsto F ( s) = \int _{\R^3} < e^{is ( H_{\rm mat} + |k| - \mu) } E(k) u, E(k) v > dk $$
  belongs to $L^1 (  \R_+)$ and one has,
  $$ < Tu, v > = \int _{ \R_+}  F( s)  ds. $$
According to the dominated convergence Theorem,
  $$ < T u, v > = \lim _{\varepsilon \rightarrow 0}
  \int _{ \R_+} e^{- \varepsilon s }  F( s)  ds $$
  $$ = \lim _{\varepsilon \rightarrow 0}
  \int _{\R^3 \times  \R_+} e^{- \varepsilon s }
   < e^{is ( H_{\rm mat} + |k| - \mu) } E(k) u, E(k) v > dk ds. $$
For all  $\varepsilon >0$, the function
 $$ (k, s) \rightarrow G(\varepsilon, k, s) =  e^{- \varepsilon s }
   < e^{is ( H_{\rm mat} + |k| - \mu) } E(k) u, E(k) v > $$
belongs to $L^1 (\R^3 \times \R_+)$.
For every $\varepsilon >0$ and any  $k\in \R^3$, one sees,
  $$  \int _{\R_+}  G(\varepsilon, k, s)   ds =
  i < ( H_{\rm mat} + |k| - \mu + i \varepsilon )^{-1} E(k) u, E(k) v >.$$
  From  Fubini Theorem, for all $\varepsilon >0$,
  $$ \int _{\R^3 \times  \R_+} e^{- \varepsilon s }
   < e^{is ( H_{\rm mat} + |k| - \mu) } E(k) u, E(k) v > dk ds   $$
   $$ =
   i \int _{\R^3} < ( H_{\rm mat} + |k| - \mu + i \varepsilon )^{-1} E(k) u, E(k) v >
   dk. $$
   Point iii) is then proved.

 {\it Point iv). }   For each $\lambda \in S_{\rm inf}$
and for each $u \in {\rm Ker} ( H_{\rm mat} - \lambda I)$, we have from (\ref{Tt-integ}),
  $$   < ( T _{\lambda}+ T_{\lambda} ^{\star}) u, u > =  2 \lim _{t\rightarrow \infty} \int _{\R^3 \times (0, t) }
     <  \cos (  s ( H_{\rm mat} + |k| - \lambda) )  E(k) u , E (k) v > dk ds.  $$
It is standard that, 
  $$ \lim _{t\rightarrow \infty} \int _{\R^3 \times (0, t) }
   <  \cos (  s ( H_{\rm mat} + |k| - \lambda) ) \Pi ^{\rm sup} (\lambda ) E(k) u ,
   \Pi ^{\rm sup} (\lambda ) E (k) v > dk ds  = 0. $$
  For all $\rho \in S_{\rm inf} $ with $\rho < \lambda$, it also standard  using Proposition \ref{lim-int} in  Section 3.1
  below that, 
  $$ \lim _{t\rightarrow \infty} \int _{\R^3 \times (0, t) }
   <  \cos (  s ( H_{\rm mat} + |k| - \lambda) ) \Pi  (\rho ) E(k) u ,
  \Pi  (\rho ) E(k) u > dk ds  $$
  $$ = 2 \pi \int _{|k| = \lambda - \rho } \Vert \Pi  (\rho ) E(k) u \Vert ^2 d\sigma (k). $$
Thus Point iv) is derived.

{\it Point v). }  Similarly, one gets for each $\lambda \in S_{\rm inf}$
and for each $u \in {\rm Ker} ( H_{\rm mat} - \lambda I)$,
  $$   < ( T_{\lambda}  - T_{\lambda}^{\star}) u, u > =
  2 i  \lim _{t\rightarrow \infty} \int _{\R^3 \times (0, t) }
     <  \sin (  s ( H_{\rm mat} + |k| - \lambda) )  E(k) u , E (k) u > dk ds.   $$
One knows that,
$$ \lim _{t\rightarrow \infty} \int _{\R^3 \times (0, t) }
< \sin (  s ( H_{\rm mat} + |k| - \lambda) )  \Pi ^{\rm sup} (\lambda )  E(k) u ,
\Pi ^{\rm sup} (\lambda )  E(k) u > dk ds  $$
$$ =  \int _{\R^3 } < ( H_{\rm mat} + |k| - \lambda) ) ^{-1}
\Pi ^{\rm sup} (\lambda )  E(k) u ,
\Pi ^{\rm sup} (\lambda )  E(k) u > dk ds . $$
For all $\rho \in S_{\rm inf} $ with $\rho < \lambda$, one classically has using Proposition \ref{lim-int} below that,
$$ \lim _{t\rightarrow \infty} \int _{\R^3 \times (0, t) }
< \sin (  s ( H_{\rm mat} + |k| - \lambda) )  \Pi  (\rho )  E(k) u ,
\Pi  (\rho )  E(k) u > dk ds = $$
$$ = {\rm PV} \int _{\R^3} \frac { < \Pi(\rho) E(k) u, \Pi(\rho) E(k) u > }
{ |k| + \rho - \lambda } dk, $$
where $  {\rm PV}  $ denotes the principal value of singular  integrals.
One then deduces (\ref{Bethe}).

\fpr

\section{Properties of the semigroup. }

\subsection{The Bethe-Salpeter matrix.}

Let ${\cal K}$ be the operator algebra of Definition \ref{K-P},  that is, the  algebra of
 operators  $X\in {\cal L} ( {\cal H} _{\rm inf})$
commuting with the restriction of $H_{\rm mat}$. Also let ${\cal L}$
be the operator from ${\cal K}$ into itself defined by (\ref{def-cal-L})
using the operator $L_{\infty}$ of Proposition \ref{limite-L}.
In this section, the parameter $\omega$ is 0 and will be omitted.

Our purpose here is to make explicit the map  ${\cal L}$. It is highly likely that the numerical table in
\cite{B-S} (table 15 page 266) corresponds up to the factor $g^2$ to the matrix in some basis of the operator ${\cal L}$  studied below. To do this, we shall prove inequality (\ref{LXuv}) but   the second term of the right hand side of the inequality is  this time written down.
\begin{theo}\label{Bet-Sap}  We assume that the
form factor is defined either by (\ref{stand}) or (\ref{dipol}) and that the function
$\phi$ in (\ref{stand}) or (\ref{dipol}) vanishes  the origin at order $p\geq 1$. Then we have,
for every $X\in {\cal K}$, for any $\lambda \in S_{\rm inf}$,
for all $u$ and $v$ in $E(\lambda) = {\rm Ker } ( H_{\rm mat} - \lambda I )$,
\begin{align}\label{Bet-Salp-2}  < ( {\cal L} X) u , v > &=  < ( T_{\lambda} \Pi (\lambda ) X 
+ \Pi (\lambda ) X T_{\lambda}^{\star}) u, v >\nonumber \\[4mm]
&- 2 \pi \sum _{\mu \in S _{\rm inf} \atop \mu < \lambda }
\int _{ |k| = \lambda - \mu} < X \Pi (\mu) E(k) u , \Pi(\mu) E(k) v > d\sigma (k),\end{align}
 where $\Pi (\lambda )$ denotes the orthogonal projection on $E(\lambda)$, 
 $T_{\lambda}$ is the operator of Proposition \ref{limite-T}, $ E(k)$
is the form factor  defined either by (\ref{stand}) or (\ref{dipol}),
 and $d\sigma$  is the surface measure on the sphere.
\end{theo}

The proof of Theorem \ref{Bet-Sap} uses  the following standard proposition.
 \begin{prop}\label{lim-int}  Fix any  function $F$ continuous  on $\R^3$ and rapidly decreasing at infinity. In the case
 $\lambda >0$, we have,
 $$ \lim _{t\rightarrow \infty} \int _{\R^3 \times (0, t) }
  \cos ( s(  |k| -\lambda  )) F(k) dk = \pi \int _{ |k| = \lambda }
  F(k) d\sigma (k),$$
where $d\sigma $  is the surface measure
on the sphere. In the case $ \lambda \leq  0$, we have,
 $$ \lim _{t\rightarrow \infty} \int _{\R^3 \times (0, t) }
  \cos ( s(  |k| -\lambda )) F(k) dk = 0. $$
  If $\lambda >0$, we have, for each suitable function $F$,
 $$ \lim _{t\rightarrow \infty} \int _{\R^3 \times (0, t) }
  \sin  ( s(  |k| -\lambda )) F(k) dk = PV \int _{\R^3}
  \frac { F(k)} {|k| - \lambda } dk. $$

 \end{prop}

 {\it Proof of  Theorem \ref{Bet-Sap}.}
Since $u$ and $v$ are eigenfunctions of  $H_{\rm mat}$ with the same eigenvalue $\lambda \in S_{\rm inf}$, one has
according to (\ref{integ-L-t}) with $\omega = 0$,
 $$     < ( {\cal L}  X)u , v >  =  < (L _{\infty}  X)u , v > = I_1 + I_2 $$
  with
\begin{align*}  I_1 = \lim _{t\rightarrow \infty}
\int _{\R^3 \times (0, t) }  \big ( &e^{i s |k|}
 <  E^{\rm free} (k, - s)^{\star}\  E(k)  Xu , v >  \\[4mm]
  +  &e^{-is  |k|} < X   E^{\star} (k)   E^{\rm free} (k, - s) \big )u , v >  dk ds \end{align*}
and
\begin{align*} I_2 = - \lim _{t\rightarrow \infty}
\int _{\R^3 \times (0, t) }  \big ( &e^{i s |k|}
<  E^{\rm free} (k, - s)^{\star}\ X \  E(k)  u , v >   \\[4mm]
+ &e^{-is  |k|} <    E^{\star} (k) X  \  E^{\rm free} (k, -s)  u , v > \big) dk ds.  \end{align*}
From (\ref{Tt-integ}), (where $t$ tends to infinity),
one gets $I_1 =   < ( T_{\lambda} X + X T_{\lambda}^{\star}) u, v > $.
One notes that,
$$ I_2 = - 2 \lim _{t\rightarrow \infty} \int _{\R^3 \times (0, t) }
  < X \cos ( s( H_{\rm mat} + |k| - \lambda )) E(k) u , E(k) v > dk ds. $$
With $\Pi _{\rm sup} (\lambda )$ standing for the orthogonal projection on the spectral subspace of  $H_{\rm mat}$ for the interval $[\lambda , + \infty)$, one writes $I_2$ as,
 \begin{align*} I_2 = &- 2 \lim _{t\rightarrow \infty} \int _{\R^3 \times (0, t) }
< X \cos ( s( H_{\rm mat} + |k| - \lambda ))\Pi _{\rm sup} (\lambda )  E(k) u , E(k) v > dk ds\\[4mm]
 &- 2 \lim _{t\rightarrow \infty} \sum _{\mu \in S _{\rm inf} \atop \mu < \lambda }
\int _{\R^3 \times (0, t) }
 < X \cos ( s( H_{\rm mat} + |k| - \lambda ))\Pi (\mu)    E(k) u , E(k) v > dk ds. \end{align*}
Note that the fact that $X$ commutes with $H_{\rm mat}$ is used above.
 The first term is vanishing from Proposition \ref{lim-int}.
Consequently,
$$ I_2 = - 2 \lim _{t\rightarrow \infty} \sum _{\mu \in S _{\rm inf} \atop \mu < \lambda }
\int _{\R^3 \times (0, t) }
< X \cos ( s(  |k|+ \mu  - \lambda ))\Pi (\mu)   E(k) u , E(k) v > dk ds. $$
From Proposition \ref{lim-int}, one gets,
$$ I_2 = - 2 \pi  \sum _{\mu \in S _{\rm inf} \atop \mu < \lambda }
\int _{|k| = \lambda - \mu} < X \Pi (\mu)   E(k) u , E(k) v > dk ds $$
which proves Theorem  \ref{Bet-Sap}.

\fpr

\subsection{Generator of a Markov semigroup.}

We consider a finite dimensional Hilbert space $E$    written as a direct sum decomposition of orthogonal  finite dimensional  subspaces  $E_j$ ($0\leq j \leq N$). Namely,

  $$ E =  \bigoplus _{j\geq 0 }  E_j. $$
The operator  $\Pi_j$ denotes the  orthogonal projection on $E_j$.
The space ${\cal K}$  is the set of operators  $X   \in {\cal L}(E)$ commuting with all the  $\Pi_j$ and ${\cal K}_{\R}$ denotes the self-adjoint operators belonging to ${\cal K}$.

We also consider a linear map 
 ${\cal L}$ from ${\cal K}_{\R}$
to ${\cal K}_{\R}$. Our aim in this section is to give a sufficient condition implying that  $G(t) = e^{-t {\cal L}}$ is  a Markov semigroup in ${\cal K}_{\R}$.
(Definition \ref{def-Markov}).

One can write using $\sum \Pi_j = I$,
 $$ {\cal L} (X) =  \sum_{j ,  m }   \Pi_j ( {\cal L} ( X \Pi_m) ),$$
for every  $X\in {\cal K}$.

The assumptions on the map   ${\cal L}$ are the following ones:

 \hskip 1cm{\bf (H1)} \  One has 
 $$  \Pi_j  {\cal L} ( X \Pi_m)  = 0, \  {\rm if} \ j\leq m,$$
for all $X\in {\cal K}_{\R}$.

  \hskip 1cm{\bf (H2)} \   If  $j>m$ and if  $X   \in {\cal K}_{\R}$ is nonnegative then
   $ \Pi_j  {\cal L} ( X \Pi_m) $ (which is self-adjoint since 
  $\Pi_j  {\cal L} ( X \Pi_m)  = \Pi_j  {\cal L} ( X \Pi_m)  \Pi_j$ and since  $X \Pi_m = \Pi_m X \Pi_m \in {\cal K}_{\R}$) is nonpositive.

   \hskip 1cm{\bf (H3)} \    For each  $j\geq 0$, there exists an element
   $T _j\in {\cal L}(E_j)$ satisfying  for all $X \in {\cal K}_{\R}$, 
 $$ \Pi_j  {\cal L} ( X \Pi_j)  =  T_j \Pi_j X + X \Pi_j   T_j^{\star}. $$
Here $T_j^{\star}$ is the adjoint of  $T_j$ and it is therefore also an element of 
 ${\cal L}(E_j)$.

   \hskip 1cm {\bf (H4)}\   One has  ${\cal L} (I) = 0$.

Notice that these hypotheses imply that 
 $T_j + T_j^{\star}$ is nonnegative, for each  $j$. We shall below prove the next result.

\begin{theo}\label{abstr} Under the above hypotheses, the family $G(t) = e^{-t{\cal L}}$
($t>0$) is a Markov 
semigroup  in ${\cal K}_{\R}$.

\end{theo}

We shall use the three following Lemmas for the proof of Theorem \ref{abstr}.

\begin{lemm}\label{pr-enc-LX} Fix $X\in {\cal K}_{\R}$.  If $f \in   E_j$ ($j\geq 0$)
 is satisfying $Xf = \lambda f$ with $\lambda \in \R$ then 
 \be\label{encad-LX}   ( \lambda - \Vert X \Vert ) < ( T_j + T_j^{\star} ) f , f >
 \leq < {\cal L} (X) f, f > \leq
 ( \lambda +  \Vert X \Vert ) < ( T_j + T_j^{\star} ) f , f >.  \ee 

\end{lemm}

{\it Proof of Lemma \ref{pr-enc-LX}.}  Under these hypotheses, one has  $\Vert X \Vert  I - X \geq 0$ and
$  X + \Vert X \Vert I \geq  0$. According to  (H2), if $m\not = j$ then,
$$ \Vert X \Vert  \  \Pi_j  ( {\cal L}(\Pi_m))  \leq \Pi_j ( {\cal L}( \Pi_m X ) )
\leq  - \Vert X \Vert \  \Pi_j ({\cal L}(\Pi_m)).$$
One then sums up these inequalities over $m \not = j$.
From  (H4), one sees,
$$ -  \Vert X \Vert \ \Pi_j  ( {\cal L}(\Pi_j))  \leq \Pi_j ({ \cal L} X) -
\Pi_j ( {\cal L} \Pi_j X)  \leq  \Vert X \Vert \  \Pi_j  ( {\cal L}(\Pi_j)).$$
In view of  (H3), one checks that, 
$$ -  \Vert X \Vert \  ( T_j + T_j^{\star}) \leq
\Pi_j ({ \cal L} X) -  (\Pi_j T_j X + X \Pi_j T_j^{\star})
 \leq   \Vert X \Vert \  ( T_j + T_j^{\star}).   $$
If $f\in E_j$ satisfies $Xf = \lambda f$ then one deduces  (\ref{encad-LX}). The Lemma is proved.

\fpr

We also use the following result.

 \begin{lemm}\label{lpa} Assume that hypothesis (H1) is satisfied. Take $X$ an element of ${\cal K}$ and $f$ a continuous function on 
 $[0, \infty)$ into ${\cal K}$. Fix $j\leq N$. Then the function
 $u \in C^1 ( [0, \infty) , {\cal K}) $ satisfying, 
 $$\frac { du } {dt } = - \Pi_j   ( {\cal L} u(t)) + \Pi_j f(t) $$
together with  $u(0) = \Pi_j X$ is given by,
\be\label{sol-sys} u(t) =  e^{-t T_j}\Pi_j X e^{-t T_j^{\star}}  +
 \int _0^t  \Pi_j e^{(s-t) T_j}(\Pi_j f(s)) e^{(s-t) T_j^{\star}} ds.  \ee 

 \end{lemm}

 {\it Proof of Lemma \ref{lpa}.}  One sees that the   function $u$ defined in  (\ref{sol-sys}) is satisfying,
 $$ \frac { du } {dt } = - T_j u(t) - u(t) T_j^{\star}  + \Pi_j f(t). $$
Besides $u(t) = \Pi_j u(t)$ and therefore $u(t) \Pi_m = 0$
  if $j\not = m$.  One then has according to (H1), 
   $$ {\cal L} u(t) = T_j u(t) + u(t) T_j^{\star}. $$
Thus,
   $$ \frac { du } {dt } = \Pi_j \frac { du } {dt } =
   -  \Pi_j {\cal L} u(t) +  \Pi_j f(t). $$
Finally, since $u(0) = \Pi_j X$, the Lemma holds true.

     \fpr

  The third Lemma is concerned with a "matrix" expression of the semigroup.    For that purpose, one writes, 
    \be\label{def-phi-jm} e^{- t{\cal L}} X = \sum _{jm}  \phi _{jm}  (t,X),\quad \phi _{jm}  (t, X) = \Pi_j \left ( e^{- t{\cal L}}(\Pi_m  X ) \right ), \quad  X\in{\cal K}.\ee 

     \begin{lemm}\label{matr-exp}   Assume that  hypotheses (H1)-(H4) are  satisfied. Then all the following properties hold true.
     
     i) One has $\phi _{jm}  (t, X) = 0$   if $j<m$.

     ii) One has,
    $$  \phi _{jj}  (t, X) = e^{-t T_j } \Pi_j X  e^{-t T_j^{\star}  } .$$
    iii) If $j>m$ then,
\be\label{phi-jm-int}  \phi _{jm}  (t,X)  = \Pi_j \int _0^t e^{(s-t) T_j } f_{jm}(s)  e^{(s-t) T_j^{\star}  } ds  \ee 
with
\be\label{f-j-m}  f_{jm} (t,X) = -
\sum _{p= m}^{j-1}  (\Pi_j {\cal L} (\Pi_p  \phi_{pm}  (t,X))) . \ee

     \end{lemm}

     {\it Proof of Lemma \ref{matr-exp}.} First, Point i)  is a direct consequence of  (H1).
      Next, one verifies Point ii). One notices that,
   $$ \frac { d } {dt } \phi _{jj}  (t, X) =- \Pi_j ( {\cal L} e^{-t {\cal L}}
   ( \Pi_j X))  $$
   $$ = -\sum _k \Pi_j ( {\cal L} ( \Pi_k  e^{-t {\cal L}}(  \Pi_j X)))  . $$
   In view of hypothesis (H1), one has $\Pi_j ( {\cal L} ( \Pi_k  Z) = 0 $
   for all $Z\in {\cal K}$ if $j<k$. One also sees that  $\Pi_k  e^{-t {\cal L}}
   ( \Pi_j X) = 0$ if $j>k$.  Therefore, only one term is non vanishing in the above sum and is corresponding to $k= j$. Thus, one gets,
   $$ \frac { d } {dt } \phi _{jj}  (t, X) = -\Pi_j ({\cal L} \Pi_j e^{-t {\cal L}}
   ( \Pi_j X)).  $$
That is,
    $$ \frac { d } {dt } \phi _{jj}  (t, X) =  -  \Pi_j {\cal L} \phi_{jj} (t,X).$$
 Point ii) is then proved   according to Lemma  \ref{lpa}.
  Finally, one checks  Point iii). To this end, one observes that,
$$  \phi' _{jm}  (t,X) = - \Pi_j \left ({\cal L}  e^{-t{\cal L}}(\Pi_m  X ) \right )
= - \sum _{p=0 }^N \Pi_j \left ({\cal L}\Pi_p   e^{- t{\cal L}}(\Pi_m  X ) \right ). $$
The $p$-th term in the above sum is
non vanishing only if  $m \leq p \leq j$. Thus,
$$ \phi' _{jm}  (t,X) = -  \Pi_j {\cal L} \phi_{jm}  (t,X) + f_{jm} (t,X). $$
Since  $j > m$, one has $\phi_{jm} (0,X) = 0$. By Point ii) of Lemma \ref{lpa}, 
Point iii) then follows.

\fpr

  {\it Proof of Theorem \ref{abstr}.  
   1. Contraction semigroup.} One proves that the hypotheses of 
Hille-Yosida Theorem are all satisfied. For every  $X$ in ${\cal K}_{\R}$,
 one of the two subspaces  $E_+ = {\rm Ker } ( X - \Vert X \Vert \ I)$ or
  $E_- = {\rm Ker } ( X - \Vert X \Vert \ I)$ is not reduced to $0$.
  Suppose that  $E_+$ is not $0$. Since all the $\Pi_j$ commute with $X$ then $E_+$ is invariant under  the $\Pi_j$. One of them restricted to  $E_+$ is thus not reduced to $0$. One of its eigenvalues is then non vanishing. This eigenvalue can only be equal to $1$. Therefore, there exists a normalized $f\in E_j$ satisfying
   $Xf = \Vert  X \Vert \ f$.  From Lemma  \ref{pr-enc-LX}, one has  $ < ( {\cal L} (X)) f , f >
   \geq 0$. Thus, one gets for all  $\lambda > 0$, 
   $$  \lambda \Vert X \Vert = \lambda < Xf , f > \leq 
    < ( \lambda X + ({\cal L}X)) f , f > \leq
    \Vert  \lambda X + ({\cal L}X) \Vert. $$
    The hypotheses of 
Hille-Yosida Theorem are then verified in that case.
The same proof holds in the case that    
  $E_-$  is not reduced to $0$. In both cases, $e^{-t{\cal L}}$ is a contraction semigroup in ${\cal K}_{\R}$.

  {\it  2. Conservation of positivity .}
  One has to prove that, if $X\in {\cal K}_{\R}$ is nonnegative then  $ \phi _{jm}  (t)$ 
  is also nonnegative  self-adjoint for all $j$ and $m$ and for every $t>0$.
By Point ii) of Lemma \ref{matr-exp},  $\phi _{jj}  (t)$  is indeed nonnegative  self-adjoint.
One  now proves by induction on $j>m$ that  $\phi _{jm}  (t)$
is nonnegative  self-adjoint. To this end,
suppose that this property is satisfied  for all integers $p$ with  $m\leq p <j$.
From the induction hypothesis, the operator $\phi_{pm}  (s)$
is nonnegative self-adjoint if $p < j$. From hypothesis (H2),
the operator $\Pi_j {\cal L} (\Pi_p  \phi_{pm}  (s)) $ is nonpositive self-adjoint.
Therefore $f_{jm} (t)$ defined in (\ref{f-j-m}) is a nonnegative self-adjoint
 semigroup.
This proves  the conservation of positivity. We have $G(t) I = I$ by hypothesis (H4).
Therefore the proof of Theorem \ref{abstr} is completed.

\fpr

We now turn to the exponential behavior of 
 $G(t)$.

  \begin{theo}\label{dec-exp}  Let  ${\cal K}_{\rm dec}$ be the set of  all $X\in {\cal K}$
   satisfying $\Pi_0 X = 0$.  In addition to  (H1)-(H4), we make the two following hypotheses:
   
   \hskip 1cm{\bf (H5)} \  The map  ${\cal L}$ is acting from  ${\cal K}$
   into ${\cal K}_{\rm dec}$.

     \hskip 1cm{\bf (FGR)} \  There exists $\gamma >0$ such that, if $j\not= 0$ then the operator $T_j$ in hypothesis (H3) satisfies $T_j + T_j ^{\star} \geq \gamma I$.

     Take $\delta \in (0, \gamma )$.
     Then, under hypotheses (H1)-(H5) and (FGR), there exists
     $C(\delta ) >0$  such that, for all $X$ in ${\cal K}_{\rm dec}$,
$$ \Vert   e^{- t{\cal L}}  X    \Vert
\leq C(\delta ) \Vert X \Vert e^{- \delta t}. $$

  \end{theo}

{\it Proof of Theorem \ref{dec-exp}.}  We shall show that the $\phi_{jm}(t)$ 
defined in  (\ref{def-phi-jm}) satisfy,
\be\label{majo-phi-jm}  \Vert \phi_{jm}(t) \Vert \leq C(\delta ) \Vert X \Vert e^{- \delta t}. \ee 
We have seen that, under hypothesis (H1), $\phi _{jm}  (t) = 0$   if $j<m$.
If $j=m$, one has from  Lemma \ref{matr-exp} $ (i)$, for all $u$ and $v$ in $E_j$,
$$  < \phi _{jj}  (t) u , v > = < X e^{-t T_j^{\star}} u ,  e^{-t T_j^{\star}} v >.  $$
Consequently,
$$ |  < \phi _{jj}  (t) u , v >  | \leq \Vert X \Vert \
\Vert  e^{-t T_j^{\star}} u \Vert \ \Vert  e^{-t T_j^{\star}} v \Vert. $$
We shall check that under our hypotheses,
\be\label{maj-exp}  \Vert e^{ -t T_j}  u  \Vert^2 \leq \Vert u \Vert ^2 e^{-\gamma t},
\quad   \Vert e^{ -t T_j^{\star}}  u  \Vert^2 \leq \Vert u \Vert ^2 e^{-\gamma t}.  \ee 
Indeed, one has,
$$ \frac {d} {dt } \Vert e^{ -t T_j}  u  \Vert^2  = - < (T_j + T_j^{\star} ) e^{ -t T_j}  u  ,
e^{ -t T_j}  u  >. $$
Using hypothesis (FGR),
$$ \frac {d} {dt } \Vert e^{ -t T_j}  u  \Vert^2  \leq - \gamma  \Vert e^{ -t T_j}  u  \Vert^2. $$
One then deduces the first estimate in  (\ref{maj-exp}) and the second is proved similarly. Thus, 
 $$  |  < \phi _{jj}  (t) u , v >  | \leq \Vert X \Vert \  e^ {-\gamma t}
 \Vert u \Vert \  \Vert v \Vert  .$$
Therefore,
 $$ \Vert \phi _{jj}  (t) \Vert _{{\cal L}(E_j)} \leq \Vert X \Vert \ e^{-\gamma t}.$$
 We shall prove inequality 
 (\ref{majo-phi-jm}) by induction on $j>m$. Suppose that this inequality holds  for all $\phi_{pm} (t)$ with $m\leq p \leq j-1$.
Then, the  function $f_{jm}$ defined in (\ref{f-j-m}) satisfies,
 \be\label{majo-fjm}  \Vert f _{jm}  (t) \Vert _{{\cal L}(E_j)} \leq C(\delta ) e^{-\delta t}. \ee 
From (\ref{phi-jm-int}), for all $u$ and $v$ in $E_j$,
 $$ | < \phi _{jm}  (t)u, v > | \leq \int _0^t
  \Vert f _{jm}  (s) \Vert  \ \Vert  e^{(s-t) T_j^{\star} } u \Vert  \
  \Vert e^{(s-t) T_j^{\star} } v \Vert    ds.  $$
Using (\ref{maj-exp}) and (\ref{majo-fjm}),
 $$ | < \phi _{jm}  (t)u, v > | \leq  C(\delta ) \Vert X\Vert \
 \Vert u\Vert \  \Vert v\Vert \  \int _0^t
  e^{- \delta s} e^{\gamma (s-t)} ds $$
  $$ \leq C_{jm} (\delta )  \Vert X\Vert \
 \Vert u\Vert \  \Vert v\Vert \  e^{-\delta t}. $$
  Consequently, inequality (\ref{majo-phi-jm}) holds true for all
 $j>m$.
 
 \fpr

\subsection{Semigroup generated by the Bethe Salpeter matrix. }

{\it Proof of Theorem \ref{se-gr}.} In this section, we use the results of the preceding section with
 $E = {\cal H}_{\rm inf}$
and the subsets $E_j$ as the eigenspaces  $E(\lambda ) =
{\rm Ker} (H_{\rm mat}-\lambda I)$ for the eigenvalues
$\lambda \in S_{\rm inf}$. Recall that  the map ${\cal L}$ is  defined by (\ref{def-cal-L})
using  $L_{\infty}$ in Proposition \ref{limite-L}
with $\omega = 0$
and  the operator $T_{\lambda}$  is defined in Proposition \ref{limite-T}.
Hypotheses
 (H1)(H2) and (H3) of Theorem  \ref{abstr} are satisfied according to equality (\ref{Bet-Salp-2}).
   
  In Proposition \ref{limite-L},   $L_{\infty}X$ is defined
as an element of  ${\cal L} ({\cal H}_{\rm inf})$
for either  $X\in {\cal L} ({\cal H}_{\rm inf})$
or $X\in {\cal L} ({\cal H}_{\rm sup})$. Then, if $X\in {\cal K}$,  $L_{\infty}X$ is associated with
an element  ${\cal L}X$ of ${\cal K}$ defined in (\ref{def-cal-L}).
Let the operator $I_{\rm sup}$ be given by
  $I_{\rm sup} (x) = x$ for every $x\in {\cal H}_{\rm sup}$
 and $I_{\rm sup} (x) = 0$ for all  $x\in {\cal H}_{\rm inf}$. Thus, $I_{\rm sup}\in {\cal L} ({\cal H}_{\rm sup})$. Therefore,
  $L_{\infty}I_{\rm sup} $ can be now examined.

From (\ref{integ-L-t}), we have  for all $u$ and $v$ in ${\cal H}_{\rm inf}$,
\begin{align*} <  \left (  L_{\infty}  I_{\rm sup} \right ) u, v >  =
- \lim _{t\rightarrow \infty } \int _{\R^3 \times (0, t) }  \big ( &e^{i s |k|}
<  E^{\rm free} (k, - s)^{\star}\ \Pi_{\rm sup}  E(k) u, v >\\[4mm]
+& e^{-is  |k|} < E^{\star} (k) \Pi_{\rm sup}   E^{\rm free} (k, - s)u , v >  \big ) dk ds .\end{align*}
This comes from the fact that $I_{\rm sup} u = I_{\rm sup} v = 0$.   Then, if $u$ and $v$ are eigenfunctions of  $H_{\rm mat}$ sharing the same eigenvalue $\mu \in S_{\rm inf}$,
$$ <  \left (  L_{\infty}  I_{\rm sup} \right) u, v >  =  -2  \lim _{t\rightarrow \infty } \int _{\R^3 \times (0, t) } < \cos ( s ( H_{\rm mat} + |k| - \mu )) \Pi_{\rm sup} E(k) u , E(k) v > dk ds.  $$
Thus, the limit is zero from Proposition \ref{lim-int}. That is,  ${\cal P}_{\cal K}  (  L_{\infty}  I_{\rm sup}  ) = 0$ where   ${\cal P}_{\cal K}$ is the projection
defined in (\ref{proj-P}).
Since  $L_{\infty} I = 0$ from  (\ref{L-omega}), we deduce  with $I= I_{\rm sup} + I_{\rm inf}$ that, ${\cal P}_{\cal K} \left (  L_{\infty}  I_{\rm inf} \right ) = 0$. Thus,
 ${\cal L} I_{\rm inf}  = 0$.  Therefore, hypothesis (H4) of  Theorem  \ref{abstr} is also satisfied.  
By this theorem,  the maps $G(t)$ defined in (\ref{def-cal-L}) and (\ref{def-Gt}) defines a 
Markov 
semigroup in ${\cal K} _{\R}$. The prrof of Theorem  \ref{se-gr}  is completed.  
 
 \fpr 
 
 We now investigate the exponential behavior of the semigroup.
 Let $\mu_0$ be the smallest eigenvalue of  $H_{\rm mat}$ supposed non degenerate  and  $u_0$ be a corresponding unit eigenvector.

{\it Proof of Theorem \ref{expo}. Point (i)}
  Let us show that for all $X\in{\cal K}$, one has  ${\cal L} X\in {\cal K}_{\rm dec}$ from (H5)
that is to say,   $< ( L_{\infty} X ) u_0 ,  u_0 > = 0$. According to Theorem \ref{Bet-Sap}, one has that,
$$   < ( L_{\infty} X) u_0 , u_0 > =  < ( TX + X T^{\star}) u_0, u_0 >.  $$
Indeed, the second term in the right hand side  of (\ref{Bet-Salp-2})
vanishes since  $u_0$ is the ground state implying that sum runs on the empty set.
Since  $X$ lies in ${\cal K}$ and thus commutes  with
$H_{\rm mat}$, using that $u_0$ is  non degenerate,
there is $a \in \C $ satisfying $Xu_0 = a u_0$. Thus,
$$  < ( L_{\infty} X) u_0 , u_0 > = a < ( T + T^{\star}) u_0 , u_0 >. $$
The above  right hand side is zero  from (\ref{gamma}) since again,   the sum in the right hand side of  (\ref{gamma}) is running on the empty set as $u_0$ denotes  the ground state. Therefore, one  has ${\cal L} X\in {\cal K}_{\rm dec}$
for every $X\in {\cal K}$. Besides, $X$ is a multiple of
  the identity for every $X\in{\cal K}_{\rm inv}$, thus $ G(t) X = X$ from Markov properties.

{\it Point ii)} This point   directly follows from Theorem \ref{dec-exp}.

 {\it Point iii) }  If $X \in {\cal K} _{\rm inv} \cap {\cal K} _{\rm dec}$ then one has 
 $ < X u_0 , u_0 > = 0$ and  $X = \lambda I$, with $\lambda \in \C$. Therefore $X= 0$, 
  proving that  ${\cal K} _{\rm inv} \cap {\cal K} _{\rm dec} = \{0\}$. 
  For all $X$ in ${\cal K}$, we  have $X = X' + X''$, with $X' = < X u_0, u_0> I$ and 
  $X'' = X - X'$. We have $X' \in {\cal K}_{\rm inv}$ and $ X'' \in {\cal K}_{\rm dec} $. 
  Therefore  ${\cal K} = {\cal K} _{\rm inv} \oplus {\cal K} _{\rm dec}$.
 For all $X$ in ${\cal K}$, one gets,
 $$ \Vert G(t) X - \pi _{\rm inv} X \Vert = \Vert G(t)( X - \pi _{\rm inv} X) \Vert
 = \Vert G(t)(\pi _{\rm dec} X) \Vert \leq C e^{- \delta t} \Vert X \Vert.$$
which proves  Theorem \ref{expo}.

\fpr

 \section{Markov approximation (by the  semigroup). }

The two error terms in Proposition \ref{deriv-prelim} are bounded below
in Proposition \ref{P-majo-R1} and Proposition \ref{P-majo-R2}. For each term,
 we shall give two bounds. One is using dipolar approximation whereas the other is not. Bounds using dipolar approximation are more precise.
We first recall the following points (see also Section 2.2). If  the form factor  is defined by (\ref{dipol}), we get,
 \be\label{deriv-Ek-dip}  \Vert(  \partial_{\rho}  ^{\alpha }\rho^{1/2}    E (\rho \omega )  )
 \Vert _{{\cal L} ( W_{m+1}^{\rm mat} ,  W_m^{\rm mat}) }  \leq C_{m N}  (1+ \rho)^{-N} ,\ee
  for all integers $m$ and $N$. One gets using Propostion \ref{norme-tenso},
 \be\label{deriv-Ek-dip-tens}  \Vert(  \partial_{\rho}  ^{\alpha }\rho^{1/2}
 ( I \otimes   E (\rho \omega )  ))
 \Vert _{{\cal L} ( W_{m+1}^{\rm tot} ,  W_m^{\rm tot}) }  \leq C_{m N}  (1+ \rho)^{-N}.\ee

We note the following distinction between  (\ref{deriv-Ek}) and
 (\ref{deriv-FF}). Inequality  (\ref{deriv-FF}) can only be applied with functions $u\in {\cal S} (\R^3)$ and in particular with
functions $u\in {\cal H}_{\rm inf}$, from Agmon inequalities, whereas inequality (\ref{deriv-Ek})
 can be applied with any function $u \in W_{m+1}^{\rm mat}$.

 \subsection{First error term in Proposition \ref{deriv-prelim}. }

 \begin{prop}\label{P-majo-R1}  Set $X\in {\cal K}$.  Let $ R_1 (t,  g, \omega , X )$ be the operator defined in (\ref{defi-R1}).
  Take  $\lambda $ and $\mu$ in $S_{inf}$. Let $\omega = \mu - \lambda$.
Fix $u$  in  ${\rm Ker} ( H_{\rm mat}- \lambda I)$ and $v$
in ${\rm Ker} ( H_{\rm mat}- \mu I)$.  Then,

 i) If the form factor  is defined by (\ref{stand}) then,
 \be\label{f-majo-R1-st} | < R_1 (t,  g, \omega , X ) u, v > | \leq
 C g^3 (1+ t^2)  \Vert X \Vert \   \ \Vert u \Vert \ \Vert v \Vert. \ee

 ii) Take the form factor   (\ref{dipol}) and suppose that the function $\phi$ in (\ref{dipol})  is vanishing at the origin at the order $p\geq 1$. Then,
\be\label{f-majo-R1} | < R_1 (t,  g, \omega , X ) u, v > | \leq
C g^3 \Vert X \Vert \ \Vert u \Vert \ \Vert v \Vert. \ee
 \end{prop}

In order to prove Proposition \ref{P-majo-R1}, we shall give two successive integral representations of the operator $R_1 (t,  g, \omega , X )$. The first one 
  (Proposition \ref{R1-A}) is sufficient without dipolar approximation.  The second one (Proposition \ref{R1-B}), which is deduced from the first one,  gives a more precise bound of $R_1 (t,  g, \omega , X )$ which is an error term, but requires dipolar approximation.

\begin{prop}\label{R1-A}  Under the hypotheses of  Proposition \ref{P-majo-R1},  one has,
\be\label{f-R1A}  <  R_1 (t,  g, \omega , X )  u, v> = \ee
$$ (ig)^2 \int _{\R^3 \times (0, t)} \Big (
< ( I \otimes e^{ i (t-s) ( H_{\rm mat} + |k| - \lambda )})  [ H_{\rm int} ,
[ (a(k) \otimes I) , W (s) ] ] (\Psi_0 \otimes u) ,
   (\Psi_0 \otimes E(k) v) >  $$
$$ -
< [ [ (a^{\star} (k) \otimes I) , W(s)  ] , H_{\rm int}] ( I \otimes e^{ i (s-t) ( H_{\rm mat}
+ |k| - \mu)} )   (\Psi_0 \otimes E(k) u) , (\Psi_0 \otimes v ) > \Big )  dk ds, $$
where
$$ W(s) = S^{\rm tot} (s, g)X. $$

\end{prop}

{\it Proof of Proposition \ref{R1-A} .}  We start from (\ref{defi-R1}) and we investigate the function in the integral.
We use inequality  (\ref{H-int-fr}) taking account of 
$\sigma _0 ( ( a^{\star} (k) \otimes I) A ) = 0$ and $\sigma _0 (  A ( a (k) \otimes I) ) = 0$
for any operator  $A$.
We then obtain,
\begin{align*} \sigma _0 \big ( &H_{int}^{free} (s-t) [ H_{int } , (W(s) - I \otimes \sigma_0 W(s) ) ]\big)\\[4mm]
& =
\int _{\R^3 } e^{i(t-s) |k|} \sigma _0 \big ( ( I \otimes E^{\rm free} (k, s-t) ^{\star} )
( a(k)  \otimes I) [ H_{int } , (W(s) - I \otimes \sigma_0 W(s) ) ] \big )  dk \\[4mm]
& =   \int _{\R^3 } e^{i(t-s) |k|} \sigma _0 \big ( ( I \otimes E^{\rm free} (k, s-t) ^{\star} )
  \big [ ( a(k)  \otimes I),  [ H_{int } , (W(s)  - I \otimes \sigma_0 W(s) ) ] \big ] \big )  dk \\[4mm]
  &= -   \int _{\R^3 } e^{i(t-s) |k|} \sigma _0 \big ( ( I \otimes E^{\rm free} (k, s-t) ^{\star} )
  \big [ H_{int } , [ (W(s)  - I \otimes \sigma_0 W(s) ) , ( a(k)  \otimes I) ]  \big ]\big )  dk\\[4mm]
  &\ \ \ -   \int _{\R^3 } e^{i(t-s) |k|} \sigma _0 \big ( ( I \otimes E^{\rm free} (k, s-t) ^{\star} )
  \big [ (W(s) - I \otimes \sigma_0 W(s) ) , [   ( a(k)  \otimes I) ,  H_{int } ] \big ]\big )  dk. \end{align*}
 One notes that $[   ( a(k)  \otimes I) ,  H_{int } ] $ is written as  $I \otimes V$ for some $V$. One also checks that,
  $\sigma _0 ( ( I\otimes U) (  W(s)  - (I\otimes \sigma _0W(s) )) (I\otimes V) = 0$ for every  operators  $U$ and $V$ in ${\cal H}_{\rm mat}$. Then, the last term above is vanishing. Besides, one has $[ (I \otimes \sigma_0 W(s)) , (a(k) \otimes I) ] = 0$.
Consequently,
\begin{align*}  \sigma _0 \big ( &H_{int}^{free} (s-t) [ H_{int } , (W(s) - I \otimes \sigma_0 W(s)) ] \big ) \\[4mm]
 &=  \int _{\R^3 } e^{i(t-s) |k|} \sigma _0 \big ( ( I \otimes E^{\rm free} (k, s-t) ^{\star} )
 \big [ H_{int } , [ ( a(k)  \otimes I) , W(s) ] \big ] \big ) dk. \end{align*}
Similarly,
\begin{align*}   \sigma _0 \big ( &[ H_{int } , (W(s) - I \otimes \sigma_0 W(s) ) ]  H_{int}^{free} (s-t) \big ) \\[4mm]
& =  \int _{\R^3 } e^{i(s-t) |k|} \sigma _0 \big ( \big [ [(a^{\star}(k) \otimes I), W(s) ],
H_{int} \big ] ( I \otimes E^{\rm free} ( k , s-t) ) \big ) dk. \end{align*}
Consequently,
\begin{align*}    \sigma _0 \big (  A(s-t)&A(0)  ( W(s) - I \otimes \sigma_0 W(s)) \big ) \\[4mm]
 =  \sigma _0  \int _{\R^3 }\big ( & e^{i(t-s) |k|} ( I \otimes E^{\rm free} (k, s-t) ^{\star} )
 \big [ H_{int } , [ ( a(k)  \otimes I) , W(s) ] \big ]  \\[4mm]
 -  &e^{ i(s-t) |k|} \big [ [(a^{\star}(k) \otimes I), W(s) ],
H_{int} \big ] ( I \otimes E^{\rm free} ( k , s-t) ) \big ) dk.
\end{align*}
If  $u$,$v$, $\lambda $, $\mu$ and  $\omega$ are taken as in Proposition \ref{P-majo-R1} then,
$$ e^{i\omega (t-s) } < \sigma _0 \big (  A(s-t)A(0)  ( W(s) - I \otimes \sigma_0 W(s)) \big ) u, v > =  $$
$$ = \int _{\R^3 } \Big (
< ( I \otimes e^{ i (t-s) ( H_{\rm mat} + |k| - \lambda )})  [ H_{\rm int} ,
[ (a(k) \otimes I) , W (s) ] ] (\Psi_0 \otimes u) ,
   (\Psi_0 \otimes E(k) v) >  $$
$$ -
< [ [ (a^{\star} (k) \otimes I) , W(s)  ] , H_{\rm int}] ( I \otimes e^{ i (s-t) ( H_{\rm mat}
+ |k| - \mu)} )   (\Psi_0 \otimes E(k) u) , (\Psi_0 \otimes v ) > \Big )  dk. $$
Proposition \ref{R1-A}  is thus completred.

\fpr

We need Lemma \ref{las} below before beginning the proof of Point i)  of Proposition \ref{P-majo-R1}.

\begin{lemm}\label{las} For any $X\in{\cal K}$, for all
$g>0$ small enough and any $t>0$, one has,
$$ \Vert   [ (a(k) \otimes I) , S^{\rm tot} (s , g) X ] \Vert _{ {\cal L}
( W_{m+1}^{\rm tot} ,  W_m^{\rm tot}) } \Vert \leq C g s \Vert X \Vert
|k|^{-1/2} (1+ |k|)^{-N} .$$
\end{lemm}

{\it Proof of Lemma \ref{las}.}
 For each   bounded operator $X$  in  ${\cal H}_{\rm mat}$,
for all $k\in \R^3 $ and every $s>0$, one has,
 \be\label{ak-prop-2} [ (a(k) \otimes I) , S^{\rm tot} (s , g) X ] = ig \int _0^s
 e^{i\sigma |k| } [ S^{\rm tot} ( \sigma , g) E(k) ,
  S^{\rm tot} ( s , g)X ] d\sigma \ee
  and
  \be\label{a-star-k-prop-2} [ (a^{\star}(k) \otimes I) , S^{\rm tot} (s , g) X ] = - ig \int _0^s
  e^{- i\sigma |k| } [ S^{\rm tot} ( \sigma , g)  E^{\star}(k) ,
  S^{\rm tot} ( s , g)X ]  d\sigma . \ee
These equalities  follow from
 (\ref{ak-prop}) and (\ref{a-star-k-prop})
 when noticing that  $[ e^{i s H(g)  } (a(k) \otimes I)  e^{-i s H(g)  } ,
  S^{\rm tot} (s , g) X ] = 0$ and similarly when $a (k)$ is replaced by $a^{\star } (k)$.
 The inequality of Lemma \ref{las} then follows in view of (\ref{majo-Ek}).
 
 \fpr

{\it Proof of Point i)  of Proposition \ref{P-majo-R1}.}
From Proposition \ref{R1-A}, one has,
$$ |   <  R_1 (t,  g, \omega , X )  u, v> | \leq  g^2 \int _{\R^3 \times (0, t)}
 \Big (  \Vert [ H_{\rm int}, [ (a(k) \otimes I ), W(s) ]] (\psi_0 \otimes u ) \Vert
  \ \Vert \Psi_0 \otimes E(k) v \Vert  $$
$$ + \Vert \Psi_0 \otimes E(k) v \Vert  \
\Vert [ H_{\rm int}, [ (a(k) \otimes I ), W^{\star} (s) ]] (\psi_0 \otimes v ) \Vert
 \Big ) dk ds. $$
Using (\ref{majo-Ek}),
 $$ \Vert \Psi_0 \otimes E(k) v \Vert \leq C |k|^{-1/2} (1 + |k|)^{-N}
 \Vert v \Vert _{W_1^{\rm mat}} \leq C'  |k|^{-1/2} (1 + |k|)^{-N}
 \Vert v \Vert .$$
According to Lemma \ref{las},
 $$   \Vert [ H_{\rm int}, [ (a(k) \otimes I ), W(s) ]] (\Psi_0 \otimes u ) \Vert  \leq
 C g s \Vert X \Vert \ |k|^{-1/2} |k|^{-1/2} (1+ |k|)^{-N} \Vert u\Vert _{W_3^{\rm mat}} $$
 $$  \leq
 C g s \Vert X \Vert \ |k|^{-1/2} |k|^{-1/2} (1+ |k|)^{-N} \Vert u\Vert. $$
Therefore Point i) is proved.

\fpr

Let us now turn to the second integral representation.

 \begin{prop}\label{R1-B}  Set $ R_1 (t,  g, \omega , X )$  the operator defined in (\ref{defi-R1}).
Let  $X\in {\cal K}$. Fix  $\lambda $ and  $\mu$ in $S_{inf}$. Set  $\omega = \mu - \lambda$.
Take  $u\in {\rm Ker} ( H_{\rm mat}- \lambda I)$ and $v\in {\rm Ker} ( H_{\rm mat}- \mu I)$. Then,
$$ <  R_1 (t,  g, \omega , X )  u, v>  = J_1(t) + J_2(t), $$
with
$$ J_m (t) = \int _{0< \sigma < s< t} \Psi_m (\sigma , s, t) d\sigma ds, $$
where
$$  \Psi_1 (\sigma , s, t) = (ig)^3 \int _{\R^3}
 e^ {i \sigma |k|}\hskip 6cm
 $$
 $$
\hskip 1cm < ( I \otimes e^{i (t-s) ( H_{\rm mat} + |k| - \lambda ) } )   \Big [ H_{\rm int}, [ e^{i\sigma H(g)} ( I \otimes E(k) )  e^{-i\sigma H(g)} ,
   W(s) ] \Big ] (\Psi_0 \otimes u) , 
    (\Psi_0 \otimes E(k) v  ) > dk  $$
    and
$$  \Psi_2 (\sigma , s, t) = (ig)^3 \int _{\R^3}  e^ {-i \sigma |k|} \hskip 6cm
 $$
 $$
  \hskip 1cm 
 < ( I \otimes e^{i (s-t) ( H_{\rm mat} + |k| - \mu ) } ) (\Psi_0 \otimes E(k) u) ,
  \Big [ H_{\rm int}, [ e^{i\sigma H(g)} ( I \otimes E(k) )  e^{-i\sigma H(g)} ,
   W^{\star} (s) ] \Big ] (\Psi_0 \otimes v) > dk.   $$

\end{prop}

{\it Proof of Proposition \ref{R1-B}.} It is a direct consequence of  (\ref{f-R1A})  together (\ref{ak-prop-2}) and  (\ref{a-star-k-prop-2}).

\fpr

{\it Proof of  Point ii) of Proposition \ref{P-majo-R1}.}
We shall estimate $\Psi_1 (\sigma , s, t)$ using spherical coordinates setting  $k = \rho \omega$ with $\rho >0$ and
$\omega \in S^2$. Define,
$$ E^{\alpha} (\rho, \omega ) = \partial_{\rho}  ^{\alpha }\rho^{1/2}
 ( I \otimes   E (\rho \omega )  ).$$
One has,
  $$  \Psi_1 (\sigma , s, t) = (ig)^3 \int _{\R_+ \times S^2}
    e^ {i ( t-s + \sigma ) \rho } 
    \hskip 6cm
 $$
 $$
 < ( I \otimes e^{i (t-s) ( H_{\rm mat} - \lambda ) } )
  \ \Big [ H_{\rm int}, [ e^{i\sigma H(g)}  E^{0} (\rho, \omega )  e^{-i\sigma H(g)} ,
   W(s) ] \Big ] (\Psi_0 \otimes u) , E^{0} (\rho, \omega )    (\Psi_0 \otimes  v  ) > \rho d\rho d\sigma (\omega ).  $$
With  $(2p+1)$ integrations by parts, one sees,
$$ ( t-s + \sigma )^{2p+1} \Psi_1( \sigma , s , t) = (ig)^3 \sum _{\alpha + \beta = 2p+1}
\int _{\R_+ \times S^2}  a_{\alpha \beta }  e^ {i ( t-s + \sigma ) \rho }\hskip 6cm
 $$
$$ 
< (I \otimes e^{i(t-s) ( H_{\rm mat}   - \lambda ) })   \Big [ H_{\rm int}, [ e^{i\sigma H(g)}  E^{\alpha } (\rho, \omega )  e^{-i\sigma H(g)} ,
   W(s) ] \Big ] (\Psi_0 \otimes u) ,   E^{\beta } (\rho, \omega )    (\Psi_0 \otimes  v  ) > \rho d\rho d\sigma (\omega )  $$
$$ +  (ig)^3 \sum _{\alpha + \beta = 2p}
\int _{\R_+ \times S^2}  b_{\alpha \beta }  e^ {i ( t-s + \sigma ) \rho }\hskip 6cm
 $$
$$ 
< (I \otimes e^{i(t-s) ( H_{\rm mat}   - \lambda ) })   \Big [ H_{\rm int}, [ e^{i\sigma H(g)}  E^{\alpha } (\rho, \omega )  e^{-i\sigma H(g)} ,
   W(s) ] \Big ] (\Psi_0 \otimes u) ,   E^{\beta } (\rho, \omega )    (\Psi_0 \otimes  v  ) >  d\rho d\sigma (\omega ),  $$
 where the   $a_{\alpha \beta }$ and  $b_{\alpha \beta }$ are real constants.
Since  $X$ belongs to ${\cal K}$ then it commutes with $H_{\rm mat}$. Thus, it is bounded in every  $W_p^{\rm mat}$. According to  Proposition
\ref{norme-tenso}, $I \otimes X$  is bounded in all the  $W_p^{\rm tot}$.
For any small enough $g$, $e^{is H(g)}$ is uniformly bounded in all the 
$W_p^{\rm tot}$  and  $W(s)$ is therefore uniformly bounded in every  $W_p^{\rm tot}$ with a norm smaller or equal than  $ C \Vert X \Vert$.
One knows that  $H_{\rm int}$ is bounded  from $W_{p+2}^{\rm tot}$
into $W_p^{\rm tot}$. One also knows that if   $u\in W_3^{\rm mat}$ then  $\Psi_0 \otimes u$ belongs to $W_3^{\rm tot}$. Thus, according to (\ref{deriv-Ek-dip-tens}),
$$ | \Psi_1( \sigma , s , t) | \leq \frac { Cg^3} { 1 + |t-s +\sigma |^{2p+1} }
\Vert X \Vert \ \Vert u \Vert_{ W_3^{\rm mat}}  \  \Vert v \Vert_{ W_3^{\rm mat} }. $$
One knows that the space  ${\cal H} _{\rm inf}$ is included in  $ W_3^{\rm mat}$
and that all the norms are equivalent on that finite dimensional space. Then,
$$ | \Psi_1( \sigma , s , t) | \leq \frac { Cg^3} { 1 + |t-s +\sigma |^{2p+1} }
\Vert X \Vert \ \Vert u \Vert  \  \Vert v \Vert. $$
Similarly, one get a bound on  $| \Psi_2( \sigma , s , t) | $.  Consequently,
$$| <  R_1 (t,  g, \omega , X )  u, v> | \leq C  \Vert X \Vert \ \Vert u \Vert  \  \Vert v \Vert
\int _{0 < \sigma < s <t} \frac { g^3} { 1 + |t-s +\sigma |^{2p+1} }
 d\sigma ds. $$
Point ii) of Proposition \ref{P-majo-R1} is then derived.

\fpr

  \subsection{Second  error term in  Proposition \ref{deriv-prelim}. }

The main goal of this section is  Proposition \ref{P-majo-R2} below.

\begin{prop}\label{P-majo-R2}  Fix $\lambda $ and $\mu$ in $S_{\rm inf}$. Take
$u$ in ${\rm Ker} ( H_{\rm mat} - \lambda I)$ and  $v$ in ${\rm Ker} ( H_{\rm mat} - \mu I)$. Set
$\omega = \mu - \lambda $.  Let $X\in {\cal K}$. Set $R_2 (t,  g, \omega , X ) $
the operator defined in  (\ref{defi-R2}). Then,

i) Suppose that the form factor  is given by  (\ref{stand}).
Then we have,
\be\label{f-majo-R2-st} |< R_2 (t,  g, \omega , X ) u, v > | \leq
C g^3 (1+ t^2) \Vert X \Vert \ \Vert u \Vert \ \Vert v \Vert. \ee

ii) Suppose that the form factor  is given by  (\ref{dipol})
and assume   that the function $\phi$ in (\ref{dipol})  is vanishing at the origin at the order $p\geq 1$.
Then the following inequality holds,
\be\label{f-majo-R2} |< R_2 (t,  g, \omega , X ) u, v > | \leq
C g^3 \Vert X \Vert \ \Vert u \Vert \ \Vert v \Vert. \ee
\end{prop}

\begin{prop}\label{err-dis-3}
  For all $X\in {\cal K} $ and $t>0$,  for all $m\geq 0$, $ \frac { d} { dt }  ( S^{\rm mat} (t, g) X $
  is well defined as an operator from $ W_{m+2} ^{mat}$ to $ W_m ^{mat}$. Moreover, there
  exists $C_m>0$ such that, for all $X\in {\cal K} $ and $t>0$,
 \be\label{majo-der-Smat}  \left\Vert  \frac { d} { dt }  ( S^{\rm mat} (t, g) X \right\Vert
 _{ {\cal L} ( W_{m+2}  ^{mat}, W_m ^{mat} )  } \leq C_m  g  \Vert X \Vert. \ee
   \end{prop}

   {\it Proof of Proposition \ref{err-dis-3}.}
 We begin to prove the inequality,  
 \be\label{majo-der-Stot} \Big \Vert \frac {d} {dt}  S^{\rm tot} (t, g) X \Big \Vert _{
{\cal L} (  W_{m+2} ^{tot} ,  W_{m} ^{tot} )}  \leq C_m g  \Vert X \Vert. \ee
Since $X\in {\cal K}$ then $X$ commutes  with $H_{\rm mat}$ and $I \otimes X$
commutes with $H(0)$. Therefore one observes that,
$$  \frac { d} { dt }  ( S^{\rm tot} (t, g) X )
  = i g   e^{  it H(g)}       [ H_{\rm int}, (I\otimes X)] e^{-  it H(g)}. $$
 Since $X\in {\cal K}$ then $X$ commutes with $H_{\rm mat}$ and is thus bounded in every $W_p^{\rm mat}$. As a consequence,  $I \otimes X$ is bounded in all the  $W_p^{\rm tot}$ (Proposition \ref{norme-tenso}). Also, $H_{\rm int }$ is bounded from  $W_{m+2} ^{\rm tot}$ to $W_{m} ^{\rm tot}$
and  $ e^{  it H(g)} $ is uniformly bounded in every  $W_p^{\rm tot}$ for any sufficiently small parameter $g$. Therefore, inequality  (\ref{majo-der-Stot}) is valid.
Then,  (\ref{majo-der-Smat}) holds true from Proposition
\ref{norme-sigma} and  Proposition \ref{majo-der-Stot} is thus proved.

   \fpr

We next turn to the integral representation of the error term. 

\begin{prop}\label{two-terms} Under the hypotheses of Proposition \ref{P-majo-R2}, one has,
$$ < R_2 (t,  g, \omega , X ) u, v >  = I_1(t) - I_2(t) $$
where
$$ I_j (t) = \int  _{ 0 < s < \sigma < t } \Phi_j (s , \sigma , t) ds d\sigma $$
with
\be\label{4-2-I-1} \Phi_1(s , \sigma , t) = (ig)^2 \int _{ \R^3}
  < e^{i(t-s) ( H_{\rm mat} + |k| - \lambda ) } [ E(k) , Z(\sigma) ] u , E(k) v >,
dk  \ee

\be\label{4-2-I-2} \Phi_2(s , \sigma , t) = (ig)^2 \int _{ \R^3}  < e^{i(s-t) ( H_{\rm mat} + |k| - \mu ) } E(k) u,
 [ Z(\sigma) ^{\star} , E(k) ]  v >  dk  \ee
and 
\be\label{Z-sigma}  Z ( \sigma) =  \frac { d} { d\sigma  }  ( S^{\rm mat} (\sigma , g) X.  \ee

\end{prop}

{\it Proof of Proposition \ref{two-terms}.}
 One has,
\begin{align*} < R_2 (&t,  g, \omega , X ) u, v >\\[4mm]
& = (ig)^2 \int _0^t  e^{i  \omega ( t-s)  }  < \sigma _0  ( A( s - t)
A(0)  \big ( I \otimes \big ( S^{\rm mat} (s, g) X - S^{\rm mat} (t, g) X \big )  \big )
     u, v > ds \\[4mm]
& = - (ig)^2 \int _{ 0 < s < \sigma <t } e^{i  \omega ( t-s)  }  < \sigma _0  \big ( A( s - t)
A(0)  \big ( I \otimes Z(\sigma)  \big ) \big ) u, v >  d s d \sigma,
\end{align*}
where  $Z(\sigma)$ is given in (\ref{Z-sigma}).
Therefore, Proposition \ref{two-terms} is proved using  (\ref{sig0-As-At}).

\fpr

{\it Proof of Point i) of Proposition \ref{P-majo-R2}.}
According to Proposition \ref{two-terms}, one has,
 $$ |< R_2 (t,  g, \omega , X ) u, v > | \hskip 10cm
 $$
 $$\leq   g^2 \int _{ \Delta (t)}
 \Big ( \Vert  [ E(k) , Z(\sigma) ] u \Vert \  \Vert  E(k) v \Vert
 +  \Vert  E(k) u \Vert \
\Vert  [ Z(\sigma) ^{\star} , E(k) ]  v \Vert  \Big ) dk ds d\sigma. $$
From (\ref{majo-Ek}), one sees,
$$ \Vert  E(k) v \Vert \leq  C |k|^{-1/2} (1+ |k|)^{-N} \Vert v \Vert _{W_1^{\rm mat}}.$$
In view of  (\ref{majo-Ek}) and of Proposition \ref{err-dis-3}, one learns,
$$  \Vert  [ E(k) , Z(\sigma) ] u \Vert \leq C g |k|^{-1/2} (1+ |k|)^{-N}
\Vert X \Vert \ \Vert u \Vert _{W_3^{\rm mat}}. $$
 Since ${\cal H}_{\rm inf}$ is included in $W_3^{\rm mat}$ and since the norms are equivalent in the finite dimensional space ${\cal H}_{\rm inf}$ then one obtains,
$$ |< R_2 (t,  g, \omega , X ) u, v > | \leq  g^3 \int _{ \Delta (t)}
|k|^{-1} (1+ |k|)^{-2N} \Vert X \Vert \ \Vert u \Vert
\ \Vert v \Vert dk ds d\sigma.  $$
Point  i) of  Proposition \ref{P-majo-R2} is then derived.

\fpr

{\it Proof of Point  ii) of Proposition \ref{P-majo-R2}.}
One uses Proposition \ref{two-terms}. Let us bound the term
$\Phi_1(s, \sigma , t)$ defined in  (\ref{4-2-I-1}).
 Again, one uses spherical coordinates setting
 $ k =  \rho \theta$ with $\rho >0$ and  $\theta \in S^2$.
 One obtains,
 $$ \Phi_1(s, \sigma , t) = (ig)^2 \int _{ \R_+ \times S^2}
 < e^{i(t-s) ( H_{\rm mat} + \rho  - \lambda ) }
  [ E(\rho \omega ) , Z(\sigma) ] u , E(\rho \omega ) v >
\rho ^2 d\rho  d\sigma (\omega). $$
One integrates in the   variable $\rho$. One gets,
$$ ( t-s)^{2p+1} \Phi_1(s, \sigma , t)=  (ig)^2 \sum _{\alpha + \beta = 2p+1}
\int _{\R_+ \times S^2}  a_{\alpha \beta }$$
$$
< e^{i(t-s) ( H_{\rm mat} + \rho  - \lambda ) }\Big [  \partial _{\rho}^{\alpha  }  ( \rho^{1/2}  E(\rho \omega )) , Z(\sigma ) \Big ] u,   \partial _{\rho}^{\beta  }  ( \rho^{1/2}  E(\rho \omega )) v >  \rho  d\rho d\sigma (\omega) $$
$$ +  (ig)^2 \sum _{\alpha + \beta = 2p}  \int _{\R_+ \times S^2} b_{\alpha \beta }\hskip 8 cm$$
$$
    < e^{i(t-s) ( H_{\rm mat} + \rho  - \lambda ) }\Big [  \partial _{\rho}^{\alpha  }  ( \rho^{1/2}  E(\rho \omega )) , Z(\sigma ) \Big ] u,  \partial _{\rho}^{\beta  }  ( \rho^{1/2}  E(\rho \omega )) v >   d\rho d\sigma (\omega), $$
where the   $a_{\alpha \beta }$ and $b_{\alpha \beta }$ are real constants.
  One uses (\ref{deriv-Ek-dip}). One obtains,  
$$ \Vert(  \partial_{\rho}  ^{\alpha }\rho^{1/2}    E (\rho \omega )  ) v
 \Vert   \leq C (1+ \rho)^{-N} \Vert v \Vert  _{W_{1}^{\rm mat}}.$$
From  (\ref{deriv-Ek-dip}) and Proposition \ref{err-dis-3},
$$ \Vert \Big [  \partial _{\rho}^{\alpha  }  ( \rho^{1/2}  E(\rho \omega )) , Z(\sigma ) \Big ] u \Vert
\leq C g (1+ \rho)^{-N} \Vert X \Vert \ \Vert u \Vert    _{W_{3}^{\rm mat}}. $$
The space  ${\cal H }_{\rm inf}$ being included $W_{3}^{\rm mat}$ and the norms on this finite dimensional space being all equivalent, one has, 
 $$  I_1 (t) =  C g^3 \int  _{ 0 < s < \sigma < t }
 \frac {1} { 1+ |t-s|^{2p+1} }  \Vert X \Vert \ \Vert u \Vert \ \Vert u \Vert  ds d\sigma $$
 $$ \leq C g^3  \Vert X \Vert \ \Vert u \Vert \ \Vert u \Vert. $$
Point ii) of  Proposition \ref{P-majo-R2} is thus proved.

 \fpr

  \subsection{Differential system with constant coefficients. }

Recall that Sections 4.1 and  4.2 are both concerned with Markov approximation  and Rabi cycle. In contrast, Sections 4.3 and 4.4 are only concerned with Markov approximation
  (that is,  $\omega = 0$). The proofs involving the Rabi cycle are carry on in Section 5.

In the case  $\omega = 0$, the differential system  (\ref{Sys-Dif-Init}) can be written as,
\be\label{sys-dif-PK}   \frac {d} {dt} {\cal P}_{\cal K}  (S^{\rm mat} (t , g) X) =
 (ig)^2 {\cal P}_{\cal K}  L^{0} (t)(S^{\rm mat} (t , g) X ) +
 \sum _{j=1}^2{\cal P}_{\cal K} R_j (t,  g, 0 , X ), \ee
where $R_1 $  and  $R_2 $ are defined in  (\ref{defi-R1}) and  (\ref{defi-R2}),
and are estimated  in  (\ref{f-majo-R1}) and (\ref{f-majo-R2}) with the dipolar approximation or in  (\ref{f-majo-R1-st}) and  (\ref{f-majo-R2-st}) without the dipolar approximation.

The goal of this section is to approximate 
 $ {\cal P}_{\cal K}  L^{0} (t)(S^{\rm mat} (t , g) X )$
by $ {\cal P}_{\cal K}  L^{0}_{\infty} {\cal P}_{\cal K} (S^{\rm mat} (t , g) X )$
  and therefore to prove  Proposition \ref{P-K-K} below. Observe that Proposition \ref{P-K-K} does not assume the dipolar approximation.

 \begin{prop}\label{P-K-K} Suppose that the  form factor $E(k)$ is given by either   (\ref{stand}) or (\ref{dipol}) and that the function $\phi$ is vanishing at the origin at the order $p\geq 1$.  The operator  in  ${\cal L} ({\cal H}_{\rm inf})$  defined by,
 \be\label{def-K} K (t,  g,  X ) = (ig)^2 \Big ( {\cal P}_{\cal K}  L^{0} (t)(S^{\rm mat} (t , g) X ) -
  {\cal P}_{\cal K}  L^{0}_{\infty} {\cal P}_{\cal K} (S^{\rm mat} (t , g) X )\Big )  \ee
is satisfying,
 $$ \Vert K (t,  g,  X )\Vert \leq C \Vert X \Vert \ \left ( g^3 + \frac {g^2} {1+t^2} \right ). $$

 \end{prop}

The proof of  Proposition \ref{P-K-K} uses Lemma \ref{commut-HS}.

 \begin{lemm}\label{commut-HS}  There exists $C>0$ satisfying     for all  $X\in{\cal K} $,
 $$ \Vert  [ H _{\rm mat } , S^{\rm mat} (t, g) X ] \Vert  _{ {\cal L} ( W_2^{\rm mat} , W_0^{\rm mat})} \leq C g \Vert X \Vert.$$
 \end{lemm}

 {\it Proof of Lemma \ref{commut-HS}.} One has, for all operators $Z$ in ${\cal H} _{\rm tot}$,
   $$ [ H_{\rm mat}, \sigma _0 Z ] = \sigma _0 ( [ H(0), Z] ). $$
Consequently, for  any operator $X$ in ${\cal H} _{\rm mat}$,
   $$  [ H _{\rm mat } , S^{\rm mat} (t, g) X ]  = \sigma_0  (
   [ H(0) ,  S^{\rm tot} (t , g) X ]), $$
    with the notation (\ref{S-tot}).
 One observes,
 \begin{align*} [ H(0) , S^{\rm tot} (t , g) X ] &=
   [ H(g) , S^{\rm tot} (t , g) X ]  - g [ H_{\rm int}  , S^{\rm tot} (t , g) X ]   \\[4mm]
& =
    e^{ i t H(g)}  [ H(g) , (I\otimes X) ]  e^{ - i t H(g)}   - g [ H_{\rm int}  , S^{\rm tot} (t , g) X ].    \end{align*}

 If $X\in {\cal K }$ then $[H_{\rm mat}, X ] = 0$ and also
 $[ H(0) , I \otimes X ] = 0$. Consequently,
   $$ g^{-1}  [ H(0) , S^{\rm tot} (t , g) X ] =
      e^{ i t H(g)} [ H_{\rm int} , (I\otimes X) ]  e^{- i t H(g)}  - [ H_{\rm int}  , S^{\rm tot} (t , g) X ].   $$
   The second term in the above  right hand side  is a bounded operator from  $W_2 ^{\rm tot}$ to $W_0 ^{\rm tot}$ uniformly in $t$. This fact comes from the following points.
The operator $e^{it H(g)}$  is uniformly
   bounded in $W_2 ^{\rm tot}$  and in $W_0 ^{\rm tot}$ (Theorem \ref{Kato-Rell}) and
   the operator $H_{\rm int}$ is bounded from  $W_2 ^{\rm tot}$   to $W_0 ^{\rm tot}$
   (Theorem \ref{Kato-Rell}). Besides, since $X\in {\cal K} $
commutes with $H_{\rm mat}$, $X$ is bounded in $W_2 ^{\rm mat}$, then $I \otimes X $
   is bounded in  $W_2 ^{\rm tot}$  (Proposition \ref{norme-tenso}). Consequently,
   for each $F$ in  $W_2 ^{\rm tot}$,
   $$ \Vert  [ H(0) , S(t, g) (I\otimes X) ] F \Vert _{ W_0 ^{\rm tot} }
   \leq   C g \Vert F  \Vert _{ W_2 ^{\rm tot} }. $$
From Proposition \ref{norme-sigma}, one deduces that,  for each  $f\in W_2 ^{\rm mat}$,
   $$ \Vert \sigma _0 \left (  [ H(0) , S(t, g) (I\otimes X) ]\right ) f   \Vert _{ W_0 ^{\rm mat} }
   \leq   C g \Vert f  \Vert _{ W_2 ^{\rm mat} }.$$
 Lemma \ref{commut-HS} is proved.

\fpr

We now prove 
Proposition \ref{P-K-K}.
Note that the operator  $L_{\infty}^0 ( {\cal P}_{\cal K} (S^{\rm mat} (t , g) X )  + \Pi_{\rm sup} (S^{\rm mat} (t , g) X \Pi_{\rm sup} ) ) $ is well defined  from Proposition \ref{limite-L}, Point ii).
 The operator defined in  (\ref{def-K}) can be written as,
$$  K (t,  g,  X ) =  K_1 (t,  g,  X ) +  K_2 (t,  g,  X ) +  K_3 (t,  g,  X ) $$
$$  K_1 (t,  g,  X )  = (ig)^2  {\cal P}_{\cal K}  L^{0} (t)\Big ( (S^{\rm mat} (t , g) X )
- {\cal P}_{\cal K} (S^{\rm mat} (t , g) X )  - \Pi_{\rm sup} (S^{\rm mat} (t , g) X )
 \Pi_{\rm sup} \Big )  $$
$$  K_2 (t,  g,  X )  = (ig)^2 {\cal P}_{\cal K} \Big ( L^{0} (t) - L^{0}_{\infty } \Big )
\ \Big (  {\cal P}_{\cal K} (S^{\rm mat} (t , g) X )
+ \Pi_{\rm sup} (S^{\rm mat} (t , g) X \Pi_{\rm sup} ) \Big  ) $$
$$   K_3 (t,  g,  X )  = (ig)^2 {\cal P}_{\cal K}  L^{0}_{\infty } \Big (
\Pi_{\rm sup} (S^{\rm mat} (t , g) X \Pi_{\rm sup} ) ). $$
One can also use a decomposition of  $K_1$ as $K_1 =  K_{11} + K_{12} + K_{13}$ with,
      \be\label{I-2}  K_{11} (t, g)X =  (ig)^2 {\cal P}_{\cal K}  L^{0} (t)  \sum _{\mu \in S_{\rm inf}} \Pi ( \mu)  ( S^{\rm mat} (t , g) X )   \Pi _{\rm sup}  \ee
      \be\label{I-3}   K_{12} (t, g) X =(ig)^2 {\cal P}_{\cal K}  L^{0} (t)  \sum _{\mu \in S_{\rm inf}} \Pi _{\rm sup} ( S^{\rm mat} (t , g) X )  \Pi ( \mu) \ee
    \be\label{I-4}  K_{13} (t, g)X = (ig)^2 {\cal P}_{\cal K}  L^{0} (t) \sum _{(\mu , \nu)\in I } \Pi ( \mu) ( S^{\rm mat} (t , g) X )   \Pi ( \nu) \ee
    where  $I$ denotes the set  of  $(\mu, \nu)$ in $S_{\rm inf} \times S_{\rm inf}$  with  $\mu \not= \nu$.

We shall then give a bound of these terms. We shall prove below for  $j\leq 3$ that,
 \be\label{majo-Ij}  \Vert K_{1j} (t, g)X  \Vert   \leq  C g^3 \Vert  X \Vert.  \ee
  {\it  Estimate of  $K_{11}(t , g) X $.}   Clearly, one has for every  $\mu\in S_{\rm inf}$
and all $f\in {\cal H} _{\rm mat}$,
$$ \Pi (\mu) ( S^{\rm mat}(t , g) X) \Pi _{\rm sup}  f = \Pi (\mu)  [ ( S^{\rm mat}(t , g) X) , H_{\rm mat} ]  ( H_{\rm mat } - \mu) ^{-1} \Pi_{\rm sup} f.$$
Therefore, by Lemma \ref{commut-HS},
\begin{align*} \Vert \Pi (\mu) ( S^{\rm mat}(t , g) X)  \Pi _{\rm sup}  f \Vert &\leq \Vert [( S^{\rm mat}(t , g) X) , H_{\rm mat} ]  \Vert_{{\cal L} ( W_2^{\rm mat} , W_0^{\rm mat}) }
\ \Vert  ( H_{\rm mat } - \mu) ^{-1} \Pi_{\rm sup} f \Vert _{W_2 ^{\rm mat }} \\[4mm]&\leq C \Vert [ ( S^{\rm mat}(t , g) X) , H_{\rm mat} ]  \Vert_{{\cal L} ( W_2^{\rm mat} , W_0^{\rm mat}) }
 \ \Vert f \Vert \end{align*}
$$ \leq C g \Vert X \Vert \ \Vert f \Vert. $$
Taking account of Proposition \ref{limite-L}, Point iii),
 the estimate (\ref{majo-Ij}) holds true  for $j=1$.

  {\it  Estimate of  $K_{12} (t, g)X$.}  One easily gets as for $K_{11}(t,g)X$,
$$ \Pi_{\rm sup} \Big ( S^{\rm mat} (t , g) X \Big ) \Pi(\mu)  =
  \Pi_{\rm sup} ( H_{\rm mat} - \mu) ^{-1} [ H_{\rm mat} , ( S^{\rm mat} (t , g) X ) ] \Pi(\mu). $$
Thus, by Lemma \ref{commut-HS},
\begin{align*}
\Vert \Pi_{\rm sup}  ( &S^{\rm mat} (t , g) X)   \Pi_{\rm inf}  \Vert \\[4mm] &\leq  C  \Vert  \Pi_{\rm sup} ( H_{\rm mat} - \mu) ^{-1} \Vert   \ \Vert [ ( S^{\rm mat}(t , g) X) , H_{\rm mat} ]  \Vert_{{\cal L} ( W_2^{\rm mat} , W_0^{\rm mat}) }\\[4mm] 
& \leq C g \Vert X \Vert. \end{align*}
According to Proposition \ref{limite-L}, Point iii),
 the estimate (\ref{majo-Ij}) holds true  for $j=2$.

   {\it  Estimate of   $K_{13} (t , g) $.}   One checks that,
  $$ \Vert \Pi (\mu) (S^{\rm mat} (t , g) X)  \Pi (\nu) \Vert
    \leq \frac {C } {|\mu - \nu|}   \Vert [ H_{\rm mat} , (S^{\rm mat} (t , g) X) ]
    \Vert_{{\cal L} ( W_2^{\rm mat} , W_0^{\rm mat}) }, $$
 for any $(\mu , \nu)\in I$ and
   every operator  $X$ bounded in ${\cal H} _{\rm mat}$.
Indeed,
$$ ( \mu  - \nu  ) \Pi ( \mu) (S^{\rm mat} (t , g) X) \Pi (\nu ) = \Pi ( \mu) [ H_{\rm mat}, (S^{\rm mat} (t , g) X) ] \Pi (\nu ). $$
 By Lemma \ref{commut-HS},
 $$ \Vert \Pi (\mu) (S^{\rm mat} (t , g) X)  \Pi (\nu) \Vert
    \leq \frac {C g } {|\mu - \nu|}  \Vert X \Vert. $$
 From Proposition \ref{limite-L}, Point iii),
 the estimate (\ref{majo-Ij}) holds true  for $j=3$.

 {\it  Estimate of  $K_{2} (t, g,X)$.} From  Proposition
 \ref{limite-L}, Point ii), since  ${\cal P}_{\cal K} (S^{\rm mat} (t , g) X )$
 belongs to ${\cal L} ( {\cal H}_{\rm inf})$ and since
 $ \Pi_{\rm sup} (S^{\rm mat} (t , g) X \Pi_{\rm sup} $ is in  ${\cal L} ( {\cal H}_{\rm sup})$, one has,
 $$\Vert  K_{2} (t, g,X) \Vert  \leq C \frac {g^2} {1+ t^2}
 \Big (  \Vert  {\cal P}_{\cal K} (S^{\rm mat} (t , g) X )
+ \Pi_{\rm sup} (S^{\rm mat} (t , g) X ) \Pi_{\rm sup}   \Vert \Big ).   $$
Thus,
 \be\label{majo-K2}  \Vert  K_{2} (t, g,X) \Vert  \leq C \frac {g^2} {1+ t^2}
 \ \Vert  X \Vert. \ee

  {\it  Estimate of   $K_3 (t , g)X $.} We shall derive the following inequality,
  \be\label{PKLT4}  \Vert K_3 (t , g)X  \Vert
  \leq C g^ 3 \Vert X \Vert .\ee
   First, one studies $ < (L_{\infty}   Z ) u, v>$ with
 $ Z =  \Pi _{\rm sup} (S^{\rm mat} (t , g) X ) \Pi _{\rm sup}$, and $u$ and $v$
  in ${\rm Ker} (H_{\rm mat}-\lambda I)$ ($\lambda $  in $S_{\rm inf}$).
Under the hypotheses, one has
  $Z  u = Z^{\star}   v = 0$.  Thus, in view of (\ref{LtZuv}),
$$  < (L_{\infty}   Z) u , v>  = -  \lim _{t\rightarrow \infty }  \int _{\R^3 \times (0, t) }
\Big (  < e^{ is (H_{\rm mat} + |k| - \lambda ) } Z E(k) u , E(k) v > $$
 $$  + < Z e^{ - is (H_{\rm mat} + |k| - \lambda ) }  E(k) u , E(k) v > \Big ) dk ds. $$
 Let us recall that,
 $$  \lim _{t\rightarrow \infty }  \int _{\R^3 \times (0, t) }
  < Z \cos ( s (H_{\rm mat} + |k| - \lambda ) )  E(k) u , E(k) v >  dk ds = 0. $$
Therefore,
$$  < (L_{\infty}   Z) u , v>  = i   \lim _{t\rightarrow \infty }  \int _{\R^3 \times (0, t) }
  < \Big [ Z ,  \sin ( s (H_{\rm mat} + |k| - \lambda ) ) \Big ]  E(k) u , E(k) v >  dk ds. $$
  We apply this fact with $ Z =   \Pi  _{\rm sup} (S^{\rm mat} (t , g) X ) \Pi  _{\rm sup} $.
   We remark that,
 $$  \lim _{t\rightarrow \infty }  \int _{\R^3 \times (0, t) }
   < \Big [ \Pi  _{\rm sup} (S^{\rm mat} (t , g) X ) \Pi  _{\rm sup} ,  \sin ( s (H_{\rm mat} + |k| - \lambda ) ) \Big ]  E(k) u , E(k) v >  dk ds   $$
   $$ = \int _{\R^3  } < \Pi  _{\rm sup} \Big [ (S^{\rm mat} (t , g) X ) ,
   (H_{\rm mat} + |k| - \lambda ) ^{-1} \Big ]\Pi  _{\rm sup}  E(k) u, E(k) v > dk $$
   $$ = -  \int _{\R^3  } < (H_{\rm mat} + |k| - \lambda ) ^{-1}
    [ (S^{\rm mat} (t , g) X ) , H_{\rm mat} ] (H_{\rm mat} + |k| - \lambda ) ^{-1}
     \Pi  _{\rm sup}  E(k) u , \Pi  _{\rm sup}  E(k) v > dk. $$
  Then, one deduces that,
  \be\label{Lind-proj-com}  < (L_{\infty} \Big (  \Pi  _{\rm sup} (S^{\rm mat} (t , g) X ) \Pi  _{\rm sup}  \Big )   u , v > = i\int _{\R^3}
  < [ H_{\rm mat} , (S^{\rm mat} (t , g) X) ] A (k , \mu ) u , A (k, \mu ) v > dk  \ee
where
  $$ A (k, \mu ) u = ( H_{\rm mat } + |k| - \mu ) ^{-1} \Pi_{\rm sup } E (k)u.  $$
Consequently,
  $$ \left |  < (L_{\infty}  \Pi  _{\rm sup} (S^{\rm mat} (t , g) X ) \Pi  _{\rm sup})   u , v >  \right | \leq \hskip 9cm $$
  $$
  \hskip 2cm    \Vert  [ H _{\rm mat } , S^{\rm mat} (t, g) X ] \Vert  _{ {\cal L} ( W_2^{\rm mat} , W_0^{\rm mat})}
   \int _{\R^3} \Vert A (k , \mu ) u \Vert _{W_2^{\rm mat} } \
     \Vert A (k , \mu ) v \Vert _{W_2^{\rm mat} } dk. $$
We have,
  $$  \Vert A (k, \mu ) u \Vert _{W_2 ^{\rm mat} } \leq \frac {C} { E_0 - \mu  }    \Vert u \Vert _{W_3^{\rm mat} }   (1+ |k|)^{-2}. $$
  Therefore, by Lemma \ref{commut-HS},
  $$  \left |  < (L_{\infty} \Big ( \Pi  _{\rm sup} (S^{\rm mat} (t , g) X ) \Pi  _{\rm sup} \Big )   u , v >  \right | \leq
  C g \Vert X \Vert \ \Vert u \Vert \   \Vert v \Vert. $$
  This is valid for each $u$ and $v$ in ${\rm Ker} (H_{\rm mat} - \lambda I)$,
  for each $\lambda \in S_{\rm inf}$. This is thus equivalent to
  (\ref{PKLT4}).

Proposition \ref{P-K-K} is a consequence of  (\ref{majo-Ij}) for  $j\leq 3$ together with  (\ref{majo-K2})
 and (\ref{PKLT4}).

  \fpr

\subsection{Proofs of Theorem \ref{approx-SG-prov}  and Theorem  \ref{approx-SG}. }

The differential system  (\ref{sys-dif-PK}) together with  Proposition \ref{P-K-K} is written in the case  $\omega = 0$ as,
  $$    \frac {d} {dt}  {\cal P} _{\cal K} ( S^{\rm mat} (t , g) X ) =
 (ig)^2 {\cal P} _{\cal K}  L _{\infty}^0  {\cal P} _{\cal K} ( S^{\rm mat} (t , g) X ) +
 (ig)^2    {\cal P} _{\cal K} ( H (t , g, X), $$
  with
  $$ H(t, g, X) = \sum _{j= 1} ^2 R_j (t , g) X + K(t, g , X), $$
where the $R_j (t , g, X)$ and  $K (t , g, X)$ are defined before.

 If the form factor $E(k)$ is chosen as (\ref{stand}) and if  $\phi$ 
 is vanishing at the origin at the order  $p\geq 1$ then, 
 according to  (\ref{f-majo-R1-st})(\ref{f-majo-R2-st})  and Proposition \ref{P-K-K},
 $$ \Vert {\cal P}_{\cal K}  H(t, g, X) \Vert \leq C \left  ( g^3 (1+ t^2)
  + \frac {g^2} { 1+ t^{2} } \right )
\Vert X \Vert.  $$
  If the form factor $E(k)$ is given by (\ref{dipol}) and if  $\phi$ 
  vanishes at $0$ at the order  $p\geq 1$ then, by (\ref{f-majo-R1})(\ref{f-majo-R2}) and by
  Proposition \ref{P-K-K},    
$$ \Vert {\cal P}_{\cal K}  H(t, g, X) \Vert \leq C \left  ( g^3  + \frac {g^2} { 1+ t^{2} } \right )
\Vert X \Vert. $$
With notation  (\ref{def-cal-L}), the differential system  (\ref{sys-dif-PK}) is written as,
 \be\label{Proj} \frac {d} {dt}   {\cal P}_{\cal K}  S^{\rm mat} (t , g) X   = (ig)^2
   {\cal L}  (  {\cal P}_{\cal K}   S^{\rm mat} (t , g) X )  + {\cal P}_{\cal K}  H(t, g, X ). \ee
  Besides, for all $X\in {\cal K}$,
$$  \frac {d} {dt }     G  (t  g^2)X  = - g^2  {\cal L}  ( G (t  g^2)  X ).$$
Also, $G  (0) X = {\cal P}  S^{\rm mat} (0, g) X = X $. Consequently,   Duhamel principle shows that,
$$ {\cal P}_{\cal K} S^{\rm mat}  (t, g) X  - G (tg^2 )X  = \int _0^t  G ( (t-s) g^2)
 {\cal P}_{\cal K}   H(s , g , X)    ds.  $$
The projections
$\pi_{\rm dec}$ and $\pi_{\rm inv}$ associated with the decomposition (\ref{som-dir}) can be applied on  ${\cal P}_{\cal K}  H(s, g,  X)$ (belonging to ${\cal K}$).
From (\ref{dex-Gt}), if the form factor $E(k)$ is defined by (\ref{stand}), and if $0<g<1$, then we have,
 $$  \int _0^t \Vert G ( (t-s) g^2) \pi_{\rm dec} ( {\cal P}_{\cal K}  H(s, g, X)  )  \Vert ds
 \leq C \Vert X \Vert \int _0^t e^{-\delta g^2 (t-s) } \Big (g^3 (1+s^2) +  \frac {g^2 } { 1+ s^{2}  } \Big ) ds. $$
 $$ \leq Cg \Vert X \Vert \  (1+ t^2).$$
   Let us underline at this stage  that if the form factor $E(k)$ is defined by (\ref{dipol}) instead of   (\ref{stand})  then the factor $(1+ t^2)$ in the right hand side of the above estimate will be 
 be replaced by a factor $ 1$.
 
  If the form factor $E(k)$ is defined by (\ref{dipol}), we have,
  $$  \int _0^t \Vert G ( (t-s) g^2) \pi_{\rm dec} ( {\cal P}_{\cal K}  H(s, g, X)  )  \Vert ds
  \leq Cg \Vert X \Vert. $$

Since elements of   ${\cal K} _{\rm inv}$ are left invariant under $G(t)$,
\be\label{a}  \int _0^t G ( (t-s) g^2) \pi_{\rm inv} {\cal P}_{\cal K}  H(s, g,  X )  ds  =
\int _0^t \pi_{\rm inv} {\cal P}_{\cal K}  H(s, g ,  X )  ds. \ee
One also remarks the existence of   $C>0$ verifying,
\be\label{deux}  \Vert \pi_{\rm inv} Z \Vert \leq C |< Z u_0 , u_0 > |,\  Z\in {\cal K}. \ee
Indeed, if $< Z u_0 , u_0 >  = 0$ then
$Z\in {\cal K} _{\rm dec}$ by definition, thus  $\pi_{\rm inv} Z = 0$. The above inequality is valid since  ${\cal K}$ is finite dimensional.

One will prove below  that, for both definitions of the form factor,
\be\label{trois}  \Big | \int _0^t  < ( {\cal P}_{\cal K}  H(s, g ,  X )) u_0 , u_0 >  ds \Big |
\leq C g \Vert X \Vert. \ee
From (\ref{a}), (\ref{deux}) and (\ref{trois}), this inequality will show that,
$$ \Big \Vert  \int _0^t G ( (t-s) g^2) {\cal P}_{\cal K}  H(s, g,  X )  ds \Big \Vert
\leq  C g \Vert X \Vert  $$
and thus,
$$  \Big \Vert {\cal P} _{\cal K} S^{\rm mat}  (t, g) X  - G (tg^2 )X  \Big \Vert
\leq   C g \Vert X \Vert  $$
which will end the proof of Theorem  \ref{approx-SG}.

To prove (\ref{trois}), the following result of  D. Hasler and I. Herbst \cite{HH} is recalled here.

  \begin{prop}\label{evolPsi-u} (\cite{HH}).  Suppose that  $\mu_0 = \inf \sigma (H_{\rm mat})$ is a simple eigenvalue  of $H_{\rm mat}$ and let  $u_0$ be a   corresponding unit eigenvector. Then,
  there exists  an eigenvector  $U(g)$
  of $H(g)$ satisfying,
   \be\label{distance}  \Vert U(g) - \Psi _0 \otimes u_0 \Vert _{ {\cal H} _{\rm tot}} \leq Cg. \ee

   \end{prop}

 The result of  \cite{HH}  is concerned with the  Pauli-Fierz Hamiltonian  but it is  very likely  that
 the result of  \cite{HH} is also valid with the simplified Pauli-Fierz Hamiltonian studied here
as it is written in \cite{HH} "We want to emphasize that the proof of Theorem 1 does not use any form of gauge invariance. In particular the conclusions hold if the quadratic terms (in the interaction)  are dropped from the Hamiltonian."

See also the alternative proof below.

  {\it Proof of inequality (\ref{trois}).}
From (\ref{Proj}), one has,
    $$       {\cal P}_{\cal K}   S^{\rm mat} (t, g) X - X    = (ig)^2
  \int _0^t      {\cal L}   {\cal P}_{\cal K}  S^{\rm mat} (s, g) X    ds   +
   \int _0^t   \ H(s, g , X )    ds. $$

According to Theorem \ref{expo},   one has ${\cal L} Z \in {\cal K}_{\rm dec}$ for all operators $Z$ in ${\cal K} $,
  that is,
  $$  < ({\cal L} Z) u_0 , u_0 >  = 0.  $$
Thus,
   $$ \int _0^t   < (  {\cal P}_{\cal K}  H(s, g ,  X )  u_0 , u_0 >   ds  = < ( S^{\rm mat} (t, g) X - X) u_0, u_0 >. $$
Also,
   $$  < ( S^{\rm mat} (t, g) X ) u_0, u_0 > =
    < (I\otimes X) e^{-it H(g)} ( \Psi_0 \otimes u_0) , e^{-it H(g)}  ( \Psi_0 \otimes u_0) >.$$
Let us check that,
      $$ \Vert  e^{-it H(g)}  ( \Psi_0 \otimes u_0) - e^{-it\mu_0}  ( \Psi_0 \otimes u_0)\Vert \leq C g $$
    with $C > 0$ independent of $t$.
Indeed, when $U(g)$ is an eigenvector  of  $H(g)$ with eigenvalue $\mu_0$,
   $$  e^{-it H(g)} U(g)  - e^{-it\mu_0} U(g) = 0.   $$
If $U(g)$ satisfies (\ref{distance}) then one indeed gets  (\ref{inv-u0}). Consequently,   
   \be\label{inv-u0} \Vert < ( S^{\rm mat} (t, g) X - X) u_0, u_0 >  \Vert
   \leq Cg \Vert X  \Vert  \ee
which proves (\ref{trois}).

   {\it Alternative proof.} Let us give a second proof without using the result of \cite{HH}.
    Set ${\cal H}^{(1)} = {\cal H} _{\rm ph} \otimes \C u_0$, let  ${\cal H}^{(2)}$
be the orthogonal of  ${\cal H}^{(1)}  $ in ${\cal H} _{\rm tot}$. Denote by  $\Pi ^{(1)}$ and  $\Pi ^{(2)}$
the corresponding orthogonal projections. We have for any  $f\in {\cal H} _{\rm tot}$,
$$ (\inf _{\mu \not= \mu_0} \mu - \mu_0) \Vert \Pi ^{(2)} f \Vert \leq  \Vert ( H(0) - \mu_0 ) f \Vert. $$
Thus,
\begin{align*}
(\inf _{\mu \not= \mu_0} \mu - \mu_0)  \Vert \Pi ^{(2)} e^{-it H(g)} (\Psi_0  \otimes u_0)  \Vert
&\leq C \Vert ( H(0) - \mu_0 ) e^{-it H(g)} (\Psi_0  \otimes u_0)  \Vert \\[4mm]
& \leq C \Vert ( H(g) - \mu_0 ) e^{-it H(g)} (\Psi_0  \otimes u_0)  \Vert + {\cal O} (g) \\[4mm]
& = C \Vert ( H(g) - \mu_0  ) (\Psi_0  \otimes u_0)  \Vert + {\cal O} (g)  = {\cal O}(g).
\end{align*}
Indeed,  $( H(0) - \mu_0 ) (\Psi_0  \otimes u_0)= 0$.
There exists $F(t)\in {\cal H} _{\rm ph}$ satisfying,
$$ \Pi ^{(1)} e^{-it H(g)} (\Psi_0 \otimes u_0) = F(t) \otimes u_0. $$
We have,
\begin{align*}
 1 -  \Vert F(t) \Vert^2  &= 1 - \Vert  F(t) \otimes u_0\Vert ^2 = 1 - \Vert \Pi ^{(1)} e^{-it H(g)} (\Psi_0 \otimes u_0)\Vert ^2 \\[4mm]
 & =   \Vert \Pi ^{(2)} e^{-it H(g)} (\Psi_0 \otimes u_0)\Vert ^2 = {\cal O} (g). \end{align*}
Consequently,
$$ e^{-it H(g)} (\Psi_0 \otimes u_0) = F(t) \otimes u_0  + {\cal O}(g). $$
We then obtain (\ref{inv-u0}). The end of the proof is left unchanged.

\section{Rabi cycle.}

\subsection{Proof of    Theorem \ref{approx-Rab}. }

Let  $X$ be in ${\cal K}$ and thus commuting with
$H_{\rm mat}$. Take  $u$ and $v$ eigenfunctions of  $H_{\rm mat}$
with distinct eigenvalues  $\lambda $ and $\mu$ in $S_{\rm inf}$. Set  $\omega = \mu - \lambda \not= 0$.

 We start with the system   (\ref{Sys-Dif-Init})  and we approximate
    $S^{\rm mat} (t , g) X $ by $X$. The system  (\ref{Sys-Dif-Init})
  is then,
 \be\label{sys-diff-5}       \left (  \frac {d} {dt} - i \omega \right )
  <  ( S^{\rm mat} (t , g) X  ) u, v > =
  (ig)^2 <( L^{\omega} (t) X ) u , v> +
   < K (t,  g, \omega , X )  u, v >  \ee
where
\be\label{rabi-K}   K (t,  g, \omega , X ) =  R_1 (t,  g, \omega , X ) +  R_2 (t,  g, \omega , X ) +
 (ig)^2  L^{\omega} (t) ( S^{\rm mat} (t, g) X - X ). \ee
      Solving the system  (\ref{sys-diff-5}) gives with $\omega = \mu - \lambda \not= 0$, since $<X u, v > =0$,
\begin{align*} <  ( S^{\rm mat} (t , g) X  ) u, v > &=  (ig)^2  \int_0^t e^{i \omega (t-s) }
< (L ^{\omega} (s)  X )  u , v> ds  \\[4mm]
& +   \int_0^t e^{i \omega (t-s) } < K (s,  g, \omega , X ) u, v >  ds.
\end{align*}
Using (\ref{L-omega}), we have,
$$  \int_0^t e^{i \omega (t-s) } < (L ^{\omega} (s)  X )  u , v> ds $$
$$ = \int _{0< \sigma <s <t}  e^{i\omega (t-s + \sigma)}
< \Big ( A(-\sigma ) A(0) (I\otimes X) \Big ) (\Psi_0 \otimes u) ,
 (\Psi_0 \otimes v)>   d \sigma ds $$
$$ = \frac {1} {i\omega } \int _0^t
( e^{i \omega t } - e^{i \omega \sigma } )
< \Big ( A(-\sigma ) A(0) (I\otimes X) \Big ) (\Psi_0 \otimes u) ,
 (\Psi_0 \otimes v)> d \sigma $$
$$ =  \frac {1} {i\omega } < \Big (e^{i \omega t }L^{0} (t)X - L^{\omega} (t)X\Big )
 u, v >. $$
In the aim to obtain inequality (\ref{approx-Rab-2}), we write,
 $$  <  ( S^{\rm mat} (t , g) X  ) u, v > =  \frac {(ig)^2 } {i \omega }
 < \big (e^{i \omega t }L^{0} X - L^{\omega} X\big )
 u, v >  + < R(t, g, \omega , X) u, v > $$
 $$ R(t, g, \omega , X) = \int_0^t e^{i \omega (t-s) }  K (s,  g, \omega , X ) ds +
  \frac {(ig)^2 } {i \omega }e^{i \omega t }   < \Big ( ( L^{0} (t) X - L_{\infty} ^{0}X) u, v >  $$
 $$ -  \frac {(ig)^2 } {i \omega }  <  ( L^{\omega } (t) X - L_{\infty} ^{\omega}X) u, v >. $$
 We indeed get an approximation by a  $2 \pi /\omega $ periodic function.
  Let us bound all the error terms. We first bound the function 
 $K$ defined in  (\ref{rabi-K}).
  We use  estimates (\ref{f-majo-R1-st})(\ref{f-majo-R2-st}) together with
  Point i) of Propositions \ref{P-majo-R1} and  \ref{P-majo-R2}.
 Besides, from Proposition \ref{err-dis-3},
 $$ \Vert S^{mat} (t, g) X - X \Vert _{ {\cal L} ( W_2^{\rm mat}, W_0^{\rm mat} ) }
 \leq C g t \Vert X \Vert.$$
In view of (\ref{majo-LII}) (Point iii)  of Proposition \ref{limite-L}),
   for all $u$ and $v$ in $ {\cal H}_{\rm inf}$,
  $$ \left | <   (L^{\omega}(t) ( S^{mat} (t, g) X - X) ) u, v > \right |
  \leq K g t \Vert X \Vert  \ \Vert u \Vert \ \Vert v \Vert. $$
As a consequence,
  \be\label{majo-K}  | <  K (s,  g, \omega , X )u, v > | \leq C g^3 (1+t^2) \Vert X 
  \Vert  \ \Vert u \Vert \ \Vert v \Vert. \ee
 Since $X$ lies in ${\cal K}$ and thus lies in ${\cal L}({\cal H}_{\rm inf})$ then Point ii) of Proposition
   \ref{limite-L} implies that, 
  \be\label{majo-L}  \left |  <  ( L^{0} (t) X - L_{\infty} ^{0}X) u, v > \right |
   \leq \frac {K} {1+t^{2p} }
   \Vert X  \Vert  \   \Vert u  \Vert    \Vert v  \Vert , \ee
  and similarly for  $ L^{\omega} (t) X$. Therefore,
  $$ | <  R(t, g, \omega , X) u, v > | \leq  C \Vert X \Vert \ \Vert u \Vert
  \ \Vert v \Vert \ \left ( g^3 ( t+ t^3) + \frac {g^2} {1+ t^2} \right ).
   $$
  Theorem \ref{approx-Rab} is then derived.
  
  \fpr

  \subsection{Proof of  Theorem \ref{approx-Rab-3}. }

  We solve  the system (\ref{Sys-Dif-Init})  on the interval $[t , t+h _{\lambda \mu}] $ 
  ($h_{\lambda \mu} = 2\pi /\omega$) still with approximating     $S^{\rm mat} (t , g) X $ by $X$.  We obtain in that case, 
$$ <  ( S^{\rm mat} (t+ h_{\lambda \mu} , g) X  ) u, v >
- <  ( S^{\rm mat} (t , g) X  ) u, v >  $$
$$ =  (ig)^2  \int_t^{t+ h_{\lambda \mu}}
 e^{i \omega (t+ h_{\lambda \mu} -s) }
< (L ^{\omega} (s)  X )  u , v> ds  $$
$$ +   \int_0^{t+ h_{\lambda \mu}}  e^{i \omega (t+ h_{\lambda \mu}-s) } < K (s,  g, \omega , X ) u, v >  ds.
$$
Note that,
$$ \int_t^{t+ h_{\lambda \mu}}
 e^{i \omega (t+ h_{\lambda \mu} -s) }
< (L_{\infty}  ^{\omega}  X )  u , v> ds  = 0.$$
As a consequence, we learn that,
$$ \left |<  ( S^{\rm mat} (t+ h_{\lambda \mu} , g) X  ) u, v >
- <  ( S^{\rm mat} (t , g) X  ) u, v > \right | $$
$$ \leq \sup _{t< s  < (t+ h_{\lambda \mu}}
| < K (s,  g, \omega , X ) u, v >  | + g^2
| < ( L  ^{\omega}(s)  X - L_{\infty}  ^{\omega}  X )  u , v> | .$$
From  (\ref{majo-K}) and (\ref{majo-L}), the proof is finished.

\fpr

 \subsection{$N$-body Rabi cycle. }

We consider in this section the case of a system of $N$ particles. The purpose is to highlight an interaction between particles  coming specifically from the QED set-up.

The particle Hilbert space is thus the skew-symmetric tensor product,
  \be\label{2C-esp} {\cal H} _{\rm mat } = \Lambda ^N {\cal H} _{\rm mat }^{(1)} \ee
with ${\cal H} _{\rm mat }^{(1)} = L^2(\R^3)$.

Definition \ref{dGamma} below will be   used to define several operators
   in $\Lambda ^N {\cal H} _{\rm mat }^{(1)}$.

   \begin{defi} \label{dGamma} (i) Let $A$ be an operator in a  Hilbert space ${\cal H}$.
 The operator $d \Gamma _1 (A)$ in $\Lambda ^N {\cal H}$ is defined by,
 $$ d \Gamma _1 (A) ( u_1 \wedge \cdots \wedge u_N )  =  \frac {1} {N!}
 \sum _{\varphi \in {\cal P}_N} { \rm sgn} (\varphi)
 (A u_{\varphi (1)} ) \wedge   u_{\varphi (2)} \cdots \wedge   u_{\varphi (N) }, $$
 for all  $u_1, \dots, u_N$ in ${\cal H}$ and with ${\cal P}_N$ being the  set  of bijections in $\{ 1 , \cdots N \}$.

 ii)  Set $B$ an operator in ${\cal H}\wedge {\cal H}$.
  For $N\geq 2$, define the operator $d \Gamma _2 (B)$ in $\Lambda ^N {\cal H}$
   by,
 $$ d \Gamma _2 (B) ( u_1 \wedge \cdots \wedge u_N )   = \frac {1} {N!} \sum _{\varphi \in {\cal P}_N }
  { \rm sgn} (\varphi)
 (B ( u_{\varphi (1)}\wedge u_{\varphi (2)} ) ) \wedge   u_{\varphi (3)} \cdots \wedge   u_{\varphi (N) }, $$
for every  $u_1,\dots, u_N$ in ${\cal H}$.

  \end{defi}

  The operator denoted above $d \Gamma_2 (B)$ is the same operator
called  Wick$ (B) $ and  defined in (2.13) in the work of  Z. Ammari \cite{A04}
which is in a more general framework. See also \cite{MR04}.

The particle Hamiltonian operator for the $N$-body system is,
 $$ H_{\rm mat } = d \Gamma _1 ( H_{\rm mat}^{(1)} ) $$
where $ H_{\rm mat}^{(1)} $ is the one-body particle. We do not write here the interaction potential: it should be without influence since it is multiplied by a factor $g^2$.

The interaction Hamiltonian $H_{\rm int}$ is defined as in  (\ref{H-int-def}) by,
 $$ E(k) = d\Gamma _1 ( E^{(1)} (k)),$$
where   $E^{(1)} (k)$ is the one-body form factor given by (\ref{dipol}).

The  photon  Hamiltonian $H_{\rm ph}$ is the same as before.

We denote by ${\cal K}^{(1)}$ the operator algebra of  Definition \ref{K-P}
in the case of one particule and ${\cal K}^{(N)}$ denotes its counterpart  for $N$ particles.
In this setting, we can define as in  (\ref{L-omega})
    and (\ref{Lt-LI}) an operator $L_{\infty}^{\omega}Z$ for
    every  $Z$ in ${\cal K}^{(N)}$ and any $\omega \in \R$.
 We shall compute in this situation
    the operator $L_{\infty}^{\omega}Z$ when
    $Z = d \Gamma _1X $ with  $X $
    being an operator in ${\cal K}^{(1)}$.

  \begin{theo}\label{Rabi-N}  Fix an operator
  $X $ in ${\cal K}^{(1)}$ where  ${\cal K}^{(1)}$ is the operator algebra defined in Section 1 in the case of a single particle.
 Then, the operator $L_{\infty }^{\omega} (d\Gamma _1 X)$ defined
  in  $\Lambda ^N {\cal H} _{\rm mat}$ verifies,
  $$   L_{\infty}^{\omega }   (d\Gamma _1 X   )  =
  d\Gamma _1 ( L_{\infty}^{\omega }X)  + d\Gamma _2 C, $$
where $L_{\infty}^{\omega }X $ is the operator for a single (isolated) particle and $C$ is the operator in
  $\Lambda ^2 {\cal H} _{\rm mat}^{(1)} $ given by,
  \be\label{compos}  C = \lim _{t\rightarrow \infty}   \int _{\R^3 \times (0, t) }
e^{i \omega s}  ( e^{i s |k|} A(k) -  e^{- i s |k|} B(k) ) dk ds, \ee
with
$$ A(k) ( u_1 \wedge u_2) =  ( E^{(1) free} (k, - s)^{\star} u_1  )
\wedge     ( [  E^{(1)} (k) ,  X] u_2  ) -
  ( E^{(1) free} (k, - s)^{\star} u_2  ) \wedge
   ( [  E^{(1)} (k) ,  X] u_1  )  $$
and
$$ B(k) ( u_1 \wedge u_2) =   ( [ E^{(1) \star} (k) ,  X] u_1  ) \wedge
 ( E^{(1) free} (k, - s)  u_2  ) -
 ( [ E^{(1) \star} (k) ,  X] u_2  ) \wedge   ( E^{(1) free} (k, - s)  u_1  ) . $$

  \end{theo}

When $B$ is a operator in ${\cal H}\wedge {\cal H}$,
   the operator $ d\Gamma _2 B $ reflects the two-body  interaction between particules. For example, identify ${\cal H} _{\rm mat}^{(1)}
  \wedge {\cal H} _{\rm mat}^{(1)} $ with the space of functions
  $F=F(x , y)$ in $L^2(\R^6)$ with $F(y , x) = - F(x , y)$, let
   $\Phi $ be  a bounded function  on $\R^3$ and
   set  the operator  $W$ in ${\cal H} _{\rm mat}^{(1)}
  \wedge {\cal H} _{\rm mat}^{(1)} $ defined by,
  $$  ( W ( u\wedge v)  ) (x , y) = \Phi (|x-y|) \ ( u(x) v(y) - u(y) v(x)). $$
 Then, one notices that  $d\Gamma _2(W)$ represents
  the sum of the two-body  interactions for a system of $N$ particles. The presence of the term $d\Gamma _2 C$ suggests therefore a two-body interaction, only revealed within the QED framework, from which is emerging  the Rabi cycle.

The proof of Theorem \ref{Rabi-N} uses the following standard Lemma (see also \cite{A04} for similar considerations).

\begin{lemm}\label{commut-DG}  For all operators  $A$ and $B$ in the Hilbert space
 ${\cal H}$, we have in the Hilbert space $\Lambda ^N {\cal H}$,
$$ d\Gamma _1( A ) \circ d\Gamma _1 (B ) = d\Gamma _1 ( AB)
+ d\Gamma _2  C (A , B) $$
where $C(A , B)$ is the operator in  $\Lambda ^2 {\cal H}$
defined by,
$$ C(A , B) ( u_1 \wedge u_2) = Au_1 \wedge B u_2 - A u_2 \wedge Bu_1. $$
In particular, the identity
$$ [ d\Gamma _1( A ) \ ,\  d\Gamma _1 (B )]
= d \Gamma _1 ( [A , B] ) $$
 holds true.

\end{lemm}

 {\it Proof of  Theorem \ref{Rabi-N}.}  From (\ref{integ-L-t}), one has,
\begin{align*} L_{\infty}^{\omega} d\Gamma_1  X  = \lim _{t\rightarrow \infty} \int _{\R^3 \times (0, t) }
&e^{i \omega s}  \Big ( e^{i s |k|}
 d \Gamma _1 E^{(1) free} (k, - s)^{\star}\ [ d\Gamma _1 E^{(1)} (k) , d\Gamma _1 X ] \\[4mm]
  -& e^{-is  |k|} [ d\Gamma _1 E^{(1) \star} (k) , d\Gamma _1 X]
\  d\Gamma _1 E^{(1) free} (k, - s) \Big ) dk ds. \end{align*}
With the help of Proposition \ref{commut-DG}, one gets,
$$ [ d\Gamma _1 E^{(1)} (k) \  ,\  d\Gamma _1 X] = d\Gamma _1  (
[  E^{(1)} (k) , X]  ) $$
and
 \begin{align*} L_{\infty}^{\omega} d\Gamma_1  X  = \lim _{t\rightarrow \infty}  d \Gamma _1
 \int _{\R^3 \times (0, t) }
e^{i \omega s}  \big ( &e^{i s |k|}
  ( E^{(1) free} (k, - s)^{\star}\   [  E^{(1)} (k) ,  X]  ) \\[4mm]
   - &e^{-is  |k|}  \Big (  [ E^{(1) \star} (k) ,  X]
\   E^{(1) free} (k, - s)  )  \big ) dk ds  +  d \Gamma _2 C,  \end{align*}
where $C$ is the operator given by (\ref{compos}).  Theorem \ref{Rabi-N}
is therefore proved.

\fpr

    \section{Non Markovian approximation. }

Theorem \ref{ANM} is proved in this section.   Remind that the marginal transition probability  $P(t, g, u_j, u_m)$ is defined by  (\ref{P-trans}), namely,
$$   P(t, g, u_j, v) = <  ( S^{\rm mat} (t , g) \pi_{u_m}   ) u_j , u_j >. $$
In order to examine this transition probability, we start from the system (\ref{Sys-Dif-Init}) with $\omega = 0$
(recall that the index $\omega$ is often omitted from the notation when it is zero),
$$   \frac {d} {dt}  <  ( S^{\rm mat} (t , g) \pi_{u_m}  ) u_j, u_j > =
  (ig)^2 < L (t)(S^{\rm mat} (t , g) \pi_{u_m} ) u_j , u_j>  \hskip 2cm $$
  $$
  \hskip 4cm + R_1 (t,  g,  \pi_{u_m} , u_j, u_j )  + R_2 (t,  g,   \pi_{u_m} , u_j, u_j ), $$
where $R_1$ is given by  (\ref{defi-R1}) and  $R_2$ by (\ref{defi-R2}).
 In the above right hand side, we approximate $S^{\rm mat} (t , g) \pi_{u_m} $
by $\pi_{u_m}$ and therefore we define an additional error term,
  $$  R_{10} (t, g,\pi_{u_m} , u_j, u_j ) =  (ig)^2
  < (  L  ( t) ( S^{\rm mat} (t , g)\pi_{u_m} - \pi_{u_m} )  ) u_j , u_j > ds.  $$
 The error coming from this term is certainly high for large $t$. In contrast to the Markov approximation, it will not be used as $t$ goes to infinity. Nevertheless, it can be more precise than the Markov approximation for some values of $t$ and $g$.

 The  system is now written as,
 $$ \frac {d} {dt}  <  ( S^{\rm mat} (t , g) \pi_{u_m}  ) u_j, u_j > =
  (ig)^2 < (L (t) \pi_{u_m} ) u_j , u_j>   \hskip 4cm. $$
  $$  \hskip 4cm
  + R_1 (t,  g,  \pi_{u_m} , u_j, u_j )  + R_2 (t,  g,   \pi_{u_m} , u_j, u_j )
  +   R_{10} (t, g,\pi_{u_m} , u_j, u_j ). $$
One gets using $\pi_{u_m} u_j = 0$,
 $$<  ( S^{\rm mat} (t , g) \pi_{u_m}  ) u_j, u_j > = (ig)^2 \int _0^t
   < (L (s) \pi_{u_m} ) u_j , u_j>  ds   \hskip 4cm$$
 $$   \hskip 2cm+ \int _0^t  \big ( R_1 (t,  g,  \pi_{u_m} , u_j, u_j )  + R_2 (t,  g,   \pi_{u_m} , u_j, u_j )
  +   R_{10} (t, g,\pi_{u_m} , u_j, u_j ) \big ) ds. $$
We begin with making explicit the first term in the  above right hand side.
Since $\pi _m u_j = 0$, one obtains,
$$ <  (  L^{0} (s) \pi_{u_m}  ) u_j , u_j >  = - 2  \int _{\R^3 \times (0, s) }
\cos ( \sigma ( |k| + \lambda _m - \lambda _j)) \ | < E(k) u_j , u_m > |^2 dk d\sigma,  $$
and consequently,
 \begin{align*}  \int _0^t < ( L^0 (t-s) & \pi _{u_m} ) u_j , u_j >    ds  \\[4mm]
 &=
-2   \int _{\R^3 } \int _{0 < \sigma < s < t}
\cos ( \sigma ( |k| + \lambda _m - \lambda _j)) \ | < E(k) u_j , u_m > |^2 dk d\sigma ds  \\[4mm]
 &= 2 \int _{\R^3} \frac { 1  - \cos ( t ( |k| + \lambda _j - \lambda _m )) }
  { ( |k| + \lambda _j - \lambda _m )^2 } \ | < E(k) u_j , u_m > | ^2 dk.  \end{align*}
Next, we give norm estimates of the three error terms. Recall that
since $u_m$ is an eigenvector
associated with an eigenvalue in $S_{\rm inf}$, the orthogonal
projection $\pi _{u_m}$ is in ${\cal K}$. Therefore
the terms $R_1$ and $R_2$  are estimated in (\ref{f-majo-R1-st})  and  (\ref{f-majo-R2-st}),
with $X = \pi _{u_m}$ and $\omega = 0$. We don't use the dipole approximation.
Recall that,
 \be\label{ANM-R1}  |  R_{1}(s, g, \pi_{u_m} , u_j, u_j ) |  + |  R_{2}(s, g, \pi_{u_m} , u_j, u_j ) |
 \leq C g^ 3 (1+ s^2 ) . \ee
Point iii) of Proposition \ref{limite-L} shows that,
$$  |  R_{10}(s, g, \pi_{u_m} , u_j, u_j ) | \leq  C g^2
\Vert  S^{\rm mat} (t , g)\pi_{u_m} - \pi_{u_m} \Vert _{ {\cal L}
( W_2^{\rm mat}, W_0^{\rm mat} ) } .$$
From Proposition \ref{err-dis-3},
\be\label{ANM-R10}  |  R_{10}(s, g, \pi_{u_m} , u_j, u_j ) | \leq  C g^3t .\ee
 Theorem \ref{ANM} is proved by (\ref{ANM-R1})    and (\ref{ANM-R10}).

\fpr

\appendix

\section{Standard photon  identities. }

The next identity is used in Appendix C.
\be\label{H1/2} \int _{\R^3} |k| \Vert a(k) f \Vert ^2 dk =
\Vert H_{\rm ph} ^{1/2}f \Vert ^2. \ee

Recall that,
\be\label{evol-a-k}  e^{ i t H(0)    } ( a(k) \otimes I)  e^{ - i t H(0)    }=
  e^{ - i t |k|    } ( a(k) \otimes I).
  \ee

 and that $[ H_{\rm ph} , a(k) ] = - |k| a(k)$ (\cite{DG}). In particular,
 \be\label{Hph-ak}  ( H_{\rm ph} +|k| + 1)^{\alpha} a(k) = a(k)  ( H_{\rm ph}  + 1)^{\alpha}. \ee

Also recall that the adjoint $a^{\star} (k)$ is not a priori well defined for fixed $k$, but one gives a sense when used in some integrals (see \cite{RS}, Volume II).
In that situation, we will use the following formula,   
\be\label{evol-a-star-k}  e^{ i t H(0)    } ( a^{\star}(k) \otimes I)  e^{ - i t H(0)    }=
  e^{  i t |k|    } ( a^{\star}(k) \otimes I)
  \ee
   and also the identity,
\be\label{COMM}  \big [ (a(k) \otimes I) , \int _{\R^3 } ( a^{\star} (p)  \otimes \Phi (p) ) dp
\big ] = I \otimes \Phi (k),\ee
for every  $\Phi\in {\cal S}(\R^3)$
   (see  (I.3) in \cite{B-F-S}).

The last equality (\ref{COMM}) is often simply written  $[ a(k) , a^{\star} (p) ]
= \delta (k-p)$.

\section{Sobolev spaces  $W_m^{\rm tot}$. }

The purpose of this appendix is to prove Point $i)$ of
 Theorem
\ref{Kato-Rell} and to give some properties of the  $W_m^{\rm tot}$ spaces.

\begin{theo} \label{t-dom}Assume that there is $C>0$ such that  $H_{\rm mat }+ C I >0$.
Then there exists a  semi-bounded self-adjoint  operator $( H(0), D(H(0))$ in
${\cal H} _{\rm tot}$   satisfying  for all $f$ in ${\cal H} _{\rm tot}^{\rm reg}$,
$$ H(0) f = ( H_{\rm ph} \otimes I)  f +  ( I \otimes  H_{\rm mat})f. $$
 The domain of $ ( C + H(0))^{m/2}$ endowed with its natural norm is denoted $W_m^{\rm tot}$. Then, the operator $e^{it H(0)}$ is uniformly
bounded in $W_m^{\rm tot}$. When $m$ is an even integer,
 there exists $C_m>0$ satisfying,
 \be\label{equiv-norm}  \frac {1} {C_m}\Vert f \Vert _{W_m^{\rm tot}} \leq \sum _{p+q \leq m /2}
 \Vert ( H_{\rm ph}^p \otimes ( C+  H_{\rm mat})^q) f \Vert \leq C_m  \Vert f \Vert _{W_m^{\rm tot}}. \ee

\end{theo}

{\it Proof of Theorem \ref{t-dom}. }
Consider the two  following self-adjoint operators $(H , D(H))$ and
  $(H' , D(H'))$   in ${\cal H} _{\rm tot}$,
 $$  H = H_{\rm ph}\otimes I ,\quad H'=  I \otimes H_{\rm mat}. $$
The spectral projections of $H$ and $H'$ commute, thus, we can use the results of  \cite{SCHM} and \cite{BI-SO} (see also \cite{PUT}).
Let $\mu_H $ (resp.  $\mu_{H'} $) be the spectral measure of $H$ (resp. of $H'$)
 which is a measure on  $\R$ with values in  ${\cal L} ( {\cal H} _{\rm ph})$
 (resp.  in ${\cal L} ( {\cal H} _{\rm mat})$). Then, from Theorem 1 in  \cite{BI-SO} or  Theorem 5.21 of   \cite{SCHM}, $\mu_H  \otimes \mu_{H'} $
 is a measure on $\R^2$ with values  in  ${\cal L} ( {\cal H} _{\rm tot})$.
This measure  maps any Borel set $F$ of $\R^2$ written as   $F= E \times E'$ to
 the operator $\mu_H (E) \otimes \mu _{H'} (E')$, which is an  operator
 in ${\cal H }_{tot}$.  Let $\varphi :\ \R^2 \rightarrow \R$ be  a real-valued Borel function,  not necessarily bounded.
Define  an operator $\varphi (H , H')$ by,
$$ \varphi (H,H' )=\int_{\R^2}   \varphi(\lambda _1,\lambda _2) d (\mu_H \otimes \mu_{H'} ) (\lambda _1 , \lambda_2). $$
One knows that  $\varphi (H , H')$ is self-adjoint on the domain given by all $f \in {\cal H} _{\rm tot}$ with,
$$ \int_{\R^2}  |\varphi (\lambda _1,\lambda _2)|^2   < d(\mu_H\otimes \mu_{H'} )(\lambda _1, \lambda _2) f  ,  f >
< \infty.  $$
In particular, if  $\varphi (\lambda _1 , \lambda _2)= \lambda _1 + \lambda _2$ then the  operator $\varphi  (H, H')$ will be denoted $H(0)$. This is the Pauli-Fierz operator  with $g= 0$, that is without interaction between particles and photons. It is not necessary that  $\varphi $ is defined everywhere. Set $C>0$ with
$  H_{\rm mat } + CI >0$. For each $m \in \R$,  let $\varphi_m (\lambda_1 , \lambda _2)
= ( C + \lambda _1 + \lambda _2 ) ^{m/2}$. It can be extend by  0 if  $\lambda _1 <0$
or if $C+ \lambda _2 <0$. If  $m\geq 0$, let  $W_m ^{\rm tot}$  stands for the domain
of the  operator $\varphi _m (H , H')$. It is standard that the  operator $e^{it H(0)}$ is uniformly
bounded in $W_m^{\rm tot}$. In  (\ref{equiv-norm}), the first inequality comes from
the binomial formula. The second inequality is a consequence of the fact that,
if $p+ q \leq m/2$ then the function  $\lambda _1 ^ p ( C+ \lambda _2) ^q \varphi_{-m }
(\lambda _1 , \lambda _2)$ is bounded on $\R^2$.

\fpr

\begin{prop}\label{norme-tenso}
 Fix  $m$ and $q$  two nonnegative integers.
    Take an operator $A$ bounded from  $W_{m+q} ^{\rm mat}$ to $W_m ^{\rm mat}$
     and from  $W_{q}  ^{\rm mat}$ to $W_0 ^{\rm mat}$. Then,  $I\otimes A$ is bounded from
     $W_{m+q}  ^{\rm tot}$ to $W_m ^{\rm tot}$.  In addition, there exists $K>0$ satisfying,
     $$ \Vert I \otimes A \Vert _{ {\cal L} (W_{m+q} ^{\rm tot} , W_m ^{\rm tot})}  \leq K (
     \Vert A \Vert _{ {\cal L} (W_{m+q} ^{\rm mat} , W_m ^{\rm mat})} +
     \Vert A \Vert _{ {\cal L} (W_{q}  ^{\rm mat} , W_0 ^{\rm mat})}  ). $$

      \end{prop}

 {\it Proof of Proposition \ref{norme-tenso}. }  From (\ref{equiv-norm}), one has,
 $$ \Vert ( I \otimes A) f \Vert _{ W_m ^{\rm tot}} \leq C_m
 \sum _{ \alpha + \beta \leq m/2} \Vert ( H_{\rm ph}^{\alpha}  \otimes H_{\rm mat}^{\beta} ) (I \otimes A) f \Vert. $$
 Under our hypothesis, by  interpolation, $A$ is bounded from
 $ W_{ 2 \beta + q }^{\rm mat}$ to  $ W_{ 2 \beta  }^{\rm mat}$ if  $0 \leq \beta \leq m/2$.
That is, $  H_{\rm mat}^{\beta} A H_{\rm mat}^{-\beta  - q/2}  $ is bounded in ${\cal H }_{\rm mat}$.
Therefore,
  $$ \Vert ( I \otimes A) f \Vert _{ W_m ^{\rm tot}} \leq C_m
 \sum _{ \alpha + \beta \leq m/2} \Vert ( I \otimes  H_{\rm mat}^{\beta} A H_{\rm mat}^{-\beta  - q/2}  )
 \Vert _{{\cal L}({\cal H} _{\rm tot})}\ \Vert  ( H_{\rm ph}^{\alpha}  \otimes H_{\rm mat}^{\beta + q/2} ) f
 \Vert.   $$
Using again  (\ref{equiv-norm}), one  obtains  Proposition \ref{norme-tenso}.

\fpr

\begin{prop}\label{norme-sigma}  One has,
 $$ \Vert \sigma_0 Z \Vert _{ {\cal L} ( W_{m+p}^{\rm mat},  W_{m}^{\rm mat}) }
 \leq C  \Vert Z \Vert _{ {\cal L} ( W_{m+p}^{\rm tot},  W_{m}^{\rm tot} ) }. $$

\end{prop}

{\it Proof of Proposition \ref{norme-sigma}.} For all $u$ and $v$ in ${\cal S}(\R^3)$, one has,
\begin{align*} < (\sigma _0 Z) u , (C + H_{\rm mat})^{m/2} v > &=
 < Z ( \Psi_0 \otimes u) , \Psi_0 \otimes  (C + H_{\rm mat})^{m/2} v >
 \\[4mm]
 & = < ( C + H(0))^{m/2} Z ( \Psi_0 \otimes u) , \Psi_0 \otimes v > .
 \end{align*}
Thus,
\begin{align*} | < (\sigma _0 Z) u , (C + H_{\rm mat})^{m/2} v > |  &\leq
  \Vert ( C + H(0))^{m/2} Z ( \Psi_0 \otimes u) \Vert \ \Vert v \Vert
   \\[4mm]
   & \leq \Vert  Z ( \Psi_0 \otimes u) \Vert  _{W_m ^{\rm tot} }\ \Vert v \Vert
   \leq \Vert Z \Vert _{ {\cal L} ( W _{m+p} ^{\rm tot} ,  W _{m} ^{\rm tot}
   ) } \ \Vert u \Vert _{W _{m+p} ^{\rm tot} } \ \Vert v \Vert.
    \end{align*}
Proposition \ref{norme-sigma} then follows.

\fpr

 \section{The operators $H(g)$ and $e^{itH(g)}$. }

 In this section, we prove  points $ii)$ and $iii)$  of  Theorem \ref{Kato-Rell}.

    {\it Point ii).} In view of (\ref{equiv-norm}), it suffices to check that,
   for all $f$ and $g$ in ${\cal H} _{\rm tot}^{\rm reg} $,
     for any integers  $p\geq 0$ and $q\geq 0$,
     \be\label{majo-I}  | Q_{\rm int}  (f, ( A_p  \otimes B_q ) g ) |
      \leq C \Vert f \Vert _{  W  _{p+q + 2 } ^{\rm tot} } \ \Vert g \Vert, \ee
 where
     $$ A_p = (H_{\rm ph} +1)  ^{p/2} ,\quad  B_q = ( C+ H_{\rm mat}) ^{q/2} $$
   with $C$  such that $C+ H_{\rm mat} >0$.  One has,
     $$  Q_{\rm int}  (f,  (A_p \otimes B_q) g ) = I_1 + I_2 $$
     with,
     $$ I_1 = \int _{\R^3} < ( I \otimes E(k) ) f \ , \
      ( a(k) A_p \otimes B_q  ) g > dk $$
     and
     $$ I_2 = \int _{\R^3} < ( a(k) \otimes E^{\star} (k)) f \ , \
      ( A_p  \otimes B_q) g > dk. $$
    From (\ref{Hph-ak}),  one notes that,
     $$ a(k) (H_{\rm ph} + I)^{p/2} = (H_{\rm ph} +|k| +  I)^{(p+1)/2}
     a(k) (H_{\rm ph} + I)^{-1/2}. $$
Therefore,
     $$ I_1 =  \int _{\R^3}  <  \big (  (H_{\rm ph} +|k| +  I)^{(p+1)/2} \otimes
     (C+ H_{\rm mat})^{q /2} E(k)  \big ) f \ ,\ \big ( a(k) (H_{\rm ph} + I)^{-1/2} \otimes I \big ) g > dk.  $$
  Using (\ref{H1/2}), one sees,
     $$  \int _{\R^3}  |k| \Vert  ( a(k) (H_{\rm ph} + I)^{-1/2} \otimes I  ) g \Vert ^2 dk
     \leq C \Vert g\Vert ^2. $$
Consequently,
 \begin{align*}
  |I_1 | ^2 &\leq C \Vert g\Vert ^2  \int _{\R^3}
     \Vert  \big (  (H_{\rm ph} +|k| +  I)^{(p+1)/2} \otimes
     (C+ H_{\rm mat})^{q /2} E(k)  \big ) f \Vert ^2 \frac {dk} {|k|}\\[4mm]
     &\leq  C \Vert g\Vert ^2  \int _{\R^3}
      \Vert  \big (  (H_{\rm ph} + I)^{(p+1)/2} \otimes
     (C+ H_{\rm mat})^{q /2} E(k)  \big ) f \Vert ^2 \frac {dk} {|k|}  \\[4mm]
     &+  C \Vert g\Vert ^2  \int _{\R^3} ( 1+ |k|)^{p+1}
      \Vert  \big ( I \otimes (C+ H_{\rm mat})^{q /2} E(k)  \big ) f \Vert ^2 \frac {dk} {|k|}.
       \end{align*}
    To bound the first term, one uses the operator $C_q (k) $ defined by,
     $$ C_q(k) = (C+ H_{\rm mat})^{q /2} E(k)  (C+ H_{\rm mat})^{-(q+1) /2}. $$
    Using the expression  (\ref{stand}) or (\ref{dipol}) of $E(k)$,
     one sees that the operator $C_q(k) $ is bounded in ${\cal H} _{\rm mat}$ with a norm satisfying,
     $$ \Vert C_q (k) \Vert \leq C (1+|k|)^{-N}.$$
Thus,
     $$ \int _{\R^3}
      \Vert  \big (  (H_{\rm ph} + I)^{(p+1)/2} \otimes
     (C+ H_{\rm mat})^{q /2} E(k)  \big ) f \Vert ^2 \frac {dk} {|k|}
     \leq C \Vert  (  A_{p+1}  \otimes
     B_{q+1}   ) f \Vert ^2. $$
Similarly,
     $$ \int _{\R^3} ( 1+ |k|)^{p+1}
      \Vert  \big ( I \otimes (C+ H_{\rm mat})^{q /2} E(k)  \big ) f \Vert ^2 \frac {dk} {|k|}
      \leq C  \Vert   (  I \otimes
     B_{q+1}  ) f \Vert ^2. $$
From (\ref{equiv-norm}), one deduces that,
      $$   |I_1| \leq C \Vert g\Vert \  \Vert f \Vert _{  W  _{p+q + 2 } ^{\rm tot} }. $$
Besides, one has,
     $$  | I_2 | \leq \Vert g \Vert \  \int _{\R^3} \left  \Vert \big (
      (H_{\rm ph} + I)^{p/2} a(k) \otimes ( C+ H_{\rm mat} )^{q/2} E^{\star } (k) \big )f \right \Vert dk. $$
Consequently,
      \begin{align*}
        | I_2 |^2  &\leq \Vert g \Vert^2  \   \int _{\R^3} (1+ |k|)^4 |k|  \left  \Vert  (
       (H_{\rm ph} + I)^{p/2} a(k) \otimes ( C+ H_{\rm mat} )^{q/2} E^{\star } (k)  )f \right \Vert^2  dk \\[4mm]
    &    \leq C \Vert g \Vert \  \int _{\R^3} (1+ |k|)^4 |k|  \left  \Vert  (
     (H_{\rm ph} + |k| +  I)^{p/2} a(k) \otimes ( C+ H_{\rm mat} )^{q/2} E^{\star } (k)  )f \right \Vert^2  dk  \\[4mm]
     & + C \Vert g \Vert \  \int _{\R^3} (1+ |k|) ^{(p+8) /2} |k|  \left  \Vert  (
      a(k) \otimes ( C+ H_{\rm mat} )^{q/2} E^{\star } (k)  ) f \right \Vert^2 dk.
       \end{align*}
      One uses  (\ref{Hph-ak}), (\ref{H1/2})  and the  adjoint of the operator $C_q(k)$ defined above. As before, one obtains,
     $$  | I_2 | \leq \Vert g \Vert   \  \Vert f \Vert _{  W  _{p+q + 2 } ^{\rm tot} }. $$
  Therefore  (\ref{majo-I}) is derived and Point $ii)$
       of  Theorem \ref{Kato-Rell} is then deduced.

{\it Point iii).}  One notices that the  operator  $(H(g)^{m/2} - H(0)^{m/2}) f$ is a polynomial function in $g$ with all the terms being of degree greater or equal than $1$ and with coefficients that
 are bounded   operators from  $W _ {m} ^{\rm tot}$ to ${\cal H} _{\rm tot}$.
Thus,  for every  $m\geq 1$, there  exists $C_m >0$ such that, for
  any  $f$ in ${\cal H} _{\rm tot}^{\rm reg}$ and for all $g$
  in $(0, 1)$,
  $$ \Vert (H(g)^{/2} - H(0)^{m/2}) f \Vert \leq C_m g   \Vert f \Vert _{W _ {m} ^{\rm tot}}. $$
 Point $iii) $ then follows from   Kato Rellich Theorem.

 \fpr

      laurent.amour@univ-reims.fr\newline
Laboratoire de Math\'ematiques de Reims UMR CNRS 9008,\\ Universit\'e de Reims Champagne-Ardenne
 Moulin de la Housse, BP 1039,
 51687 REIMS Cedex 2, France.

jean.nourrigat@univ-reims.fr\newline
Laboratoire de Math\'ematiques de Reims UMR CNRS 9008,\\ Universit\'e de Reims Champagne-Ardenne
 Moulin de la Housse, BP 1039,
 51687 REIMS Cedex 2, France.

\end{document}